\documentclass[a4paper,12pt,final,reqno,oneside]{amsart}
\usepackage{amsmath}
\usepackage{amsfonts}
\usepackage{amssymb}
\usepackage{amsthm}
\usepackage[matrix,arrow,curve]{xy}
\usepackage[dvips]{graphicx}
\usepackage{accents}
\CompileMatrices
\theoremstyle{plain}
\newtheorem*{classification-of-covering-spaces-theorem}{Classification of Covering Spaces Theorem}
\newtheorem*{existence-theorem}{Existence Theorem}
\newtheorem*{classification-theorem}{Classification Theorem}
\newtheorem*{conj}{Conjecture}
\newtheorem{prop}{Proposition}[section]
\newtheorem{lem}[prop]{Lemma}
\newtheorem*{cor}{Corollary}                % all corollaries are unnumbered
\theoremstyle{definition}
\newtheorem{defn}[prop]{Definition}         % numbered definition
\newtheorem*{defn*}{Definition}             % unnumbered definition
\newcommand{\defnemph}[1]{\textbf{#1}}
\newenvironment{rem}{\noindent\textsl{Remark.}}{}  % perhaps looks better than rem above?
\newtheorem*{example}{Example}
\numberwithin{equation}{section}
\newcommand{\set}[1]{\left\{#1\right\}}
\newcommand{\setc}[2]{\left\{#1 \; \left| \; #2 \right. \right\}}
\newcommand{\Real}{\mathbb R}
\newcommand{\RP}{\mathbb R \mathbb P}
\newcommand{\Complex}{\mathbb C}
\newcommand{\Quaternion}{\mathbb H}                 % Hamilton's numbers.

\newcommand{\Integer}{\mathbb Z}
\newcommand{\Field}{\mathbb F}
\newcommand{\eps}{\varepsilon}
\newcommand{\ceps}{\cc{\eps}}
\newcommand{\To}{\rightarrow}
\newcommand{\LA}[1]{\mathfrak{#1}}
\newcommand{\LAM}[2]{\mathfrak{\lowercase{#1}}(#2)}
\newcommand{\SL}{SL(2,\Complex)}
\newcommand{\LL}{SO_0(1,3)}
\newcommand{\subgroup}{\preceq}
\newcommand{\closure}[1]{\overline{#1}}
\newcommand{\cc}[1]{\overline{#1}}
\newcommand{\Ad}{\operatorname{Ad}}
\newcommand{\Aut}{\operatorname{Aut}}
\newcommand{\Hom}{\operatorname{Hom}}
\newcommand{\Ext}{\operatorname{Ext}}

\newcommand{\image}{\operatorname{image}}
\newcommand{\trace}{\operatorname{tr}}
\newcommand{\id}{\operatorname{id}}
\newcommand{\vf}[1]{\mathfrak{X}(#1)}               % vector fields
\newcommand{\tensors}[2]{\mathcal{T}\!\begin{smallmatrix}#1 \\ #2\end{smallmatrix}}
\newcommand{\spinors}[4]{\mathcal{S}\!\begin{smallmatrix}#1 & #2 \\ #3 & #4\end{smallmatrix}}
\newcommand{\gi}[1]{\mathfrak{#1}}
\newcommand{\ci}[1]{#1}                 % component index
\newcommand{\cgi}[1]{\ci{\gi{#1}}}
\newcommand{\ai}[1]{\pmb{\boldsymbol{#1}}}          % super-heavy bold indices!
\newcommand{\agi}[1]{\ai{\gi{#1}}}
\newcommand{\ggerman}{gothic }

\newcommand{\rroman}{Roman }
\newcommand{\restrict}[2]{#1{}_{\mid #2}{}}
\newcommand{\dat}[2]{\frac{d}{d#1}_{\mid_{#1=#2}}}
\newcommand{\sh}[1]{#1{}^{\sharp}{}}                % superscript sharp
\newcommand{\na}[1]{#1{}^{\natural}{}}              % superscript natural
\newcommand{\bundle}[3]{#1 \xrightarrow{#2} #3}
\newcommand{\pfbundle}[4]{#1 \rightsquigarrow #2 \xrightarrow{\smash[t]{#3}} #4}
\newcommand{\dd}{\mathbf{d}}
\newcommand{\bdelt}[1]{\mathsf{#1}}                 % formatting for an element of a pfb.
\newcommand{\valence}[2]{\left[\begin{smallmatrix}#1 \\ #2 \end{smallmatrix}\right]}
\newcommand{\valences}[4]{\left[\begin{smallmatrix}#1 & #2 \\ #3 & #4 \end{smallmatrix}\right]}
\newcommand{\cat}{{\star}}
\newcommand{\compose}{{\circ}}
\newcommand{\I}{\left[0,1\right]}
\newcommand{\J}{\left[0,\eps\right]}
\newcommand{\GLn}{GL(n,\Real)}
\newcommand{\GLncover}[1]{\smash{\widetilde{GL}}^+(#1,\Real)}   % should we have a _0 rather than a ^+, for consistency? needs to be changed in other places too.
\newcommand{\noop}[1]{}
\newcommand{\psmallmatrix}[1]{\left(\begin{smallmatrix} #1 \end{smallmatrix}\right)}
\renewcommand{\tilde}[1]{\widetilde{#1}}
\renewcommand{\imath}{\mathfrak{i}}
\renewcommand{\jmath}{\mathfrak{j}}

\newcommand{\shortversion}[1]{}     % the short version can probably lose some appendices as well!
\newcommand{\longversion}[1]{#1}

\newcommand{\finalversion}[1]{#1}
\newcommand{\draftversion}[1]{}

\author{Kim Morrison}
\title{Classifying Spinor Structures}
\date{}

\begin{document}

\pagenumbering{roman}

\thispagestyle{empty}
\begin{center}
\textsc{The University of New South Wales} \\
\textsc{School of Mathematics} \\
\textsc{Department of Pure Mathematics} \\

\vspace{1in}

{\Large\textbf{Classifying Spinor Structures}} \\
\vspace{1.5cm}
by \\
\vspace{0.75cm}
\textbf{Kim Morrison} \\

\vspace{1cm}

\begin{figure}
\includegraphics[width=\textwidth,clip=true,trim=0 0 0 35pt]{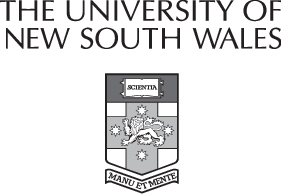}
\end{figure}

A thesis submitted for consideration in the degree of \\
Bachelor of Science with honours in pure mathematics at the \\
University of New South Wales. \\
June 2001 \\
\vspace{2cm}

Supervisor: Dr. J. D. Steele
\end{center}

\newpage
\thispagestyle{empty}

\vspace*{1in}
\begin{center}
\Large \textbf{Acknowledgements}
\end{center}
\vspace{1cm}

Firstly I would like to express my gratitude to John Steele, my
supervisor, for his guidance and assistance, and for the
considerable time he has invested in checking my drafts,
clarifying my prose, and especially in trying to make my seminar
make some sense!

A considerable number of members of the Mathematics Department
have made available their time and mathematical expertise, or
offered well considered academic advice over the past
year.\footnote{In particular Prof. M. Cowling, Dr. S. Disney,
Prof. T. Dooley and Dr. N. Wildberger, and outside the Department,
Dr J. Baez, UCR, and Dr. J. Hillman, Sydney.} Many thanks go for
this, and also for the friendly academic environment of the
Department, which has made working here a pleasure.

I would like to thank all my friends, who distracted me when I
needed to be distracted, and allowed me to work when I needed to
work. Finally, I would like to especially thank my family, without
whose unfailing support this would never have been possible.

\newpage
\thispagestyle{empty} \tableofcontents

\newpage
\setcounter{page}{1} \pagenumbering{arabic}

\begin{center}
\quote{$\Delta \accentset{\smallfrown}{\iota} \nu o \varsigma
\quad \beta \alpha \sigma \iota \lambda \epsilon \acute{\upsilon}
\epsilon \iota \quad \tau \grave{o} \nu \quad \Delta
\acute{\iota}' \quad \accentset{\supset}{\epsilon} \xi \epsilon
\lambda \eta \lambda \alpha \kappa \acute{\omega} \varsigma$

Spin has cast out Zeus and rules as king.\footnote{Aristophanes'
Clouds (l. 828). Aristophanes' inspiration here is Anaxagoras, who
held that `$\delta \accentset{\smallfrown}{\iota} \nu o
\varsigma$', meaning `spin' or `rotation', was one of the primary
effects of `$\nu o \accentset{\smallfrown}{\upsilon} \varsigma$',
the active and rational principle of the Universe
\cite{lsj:gel}.}}
\end{center}

\part*{Introduction}
\label{sec:intro}

The aim of this thesis is to investigate the mathematics of spinor
structures, and their classification. The language of principal
fibre bundles allows a thorough and coherent treatment of
pseudo-Riemannian manifolds and spinor structures. The first two
parts of this thesis give a fully geometric description of these
constructions, including classification results for inequivalent
spinor structures. The third part shows how the Dirac equation
sits naturally in the setting of spinor structures, and how spinor
structures allow us to generalise the Dirac equation to arbitrary
curved space-times. It also discusses the implications in physics
of the available choice of spinor structures. Although this
interest in the Dirac equation guides the development of the
material, we work in a more general setting. The mathematical
focus is on the classification of spinor structures, and we
consider the abstract setting both when reviewing previously known
work and when presenting new work.

Fundamentally, there are two operations on principal fibre bundles
which we are interested in. One is \emph{reducing the structure
group}. Such a reduction of the frame bundle picks out an
orthonormal structure, and so gives an alternative treatment of
pseudo-Riemannian manifolds. The details of this are given in Part
\ref{part:geometry-of-orthonormal-structures}. We first show that
pseudo-Riemannian metrics are in one to one correspondence with
appropriate reductions of the frame bundle. Thereafter, we
introduce the notion of a \emph{connection} on a principal fibre
bundle, and show that these give rise to the covariant derivatives
familiar from pseudo-Riemannian geometry.

The other fundamental operation on principal fibre bundles is
constructing the \emph{spinor bundle}. This process `unwraps' the
structure group to its simply connected covering group. This is
not always possible, and when possible, the spinor bundle need not
be uniquely defined. In Part
\ref{part:spinor-structure-classification}, we give the relevant
classifications in the general setting. This differs slightly from
the more common notion of a spinor structure, which only considers
two fold covering groups. We show how the general theory
encompasses this case.

The combination of these two processes proves fruitful. The
geometric description of pseudo-Riemannian geometry in terms of a
reduction of the frame bundle given in Part
\ref{part:geometry-of-orthonormal-structures} allows a beautifully
geometric construction of the spinor structure, in
\S\ref{sec:spinor-structure}.

To some extent the two processes are independent---for example, we
prove that the classification of the possible spinor structures
for Riemannian and Lorentzian manifolds is independent of the
particular metric structure chosen, in
\S\ref{sec:metric-independence}. On the other hand, certain
results are only available when we treat spinor structures of a
reduced orthonormal bundle. In particular, the interplay allows a
geometrical description of the calculus and algebra of spinor
structures for pseudo-Riemannian manifolds. For example, in
\S\ref{sec:lifting-connection} we see that every spinor
derivative, considered as a connection form on the spinor bundle,
is simply the pull-back of the connection form on the original
bundle. On a pseudo-Riemannian manifold there is a distinguished
connection form, and so this construction picks out a
distinguished connection on the spinor structure.  In certain low
dimensional cases, an exceptional isomorphism between the simply
connected cover of the orthogonal group and another group, such as
$\SL \cong \smash{\widetilde{SO}}_0(1,3)$, allows an explicit
development of the \emph{spinor algebra}.

We also describe a coarse classification of spinor structures,
according to the type of underlying principal fibre bundle, in
\S\ref{sec:classifying-as-bundles}. This classification extends
previous work in this direction, and we see how it allows us to
compare the spinor connections associated with different spinor
structures.

All these ideas combine in Part
\ref{part:implications-for-physics} in the analysis of the Dirac
equation. In four dimensional Minkowskian space-time the usual
presentation of the Dirac equation, using `gamma matrices', can be
rewritten using the spinor algebra and calculus as a simple pair
of covariant differential equations. This allows an immediate
generalisation to curved Lorentzian space-times. Finally, we apply
the classification of inequivalent spinor structures, and our
knowledge of how the spinor connection depends on the choice of
spinor structure, to consider the physical implications of the
choice of spinor structure for particles governed by the Dirac
equation.

%-% make sure this goes back in for the final run!
%-% on second thought - this diagram doesn't really do much :-)
\noop{ \finalversion{
\begin{figure}[!ht]
\begin{equation*}
 \xymatrix{
    \text{Principal fibre bundles}
      \ar@{=>}@(d,ul)[dd] \ar@{=>}@(r,ul)[dr] & \\
            & \genfrac{}{}{0pt}{0}{\text{Connections and}}{\text{tensor calculus}}
                \ar@{=>}@(dr,u)[dd] \ar@{.>}@(l,u)[dl]                           \\
    \text{\textbf{Spinor structures}} \ar@{=>}@(d,l)[dr] &                         \\
            & \genfrac{}{}{0pt}{0}{\text{$\SL$ and the }}{\text{Dirac equation}}  \\
 }
\end{equation*}
\caption{An imagined schematic structure.}
\end{figure}}}

\subsection*{Conventions used throughout}
All our manifolds are considered to be Hausdorff, paracompact, and
smooth. For these and other notions of topology and basic
differential geometry, refer to \cite{boo:idmrg} or the more
abstract but more comprehensive exposition in
\cite{kob:fdg1,kob:fdg2}.

We use a subscripted asterisk to indicate the derivative of a
function. Thus if $f: X \To Y$ is a smooth map, $f_* : T X \To T
Y$, and at a point $x \in X$, $f_{* x} : T_x X \To T_{f(x)} Y$.
Later we will also use this notation to indicate the induced map
$f_* : \pi_1(X) \To \pi_1(Y)$ between the fundamental groups of
$X$ and $Y$, but it will always be clear from context which sense
is intended.

We will write $K \subgroup L$ to indicate that $K$ is a subgroup
of $L$.

If $G$ is a Lie group, $\LA{G}$ denotes its Lie algebra. The Lie
algebras of matrix groups will be denoted in the conventional
manner. Thus, for example, $\LAM{so}{n}$ is the Lie algebra of the
$n$ dimensional special orthogonal group $SO(n)$. The adjoint
representation of $G$ on $\LA{G}$ is written $\Ad(g)$, and defined
as the derivative of the inner automorphism of $G$, $g' \mapsto
I_g(g') = g g' g^{-1}$, at the identity. Thus $\Ad(g) = I_{g *
e}$.

\newpage
\part{Geometry of Orthonormal Structures}
\label{part:geometry-of-orthonormal-structures}

In the following sections, we review the theory of principal fibre
bundles, and explain how pseudo-Riemannian geometry appears in
this context. This has a dual purpose. Firstly, we wish to
understand from an abstract point of view the nature of principal
fibre bundles, because later, in Part
\ref{part:spinor-structure-classification}, this will be
fundamental to understanding spinor structures. Secondly, spinor
structures for pseudo-Riemannian manifolds are the most
interesting variety of spinor structures, and so we need to place
pseudo-Riemannian geometry in this framework.

The discussion of pseudo-Riemannian geometry consists of two main
points. Firstly, every pseudo-Riemannian metric on a manifold
corresponds to a certain reduction of the frame bundle of that
manifold. Secondly, the covariant derivatives on such manifolds
correspond exactly to connections on the reduced bundle. These
facts are established in \S\ref{ssec:bundle-metric-equivalence}
and \S\ref{ssec:metric-connections} respectively.

In the process of covering this material, we also give an
introduction to the tensor algebra associated with a principal
fibre bundle. This is useful in the proofs of this section, and
will be vital in Part \ref{part:implications-for-physics} in our
discussion of the Dirac equation.

\section{The theory of principal fibre
bundles}\label{sec:principal-fibre-bundles} We now give a brief
introduction to the fundamental geometric objects underlying the
rest of this work. These are \emph{principal fibre bundles}. The
definitions here follow \cite{cho:amp}, \cite{hus:fb},
\cite{ish:mdgfp} and \cite{mil:gc}. A popular account of fibre
bundles in physics appears in \cite{ber:fbqt}. We first define a
\emph{locally trivial fibre bundle}.

\subsection{Fibre bundles}
A \defnemph{bundle} $\xi = \bundle{P}{\pi}{M}$ consists of a pair
of smooth manifolds, $P$ and $M$, respectively called the
\emph{total space} and the \emph{base space}, and a surjective map
$\pi : P \To M$ called the \emph{projection map}.

A \defnemph{fibre bundle} $\xi = \bundle{P}{\pi}{M}$ with fibre
$F$ is a bundle such that for each $m \in M$, $\pi^{-1}(m)$ is
diffeomorphic to $F$. This partitioning of $P$ into $\bigcup_{m
\in M} \pi^{-1}(m)$ is referred to as the fibration.

A \defnemph{fibre bundle morphism} from a fibre bundle $\xi =
\bundle{P}{\pi}{M}$ with fibre $F$ to a fibre bundle $\eta =
\bundle{P'}{\smash{\pi'}}{M'}$ with fibre $F'$ is a pair of maps
$(\phi,f)$ so $\phi:P \To P'$, $f:M \To M'$, and $\pi' \compose
\phi = f \compose \pi$, such that the following diagram commutes.
\[
 \xymatrix{
    P \ar[r]^\phi \ar[d]^\pi & P' \ar[d]^{\pi'} \\
    M \ar[r]^f               & M'
 }
\]
This ensures that the maps respect the fibre structure.

Morphisms can be composed, as $(\phi',f') \compose (\phi,f) =
(\phi' \compose \phi, f' \compose f)$. If $M = M'$, we say $\xi$
and $\eta$ are $M$-isomorphic, or equivalent, if there are
morphisms $(\phi,f):\xi \To \eta$, $(\phi',f'):\eta \To \xi$ so
$(\phi',f') \compose (\phi,f) = (\id_{P},\id_M)$ and $(\phi,f)
\compose (\phi',f') = (\id_{P'},\id_M)$. The bundle $\xi$ is said
to be \emph{trivial} if it is $M$-isomorphic to $\bundle{M \times
F}{\pi}{M}$, the product fibre bundle. Henceforth we will nearly
always consider only morphisms between bundles over the same base
space, so $M = M'$, and $f=\id_M$.

We can also restrict bundles. If $N$ is a submanifold of $M$,
define
\[\restrict{\xi}{N} = \bundle{\pi^{-1}(N)}{\restrict{\pi}{\pi^{-1}(N)}}{N}.\]
With this idea, we can say that bundles $\xi$ and $\eta$ over $M$
are locally isomorphic if there is an open covering
$\bigcup_\alpha U_\alpha$ of $M$ so for each $\alpha$,
$\restrict{\xi}{U_\alpha}$ and $\restrict{\eta}{U_\alpha}$ are
${U_\alpha}$-isomorphic. We can now define locally trivial as
meaning locally isomorphic to the product fibre bundle $M \times
F$. Each bundle morphism is of the form $\varphi:U \times F \To
\pi^{-1}(U)$, where $\pi(\varphi(m,f)) = m$ for all $m \in U$ and
$f \in F$, and is called a
\defnemph{local trivialisation} of the fibre bundle. Generally no
particular trivialisations are distinguished.

A \defnemph{section} of a bundle is a smooth map $\sigma:M \To P$
such that $\pi \compose \sigma = \id_M$. It assigns to each point
$m \in M$ a point in the fibre of $m$. A local section is simply a
section defined only on some open set of $M$.

\subsection{Principal fibre bundles}\label{ssec:principal-fibre-bundles}
We now reach the definition of a principal fibre bundle.
\begin{defn}\label{defn:principal-fibre-bundle}
A bundle $\bundle{P}{\pi}{M}$ is a \defnemph{principal fibre
bundle} with
\defnemph{structure group} $G$ if
\begin{enumerate}
\item The group $G$ is a Lie group, and $G$ acts on the right on
$P$: \[\bdelt{p} \mapsto \bdelt{p} g.\]
%\item The following two equivalent conditions hold.
%\begin{enumerate}
\item The $G$ action preserves the fibres of $P$, and is transitive on
fibres.
%\item The bundle $\bundle{P}{\pi}{M}$ is isomorphic to the bundle
%$\bundle{P}{\rho}{P/G}$, where $P/G$ denotes the quotient, and
%$\rho$ is the quotient map.
%\end{enumerate}
\item The $G$ action is free. That is, if $\bdelt{p} g =
\bdelt{p}$ for some $\bdelt{p} \in P$, then $g=e$.
\item There are local trivialisations compatible with the $G$
action. That is, for each $m_0 \in M$, there is an open set $U$
with $m_0 \in U \subset M$ and a map $\varphi : U \times G \To
\pi^{-1}(U)$ so $\pi(\varphi(m,g)) = m$ and $\varphi(m,g) =
\varphi(m,e) g$ for all $m \in U$ and $g \in G$.
\end{enumerate}
\end{defn}

%The equivalence of the two conditions in 2. above is not hard to
%prove.
It is clear from conditions 2. and 3. that the fibres of $P$ are
diffeomorphic to $G$. We write $\pfbundle{G}{P}{\pi}{M}$ to
indicate this situation, where $G$ is the structure group acting
on $P$.

Condition 4. is in fact guaranteed if local trivialisations exist
at all, in accordance with the following result.

\begin{lem}\label{lem:sections-trivialisations}
If $\bundle{P}{\pi}{M}$ is a bundle satisfying the first three
parts of Definition \ref{defn:principal-fibre-bundle}, then there
is a one to one correspondence between local sections $\sigma : U
\To P$ on $U$ and local trivialisations $\varphi:U \times G \To
\pi^{-1}(U)$ over $U$ compatible with the $G$ action.
\end{lem}
\begin{proof}
Clearly a local trivialisation (compatible with the $G$ action or
not) defines a section, via $\sigma(m) = \varphi(m,e)$. Given a
section, define $\psi$ by $\psi(m,g) = \sigma(m) g$. This is
clearly compatible with the $G$ action. If we began with a
trivialisation compatible wiht the $G$ action, these constructions
are mutual inverses, establishing the correspondence.
\end{proof}

As the group $G$ acts transitively and freely on each fibre, if
$\pi(\bdelt{p}) = \pi(\bdelt{p'})$ there is a unique $g \in G$ so
$\bdelt{p} g = \bdelt{p'}$. We use this to define a function
$\tau: \pi^{-1}(m) \times \pi^{-1}(m) \To G$ for each $m\in M$, so
$\bdelt{p} \tau(\bdelt{p},\bdelt{p'}) = \bdelt{p'}$. We call this
the \defnemph{translation function} for the principal fibre
bundle.

A principal fibre bundle morphism is a fibre bundle morphism that
commutes with the group action. Thus if $\pfbundle{G}{P}{\pi}{M}$
and $\pfbundle{G}{P'}{\pi'}{M}$ are principal fibre bundles, then
a smooth map $u:P \To P'$ is a principal fibre bundle morphism if
$\pi' \compose u = \pi$ and $u(\bdelt{p} g) = u(\bdelt{p})g$ for
all $\bdelt{p} \in P$ and $g \in G$.

It turns out that for any fibre bundle there is a related
principal fibre bundle, where, roughly speaking, the structure
group is the group of transformations of the fibre \cite[\S
3.3]{ish:mdgfp}. Any fibre bundle can then be derived from its
principal fibre bundle by the associated bundle construction. This
is given for vector space fibres in
\S\ref{ssec:associated-vector-bundles}, and it is used
subsequently to construct the tensor algebra associated with a
representation of the structure group of a principal fibre bundle.

\subsection{An example: the frame bundle of a manifold}\label{ssec:frame-bundle}
The primary motivating example of a principal fibre bundle is the
\emph{frame bundle} of a manifold. Given a smooth $n$ dimensional
manifold $M$, at each point $m$ the tangent space $T_m M$ is
defined as the vector space of tangent vectors\footnote{Tangent
vectors are in turn defined as derivations of the germs of smooth
functions, although we shall not need this.} at that point. The
collection of all the tangent spaces is called the tangent bundle,
and denoted $T M$. A frame is simply a basis for the tangent space
at a point. We might write a frame as $\bdelt{p} =
(e_1,\ldots,e_n)$, where the $e_i$ are tangent vectors. The frame
bundle, as a set, is the collection of frames at every point of
the manifold. We denote the frame bundle of $M$ by $F M$. It has a
projection $\pi$ taking a frame to the point at which that frame
lies.  We give it a smooth structure as an $n^2 + n$ dimensional
manifold in the obvious way.\footnote{We induce the smooth
structure for the frame bundle from the smooth structure for the
manifold itself. A coordinate chart $\varphi: U \To V$, where $U
\subset M$, and $V \subset \Real^n$ are open sets, induces a map
$\varphi_*: F U \To F V$ by $\varphi_*(e_1,\ldots,e_n) =
(\varphi_* e_1, \ldots \varphi_* e_n)$. That is, $\varphi$ pushes
forward a frame on $M$ to a frame on $\Real^n$. Now, the frame
bundle of $\Real^n$, $F \Real^n$ has an obvious smooth structure,
since the tangent space to $\Real^n$ is canonically identified
with $\Real^n$. Thus $F V \cong V \times \Real^{n^2}$, and if we
say that $\varphi_* :F U \To V \times \Real^{n^2}$ is a chart, for
each coordinate chart $\varphi$, we obtain an atlas for $F M$, and
so a smooth structure.}

Next we see that $F M$ really is a principal fibre bundle. To do
this, we must describe the group action. The general linear group
$\GLn$ acts on the right on frames in the following way. If $g \in
\GLn$, and $\bdelt{p} = (e_1,\ldots,e_n) \in F M$, then we have
\begin{align}\label{eq:frame-bundle-action}
g & = \begin{pmatrix}
g^{1}{}_{1} & \cdots & g^{1}{}_{n} \\
\vdots & \ddots & \vdots \\
g^{n}{}_{1} & \cdots & g^{n}{}_{n}
\end{pmatrix}
&& \text{and} & \bdelt{p} g & = (e_1,\ldots,e_n)
\begin{pmatrix}
g^{1}{}_{1} & \cdots & g^{1}{}_{n} \\
\vdots & \ddots & \vdots \\
g^{n}{}_{1} & \cdots & g^{n}{}_{n}
\end{pmatrix}.
\end{align}

We can now check that the principal fibre bundle axioms from
Definition \ref{defn:principal-fibre-bundle} are satisfied. All
are in fact immediately obvious, except perhaps the existence of
local trivialisations, which are provided by the coordinate charts
of $M$.

Throughout later discussions, in which we discuss theorems dealing
with abstract principal fibre bundles, it may be useful to keep in
mind this concrete and intuitive example.

\section{Tensor algebras}\label{sec:tensor-algebra}

We now begin our discussion of \emph{tensors}. Our aim is to
define the \emph{global tensor algebra} associated with a $G$
principal fibre bundle and a representation of the group $G$. We
will also present a powerful formalism for calculations in the
global tensor algebra, called the \emph{abstract index notation}.
We will use this throughout our discussion of covariant
differentiation and tensor calculus in
\S\ref{sec:tensor-calculus}, and eventually in the exposition of
the Dirac equation in \S\ref{sec:dirac-equation}. It is worth
asking why we decide to present this material in the completely
general setting, allowing an arbitrary (finite dimensional)
representation of an arbitrary Lie group. Later, we treat in
detail two tensor algebras, one associated with the group $\GLn$,
and the other associated with $\SL$. Having the general framework
available avoids unnecessary duplication.

To start, we need to describe the local tensor algebra associated
with a representation $\lambda$ of a Lie group $G$. This is
straightforward and familiar. Although the abstract index notation
is irrelevant for local tensor algebras, we introduce it in this
context in order to streamline the development of the global
tensor algebra.

Following this, we construct the global tensor algebra. The data
required are a local tensor algebra based upon a representation
$\lambda$ of a Lie group $G$, and a principal fibre bundle
$\pfbundle{G}{P}{\pi}{M}$, with structure group $G$, over a
manifold $M$. Using the associated vector bundle construction,
given below in \S\ref{ssec:associated-vector-bundles}, we define
tensors on the base manifold. In the particular case of the $\GLn$
frame bundle over a manifold, this process gives the world tensor
algebra, in terms of the tangent vectors to the manifold.

\subsection{Local tensor algebras}\label{sec:local-tensor-algebra}
To begin, we introduce the most primitive type of tensor algebra.
It is a \emph{local} tensor algebra in the sense that there is a
singled fixed underlying representation on a fixed vector space.
%Later we will introduce \emph{global} tensor algebras defined over
%manifolds, where at each point of the manifold there is a
%collection of representations, but representations at different
%points of the manifold are not canonically identified.
The purpose of this section is not only to define tensors---which,
it is hoped, will be fairly familiar in any case---but to describe
the \emph{abstract index tensor algebra}, and distinguish between
the objects of this algebra and the underlying geometrical
objects.

We first introduce the geometric tensor algebra. To this end,
suppose $G$ is an arbitrary Lie group. Suppose $\lambda$ is a
representation of $G$ on the $n$ dimensional vector space $V =
\Field^n$ over the field $\Field = \Real$ or $\Complex$. A typical
example might be the matrix representation of $\GLn$ acting on
$\Real^n$ with the standard basis. The elements of $V$ are
geometrical objects. We might denote such an element by $v$. Since
we have fixed a basis, we can consider the components of $v$,
writing these as the kernel symbol along with a numerical
superscript index, $v^1, \ldots, v^n \in \Field$.

Next, we consider the dual vector space $V^*$, which is
canonically isomorphic to $\Field^n$ also, since we have selected
a basis for $V$.  Specific components of $u \in V^*$ are indicated
with numerical indices, as in $u_1, \ldots, u_n \in \Field$.

The pairing between the vector space and its dual, $V^* \times V
\To \Field$ is written $(u,v) \mapsto \{u,v\}$. In terms of
components, this is \[\{u,v\} = \sum_{k=1}^n u_k v^k.\]
%We can
%write this more simply as \[\{u,v\} = u_k v^k,\] implicitly
%understanding the summation over a pair of indices, one subscript,
%one superscript.

The representation $\lambda$ on $V$ gives rise to the dual
representation $\lambda^*$ on $V^*$, defined so
\begin{equation}\label{eq:dual-representation}
\{(\lambda^*(g) u),v\} = \{u,(\lambda(g^{-1}) v)\}
\end{equation}
for all $u \in V^*$ and $v \in V$.

With these two fundamental representations established, we
generate all the tensor representations. The underlying vector
space for the valence $\valence{k}{l}$ tensor representation is
the collection of multilinear maps \[\underbrace{V^* \times \dots
\times V^*}_{k \text{ times}} \times \underbrace{V \times \dots
\times V}_{l \text{ times}} \To \Field.\] We denote this vector
space as $\tensors{k}{l}$. In particular $V = \tensors{1}{0}$ and
$V^* = \tensors{0}{1}$. The action of $G$ on the vector space
$\tensors{k}{l}$ is such that for $S \in \tensors{k}{l}$,
\begin{multline*}
(g(S))(x^1,\ldots,x^k,y_1,\ldots,y_l) = \\ = S(\lambda^*(g^{-1})
x^1, \ldots, \lambda^*(g^{-1}) x^k, \lambda(g^{-1}) y_1, \ldots,
\lambda(g^{-1}) y_l),
\end{multline*}
where $x^1,\ldots, x^k \in V^*$ and $y_1, \ldots y_l \in V$. This
defines a representation of $G$ on $\tensors{k}{l}$.

We now outline three operations on tensors. Firstly, we can take
the tensor product of two tensors. This is a map $\tensors{j}{k}
\times \tensors{l}{m} \To \tensors{j+l}{k+m}$. The tensor product
of $S \in \tensors{j}{k}$ and $T \in \tensors{l}{m}$ is defined by
\begin{multline*}
(S \otimes T)(x^1, \ldots, x^j, x^{j+1},\ldots,x^{j+l}, y_1,
\ldots , y_k, y_{k+1},\ldots,y_{k+m}) = \\ = S(x^1 \ldots,
x^j,y_1, \ldots , y_k)
T(x^{j+1},\ldots,x^{j+l},y_{k+1},\ldots,y_{k+m}),
\end{multline*}
where $x^1, \ldots, x^{j+l} \in V^*$, and $y_1, \ldots, y_{k+m}
\in V$. Further, it is easy to see that this map
\emph{intertwines} the representations.

Secondly, we can perform `index permutation'. This name will
become clearer later. Given a tensor $S$ in $\tensors{k}{l}$, we
can obtain $k! l!$ new tensors in $\tensors{k}{l}$, all of which
will in general be different, by permuting its arguments. For
example, if $S \in \tensors{2}{0}$, then there is another tensor,
which we might call for a moment $\tilde{S}$ in $\tensors{2}{0}$,
given by $\tilde{S}(w,z) = S(z,w)$, for all $w,z \in V^*$. Again,
it is easy to see that this operation commutes with the action of
$G$ via the tensor representation, and so the index permutation
maps intertwine the $\tensors{k}{l}$ representation with itself.

Finally, we can contract a tensor. Given an tensor $S$ in
$\tensors{k}{l}$, this produces a tensor in $\tensors{k-1}{l-1}$,
which we for the moment call $\hat{S}$, which acts as
\[ \hat{S}(x^1,\ldots x^{k-1},y_1,\ldots,y_{l-1}) = \sum_{i=1}^n
T(x^1,\ldots x^{k-1},e^i,y_1,\ldots,y_{l-1},e_i),\] where
$x^1,\ldots, x^{k-1} \in V^*$ and $y_1, \ldots y_{l-1} \in V$, and
$e^i$ and $e_i$ are the basis vectors for $V^*$ and $V$
respectively. We lose no generality by only discussing contraction
over the last argument, because by combining this operation with
index permutation, we can contract with respect to any pair of
arguments, one in $V$, the other in $V^*$. Contraction also
commutes with the action of $G$. However, we will not prove this
now, as it is more transparent in index notation.

These comments complete our description of the geometric tensor
algebra, in that we have specified the objects and algebra
operations. This presentation is, however, rather unsatisfactory
for working with these tensors, because its notation is so
cumbersome. Firstly, we cannot see from the symbol for an element
of the tensor algebra which representation it lies in, and the
operations of permutation and contraction require specialised
notation for each possible pair of indices involved.

Thus, we now introduce the abstract index tensor algebra. At first
it seems more mathematically cumbersome, but it has great
notational convenience. When we come to global tensor algebras,
the abstract index notation offers a powerful formalism without
reference to local coordinates or components. The principal
difference between abstract index notation and conventional tensor
index notation is that objects indicated, for example, as
$V^{\cgi{a}}{}_{\cgi{b c}}$ do not denote the \emph{components} of
a tensor, but the tensor itself, with the indices serving as
labels to indicate the valence. A further useful discussion on the
motivation for abstract index notation is in \cite[pp.
23--26]{wal:gr}. The idea of abstract index tensors is due to
Penrose, and they are described in his works
\cite{pen:ss-t,pen:ss-t1}. A thorough axiomatic development is
given in \cite[pp. 76--91]{pen:ss-t1}, and a simple presentation
of the formalism is in \cite[\S 3]{pen:ss-t}.

We now give an explicit description of the abstract index algebra
and its operations. We introduce an \emph{index set}, denoted
$\mathcal{L}$. For our purposes now, it will be $\mathcal{L} =
\set{\cgi{a}, \cgi{b}, \cgi{c},
\ldots,\cgi{a}_1,\cgi{b}_1,\dots,\cgi{a}_2,\ldots}$. The \ggerman
font will be used in the index set when we are referring to the
tensor algebra associated to some arbitrary group. Later, we will
use lowercase or uppercase \rroman indices to refer specifically
to the tensors associated to the groups $\GLn$ and $\SL$
respectively. The labels in the index set at this point are all
lightface, to emphasise that this is a local tensor algebra.
Later, global tensor algebras will use boldface indices. The
elements of the abstract index algebra are pairs, the first part
of which is an element from the geometric tensor algebra, while
the second part is an appropriate sequence of indices from
$\mathcal{L}$. For a tensor $S$ in $\tensors{k}{l}$, this is a
sequence of $k + l$ distinct indices from $\mathcal{L}$. We never
write this pair explicitly as
$(S,\{\cgi{a}_1,\ldots,\cgi{a}_k,\cgi{b}_1,\ldots,\cgi{b}_l\})$,
but as \[S^{\cgi{a}_1 \dots \cgi{a}_k}{}_{\cgi{b}_1 \dots
\cgi{b}_l}.\] Thus corresponding to each geometrical object in the
tensor algebra, there are a collection of objects in the abstract
index algebra. We write $\mathcal{T}^{\cgi{a}_1 \dots
\cgi{a}_k}{}_{\cgi{b}_1 \dots \cgi{b}_l}$ for the vector space of
elements of the abstract index tensor algebra of the form
$S^{\cgi{a}_1 \dots \cgi{a}_k}{}_{\cgi{b}_1 \dots \cgi{b}_l}$ for
some $S \in \tensors{k}{l}$.

\begin{example}
If $v \in V$ is a vector, then there are elements $v^{\cgi{a}},
v^{\cgi{b}}, v^{\cgi{c}}$ and so on in the abstract index algebra.
The elements $v^{\cgi{a}}$ and $v^{\cgi{b}}$ correspond to the
same geometrical object, $v$, but are \emph{not equal} in the
abstract index tensor algebra. Similarly, a tensor $S \in
\tensors{2}{0}$ has representatives $S^{\cgi{a b}}$, $S^{\cgi{e f
}}$, etc.
\end{example}

Next, we describe the tensor algebra operations in terms of
abstract index notation. The tensor product appears as a map
\[\mathcal{T}^{\cgi{a}_1 \dots \cgi{a}_j}{}_{\cgi{b}_1 \dots
\cgi{b}_k} \times \mathcal{T}^{\cgi{c}_1 \dots
\cgi{c}_l}{}_{\cgi{d}_1 \dots \cgi{d}_m} \To
\mathcal{T}^{\cgi{a}_1 \dots \cgi{a}_j \cgi{c}_1 \dots
\cgi{c}_l}{}_{\cgi{b}_1 \dots \cgi{b}_k \cgi{d}_1 \dots
\cgi{d}_m}.\] All of the indices appearing as labels of the above
vector spaces must be distinct. For example, the tensor product of
$s^{\cgi{a}}$ and $t^{\cgi{b}}$ is an element of
$\mathcal{T}^{\cgi{a b}}$, written simply as $s^{\cgi{a}}
t^{\cgi{b}}$. The corresponding geometric tensor is the map $(w,z)
\mapsto \{w,s\} \{z,t\}$, for $w$ and $z$ in $V^*$. In the
abstract index formulation the tensor product is in fact
\emph{commutative}, because we define $t^{\cgi{b}} s^{\cgi{a}} \in
\mathcal{T}^{\cgi{a b}}$ to correspond to exactly the same
underlying geometric tensor. The indices indicate the order of
arguments. On the other hand $t^{\cgi{a}} s^{\cgi{b}} \in
\mathcal{T}^{\cgi{a b}}$ corresponds to a different geometric
tensor, $(w,z) \mapsto \{w,t\} \{z,s\}$. Index permutation has a
simple appearance now, and the name becomes clear. If, for example
$S^{\cgi{a b}} \in \mathcal{T}^{\cgi{a b}}$, then $S^{\cgi{b a}}$
is the element of $\mathcal{T}^{\cgi{a b}}$ (not
$\mathcal{T}^{\cgi{b a}}$---but this will always be clear from
context) corresponding to the tensor $(w,z) \mapsto S(z,w)$ for
$w$ and $z$ in $V^*$. Thus the operation of permuting arguments is
indicated clearly by permuting the indices of the abstract index
tensor. We must be careful however only ever to permute
superscript indices, or to permute subscript indices. Finally,
contraction is indicated by a repeated index, one superscript, one
subscript. Thus $S^{\cgi{a}_1 \dots \cgi{a}_{k-1}
\cgi{c}}{}_{\cgi{b}_1 \dots \cgi{b}_{l-1} \cgi{c}}$ is the element
of $\mathcal{T}^{\cgi{a}_1 \dots \cgi{a}_{k-1}}{}_{\cgi{b}_1 \dots
\cgi{b}_{l-1} }$ which corresponds to the contraction of $S \in
\tensors{k}{l}$. Contraction on other pairs of indices is defined
similarly, by contraction of the underlying geometric tensor on
the corresponding pair of arguments.

Because each of the tensor algebra operations is defined in terms
of operations on the underlying geometric tensors, an equation
between abstract index tensors remains true if one index is
replaced throughout the equation by a new one.

Now, $\lambda(g)$ maps $V$ to $V$, and is thus equivalently a map
of $V^* \times V$ to $\Field$. Then $\lambda(g)$ is a valence
$\valence{1}{1}$ tensor, and we denote it in abstract index
notation as $g^{\cgi{a}}{}_{\cgi{b}}$. Recall that since we are
working over a fixed vector space, we can also consider the actual
components of a tensor. The numbers $g^{1}{}_{1}, g^{2}{}_{1}$ and
so on are exactly the entries of the matrix $g$. The action of the
representation can now be written out. In the tensor algebra
$\lambda(g) v$ corresponds to $g^{\cgi{a}}{}_{\cgi{b}}
v^{\cgi{b}}$ in $\mathcal{T}^{\cgi{a}}$. Since $\lambda$ is a
representation, $(gh)^{\cgi{a}}{}_{\cgi{b}} =
g^{\cgi{a}}{}_{\cgi{c}} h^{\cgi{c}}{}_{\cgi{b}}$. Similarly the
dual representation is simply written in index notation. According
to Equation \eqref{eq:dual-representation},
\begin{align*}
(\lambda^*(g)u)_{\cgi{b}} v^{\cgi{b}} & = \{\lambda^*(g)u,v\} \\
            & = \{u,\lambda(g^{-1})v\} \\
            & = u_{\cgi{a}} (\lambda(g^{-1}) v)^{\cgi{a}} \\
            & = u_{\cgi{a}}
            (g^{-1})^{\cgi{a}}{}_{\cgi{b}} v^{\cgi{b}},
\end{align*}
and so $\lambda^*(g) u$ corresponds to
$(g^{-1})^{\cgi{a}}{}_{\cgi{b}} u_{\cgi{a}}$ in
$\mathcal{T}_{\cgi{b}}$. We use these expressions for the
representations to express the action of $G$ in the tensor
representation on $\tensors{k}{l}$. An element $g \in G$ acting on
a tensor $S$ in the $\valence{k}{l}$ tensor representation gives a
valence $\valence{k}{l}$ tensor which in abstract index notation
is
\begin{equation}\label{eq:g-action-on-tensor}
 g^{\cgi{a}_1}{}_{\cgi{c}_1} \dotsm
g^{\cgi{a}_k}{}_{\cgi{c}_k}
 (g^{-1})^{\cgi{d}_1}{}_{\cgi{b}_1} \dotsm
 (g^{-1})^{\cgi{d}_l}{}_{\cgi{b}_l}
 S^{\cgi{c}_1 \dots \cgi{c}_k}{}_{\cgi{d}_1 \dots
\cgi{d}_l}.
\end{equation}

It is straightforward to check that this is in fact a
representation, using $(gh)^{\cgi{a}}{}_{\cgi{b}} =
g^{\cgi{a}}{}_{\cgi{c}} h^{\cgi{c}}{}_{\cgi{b}}$.

Finally, we use the abstract index presentation of the
representations to show that contraction commutes with the group
action. A simple case suffices, so we avoid a profusion of
indices. Suppose $S^{\ci{a}}{}_{\ci{b}} \in
\mathcal{T}^{\ci{a}}{}_{\ci{b}}$. Then $g \in G$ acts on $S$ to
give $(g)^{\ci{a}}{}_{\ci{c}} (g^{-1})^{\ci{d}}{}_{\ci{b}}
S^{\ci{c}}{}_{\ci{d}}$. Contracting on the indices $\ci{a}$ and
$\ci{b}$, we obtain
\begin{align*}
(g)^{\ci{a}}{}_{\ci{c}} (g^{-1})^{\ci{d}}{}_{\ci{b}}
S^{\ci{c}}{}_{\ci{d}} & = (g^{-1}g)^{\ci{d}}{}_{\ci{c}} S^{\ci{c}}{}_{\ci{d}} \\
    & = (e)^{\ci{d}}{}_{\ci{c}} S^{\ci{c}}{}_{\ci{d}} \\
    & = S^{\ci{d}}{}_{\ci{d}}.
\end{align*}
This is exactly the result we would obtain contracting first and
then acting by $g$. The general case, with arbitrarily many
indices, is much the same.

\noop{Specifically, if $g$ acts on a tensor $S^{\cgi{a}_1 \dots
\cgi{a}_k}{}_{\cgi{b}_1 \dots \cgi{b}_l}$, we obtain the
expression in Equation \eqref{eq:g-action-on-tensor}. Contracting
on, say, the indices $\cgi{a}_k$ and $\cgi{b}_l$, we obtain
\begin{align*}
 &\left(\prod_{i=1}^{k-1}
 g^{\cgi{a}_i}{}_{\cgi{c}_i} \right) g^{\cgi{f}}{}_{\cgi{c}_k}
 \left( \prod_{j=1}^{l-1} (g^{-1})^{\cgi{d}_j}{}_{\cgi{b}_j} \right) (g^{-1})^{\cgi{d}_l}{}_{\cgi{f}}
 S^{\cgi{c}_1 \dots \cgi{c}_k}{}_{\cgi{d}_1 \dots \cgi{d}_l} = \\
& \qquad  = (g^{-1} g)^{\cgi{d}_l}{}_{\cgi{c}_k}
\left(\prod_{i=1}^{k-1}
 g^{\cgi{a}_i}{}_{\cgi{c}_i} \right)
 \left( \prod_{j=1}^{l-1} (g^{-1})^{\cgi{d}_j}{}_{\cgi{b}_j} \right)
 S^{\cgi{c}_1 \dots \cgi{c}_k}{}_{\cgi{d}_1 \dots \cgi{d}_l} \\
 & \qquad  = \delta^{\cgi{d}_l}{}_{\cgi{c}_k}
\left(\prod_{i=1}^{k-1}
 g^{\cgi{a}_i}{}_{\cgi{c}_i} \right)
 \left( \prod_{j=1}^{l-1} (g^{-1})^{\cgi{d}_j}{}_{\cgi{b}_j} \right)
 S^{\cgi{c}_1 \dots \cgi{c}_k}{}_{\cgi{d}_1 \dots \cgi{d}_l} \\
 & \qquad =
\left(\prod_{i=1}^{k-1}
 g^{\cgi{a}_i}{}_{\cgi{c}_i} \right) \left( \prod_{j=1}^{l-1}
(g^{-1})^{\cgi{d}_j}{}_{\cgi{b}_j} \right)
 S^{\cgi{c}_1 \dots \cgi{c}_{k-1} \cgi{f}}{}_{\cgi{d}_1 \dots \cgi{d}_{l-1}
 \cgi{f}}
\end{align*}%! this is pretty hideous. explain delta, perhaps just do an exmaple.
which is exactly the result of $g$ acting on the contraction of
the tensor $S$ over the indices $\cgi{a}_k$ and $\cgi{b}_l$. }

\subsection{Associated vector
bundles}\label{ssec:associated-vector-bundles} %-% this is all rather sudden
Fundamental to the idea of a global tensor algebra is the notion
of an associated vector bundle, which we will develop here,
following \cite[\S3.3]{ish:mdgfp}. Say $\pfbundle{G}{P}{\pi}{M}$
is a principal fibre bundle, and $\lambda$ is a finite dimension
representation of the group $G$ on a vector space $V$. We will
write this action of $G$ on $V$ as $(g,v) \mapsto \lambda(g)v$ for
$g \in G$ and $v \in V$. We consider the product space $P \times
V$. Define on this an equivalence relation $\sim$, so that
\[ (\bdelt{p},v) \sim (\bdelt{p} g, \lambda(g^{-1}) v)\] or
equivalently \[ (\bdelt{p} g, v) \sim (\bdelt{p}, \lambda(g) v).\]
We call the set of equivalence classes $(P \times V)/\sim$ the
associated vector bundle for $\lambda$. The vector bundle is also
denoted as $P \times_G V$. It is given the quotient topology, and
so in particular if $\bdelt{p}_{\alpha} \xrightarrow{\alpha}
\bdelt{p}$ and $v_{\alpha} \xrightarrow{\alpha} v$, then
$[\bdelt{p}_{\alpha},v_{\alpha}] \xrightarrow{\alpha}
[\bdelt{p},v]$.\footnote{A smooth structure is determined as
follows. Given a coordinate chart $\phi:U \subset M \To \Real^p$,
and a local section $\sigma:U \To P$, define $\psi : \pi^{-1}(U)
\times_G V$ by $\psi([\sigma(m),v]) = (\phi(m), v)$. The
collection of all of these provide an atlas for $P \times_G V$.}

Such a vector bundle is clearly a fibre bundle, with fibre $V$,
and locally trivialisable. Since the fibre is the vector space
$V$, we can perform the usual vector space operations on elements
of the vector bundle lying over the same point of the base
manifold. Suppose for example that $\bdelt{p},\bdelt{p'} \in P$,
$\pi(\bdelt{p})=\pi(\bdelt{p'})$, and $\bdelt{p'} = \bdelt{p} g$
for some $g \in G$. Then $[\bdelt{p}, v] + [\bdelt{p'}, u] =
[\bdelt{p}, v + \lambda(g) u]$.

%-% more detail.. abstract definition of a vector bundle?

\subsection{General construction of a global tensor algebra}
\label{ssec:global-tensor-algebra} Equipped with this
construction, we can describe the \emph{global abstract index
tensor algebra} associated with a principal fibre bundle and a
particular representation of the structure group. Firstly, we
construct the local abstract index tensor algebra, which is
generated by the representation, as in
\S\ref{sec:local-tensor-algebra}. The global tensor algebra then
arises as a collection of associated bundles. Conventional
developments differ in that they emphasise the algebraic
properties of the global tensor algebra, and consider it central.
On the other hand, we consider the principal fibre bundle as
primary, and the global tensor algebra as secondary.

For each abstract index tensor representation
$\mathcal{T}^{\cgi{a}_1 \dots \cgi{a}_k}{}_{\cgi{b}_1 \dots
\cgi{b}_l}$, define the associated vector bundle
\[\mathcal{T}^{\agi{a}_1 \dots \agi{a}_k}{}_{\agi{b}_1 \dots \agi{b}_l} = P \times_G \mathcal{T}^{\cgi{a}_1 \dots
\cgi{a}_k}{}_{\cgi{b}_1 \dots \cgi{b}_l}.\] Thus a typical element
of $\mathcal{T}^{\agi{a}_1 \dots \agi{a}_k}{}_{\agi{b}_1 \dots
\agi{b}_l}$ is \[S^{\agi{a}_1 \dots \agi{a}_k}{}_{\agi{b}_1 \dots
\agi{b}_l} = [\bdelt{p}, S^{\cgi{a}_1 \dots
\cgi{a}_k}{}_{\cgi{b}_1 \dots \cgi{b}_l}]\] for some $S \in
\tensors{k}{l}$. These objects are the \emph{global tensors}, and
take indices from the same labelling set as the local tensor
algebra, but with boldface indices, to distinguish them from the
local tensors. Such a tensor is only defined at a single point,
the point $\pi(\bdelt{p}) \in M$---a tensor field is a cross
section of this associated bundle.

Combining in this fashion the notational convenience of the local
abstract index algebra and the geometric construction of an
associated vector bundle, we obtain an extremely useful
description of the tensors on a manifold. The tensor operations of
forming tensor products, performing index permutations, and taking
contractions, all have simple presentations. Specifically, to
perform any of these operations on elements of the global tensor
algebra, we simply perform the operation on the corresponding
element of the local tensor algebra. As we have seen, the tensor
operations in the local tensor algebra all commute with the group
action, ensuring that this prescription for the tensor operations
in the global tensor algebra is well defined. %-%this is quite fast as well

\begin{example}
Suppose $S_{\agi{a b}} = [\bdelt{p}, S_{\cgi{a b}}]$ and
$y^{\agi{c}} = [\bdelt{p'},y^{\cgi{c}}]$. Then there is some $g
\in G$ so $\bdelt{p'} = \bdelt{p} g$, and we can define
$x^{\cgi{c}}$ by $x^{\cgi{c}}= g^{\cgi{c}}{}_{\cgi{d}}
y^{\cgi{d}}$, so $y^{\agi{c}} = [\bdelt{p}, x^{\cgi{c}}]$. In this
case, we give examples of tensor operations. In each case the
expression on the left is defined by that on the right.
\begin{align*}
 S_{\agi{b a}} & = [\bdelt{p}, S_{\cgi{b a}}], \\
 S_{\agi{a b}} y^{\agi{c}} & = [\bdelt{p}, S_{\cgi{a b}} x^{\cgi{c}}], \\
 \intertext{and}
 S_{\agi{a b}} y^{\agi{b}} & = [\bdelt{p}, S_{\cgi{a b}} x^{\cgi{b}}].
\end{align*}
\end{example}

\subsection{World tensors}\label{ssec:world-tensors}
We now specialise this machinery to deal with the world
tensors---that is, tensors defined in terms of tangent vectors to
a manifold. The tangent bundle has a direct and geometrical
interpretation, and need not be described as a vector bundle
associated to a principal fibre bundle, in this case the $\GLn$
frame bundle. However, when we later come to define spinors, there
is no analogous direct interpretation. They \emph{must be}
constructed geometrically as an associated vector bundle.
Preempting this, we show how that tangent bundle, and its related
tensor bundles, are generated from the frame bundle, applying the
theory of local tensor algebras and associated vector bundles.

The relevant Lie group is $\GLn$, acting on $\Real^n$. As in
\S\ref{sec:local-tensor-algebra}, there is an abstract index local
tensor algebra. The index set will consist of lowercase \rroman
letters. The relevant principal fibre bundle is the frame bundle
described in \S\ref{ssec:frame-bundle}.

We can now reobtain the tangent bundle, as an associated vector
bundle. Specifically, $T M \cong F M \times_{\GLn} \Real^n$, as
follows. If $\bdelt{p} = (e_1,\ldots,e_n) \in F M$, and $v \in
\Real^n$, then
\[[\bdelt{p}, v] = \sum_{i=1}^n e_i v^i =
(e_1,\ldots,e_n) \begin{pmatrix} v^1 \\ \vdots \\
v^n \end{pmatrix}.\] This is well defined, as
\begin{align*}
 [\bdelt{p} g, v] = (e_1,\ldots,e_n) \begin{pmatrix}
g^{1}{}_{1} & \cdots & g^{1}{}_{n} \\
\vdots & \ddots & \vdots \\
g^{n}{}_{1} & \cdots & g^{n}{}_{n}
\end{pmatrix} \begin{pmatrix} v^1 \\ \vdots \\
v^n \end{pmatrix} = [\bdelt{p}, g v].
\end{align*}

Equipped with this isomorphism, we henceforth always consider the
frame bundle as primary, and the tangent bundle a derived object.

Producing the world tensor algebra is now simply a matter of
stating that it is the global abstract index algebra associated
with the frame bundle, and the representation of $\GLn$ on
$\Real^n$. Thus for example we have tensor bundles
$\mathcal{T}^{\ai{a b}}{}_{\ai{c d}}$, etc. Tensor operations all
have a simple appearance in abstract index notation, but we are
assured that no reference is made to local coordinates or
components. That is, abstract index tensor equations are true
equations between tensors.

\subsection{Product bundles}\label{ssec:product-bundles}
Later, we will deal with two principal fibre bundles at once, with
one generally the frame bundle. In this case, we can have vectors
and tensors associated with either the frame bundle or the
abstract principal fibre bundle. If we wish to emphasis that
tensors are associated to the frame bundle, we call them world
tensors, as above. Tensors associated to a principal fibre bundle
other than the frame bundle will use special indices, either a
\ggerman script for an abstract principal fibre bundle, or
uppercase \rroman characters for an $\SL$ spinor structure,
defined later. Often, especially when using covariant derivatives
in \S\ref{ssec:covariant-derivatives}, we will need tensors with
indices associated with both of the bundles. This can be
formalised by considering these tensors as tensors in a vector
bundle associated to the product bundle, which we mention now.

\begin{defn*}[Product bundle over a base space]

Suppose \[\xi = \pfbundle{G}{P}{\pi_P}{M} \textrm{ and } \eta =
\pfbundle{H}{Q}{\pi_Q}{M}\] are principal fibre bundles defined
over the same base manifold. Define \[P \times_M Q =
\setc{\bdelt{(p,q)} \in P \times Q}{\pi_P(\bdelt{p}) =
\pi_Q(\bdelt{q})},\] and $\pi : P \times_M Q \To M$ by
$\pi(\bdelt{p},\bdelt{q}) = \pi_P(\bdelt{p})$. The product group
$G \times H$ acts on $P \times_M Q$ by $(\bdelt{p},\bdelt{q})(g,h)
= (\bdelt{p} g ,\bdelt{q} h)$. Then the principal fibre bundle
$\pfbundle{G \times H}{P \times_M Q}{\pi}{M}$ is called the
product principal fibre bundle of $\xi$ and $\eta$ over $M$.
\end{defn*}

Given an associated vector bundle for each of the two principal
fibre bundles, we can form an associated vector bundle for the
product bundle, using the tensor product of the underlying
representations. This means we can use equations with tensors with
two types of indices unambiguously.

\section{The special orthogonal groups}\label{sec:SO-groups}
In order to discuss the special orthogonal group $SO(p,q)$, we
return to local tensor algebras, and specialise to the
representation of $\GLn$ acting on $\Real^n$. Again, we use
lowercase \rroman indices for the abstract index labelling set.

\subsection{The indefinite inner product, and $SO(p,q)$ as a
subgroup of $\GLn$} We introduce an \emph{inner product} on
$\Real^n$, where $p+q=n$. The inner product is a symmetric valence
$\valence{0}{2}$ tensor, written $\eta_{\ci{a} \ci{b}}$, and is
not necessarily positive definite. In fact, if $q \neq 0$ then it
will not be positive definite, so the term inner product is used
only loosely here. We define $\eta_{\ci{a} \ci{b}}$ so for any $x,
y \in \Real^n$,
\begin{equation}
\label{eq:innerproduct}
 \eta_{\ci{a} \ci{b}} x^{\ci{a}} y^{\ci{b}}
 =
  \sum_{\ci{i}=1}^{p} x^{\ci{i}} y^{\ci{i}} -
 \sum_{\ci{j}=p+1}^{p+q} x^{\ci{j}} y^{\ci{j}}.
\end{equation}
Since $\eta_{\ci{a b}}$ is nondegenerate, it has an inverse as a
map from $\Real^n$ to its dual, in the sense that there is a
valence $\valence{2}{0}$ tensor $\eta^{\ci{a b}}$ such that
$\eta^{\ci{a c}} \eta_{\ci{c b}} = \delta^{\ci{a}}_{\ci{b}}$. %-% compare this with spinor algebra...

The orthogonal group is then the subgroup of $\GLn$ preserving
$\eta_{\ci{a} \ci{b}}$, and is denoted $O(p,q)$. An element $k \in
\GLn$ acts on $\Real^n$ by $x^{\ci{a}} \mapsto
{k^{\ci{a}}}_{\ci{b}} x^{\ci{b}}$. Thus the action of $k^{-1}$ on
the inner product, a valence $\valence{0}{2}$ tensor, is given by
\[\eta_{\ci{a} \ci{b}} \mapsto {k^{\ci{c}}}_{\ci{a}}
{k^{\ci{d}}}_{\ci{b}} \eta_{\ci{c} \ci{d}},\] and so $O(p,q)$ is
the subgroup of all $k^{-1} \in \GLn$ such that
\begin{equation}\label{eq:k-in-O}
\eta_{\ci{a} \ci{b}} = {k^{\ci{c}}}_{\ci{a}} {k^{\ci{d}}}_{\ci{b}}
\eta_{\ci{c} \ci{d}}.
\end{equation}
It is clear that $O(p,q)$ does actually form a subgroup, and so
equivalently $O(p,q)$ is the collection of all $k \in \GLn$ so
Equation \eqref{eq:k-in-O} holds.

\subsection{Index manipulations in the tensor
algebra}\label{ssec:index-manipulation} Once we have fixed this
inner product, we use it to introduce \emph{index raising} and
\emph{index lowering} conventions for the tensors over $\Real^n$.
Specifically, given a tensor ${T^{\ci{a}_1 \dots
\ci{a}_k}}_{\ci{b}_1 \dots \ci{b}_l}$, define
\[T^{\ci{a}_1 \dots \ci{a}_{i-1}}{}_{\ci{a}_i}{}^{\ci{a}_{i+1} \dots \ci{a}_k}{}_{\ci{b}_1 \dots \ci{b}_l} = T^{\ci{a}_1 \dots \ci{a}_{i-1}}{}^{\ci{c}_i}{}^{\ci{a}_{i+1} \dots \ci{a}_k}{}_{\ci{b}_1 \dots \ci{b}_l} \eta_{\ci{c}_i \ci{a}_i}\]
and
\[T^{\ci{a}_1 \dots \ci{a}_k}{}_{\ci{b}_1
\dots \ci{b}_{j-1}}{}^{\ci{b}_j}{}_{\ci{b}_{j+1} \dots \ci{b}_l} =
T^{\ci{a}_1 \dots \ci{a}_k}{}_{\ci{b}_1 \dots
\ci{b}_{j-1}}{}_{\ci{d}_j}{}_{\ci{b}_{j+1} \dots \ci{b}_l}
\eta^{\ci{d}_j \ci{b}_j}.\]

Thus given a valence $\valence{k}{l}$ tensor, we obtain a number
of other tensors, all denoted with the same kernel letter, but
with different arrangements of indices. Within these conventions,
it is important to keep track of the order of superscript and
subscript indices, because, for example, if $T^{\ci{a} \ci{b}}$ is
a valence $\valence{2}{0}$ tensor, then
\[T_{\ci{a}}{}^{\ci{b}} = T^{\ci{c} \ci{b}}
\eta_{\ci{c} \ci{a}} \neq T^{\ci{b} \ci{c}} \eta_{\ci{c} \ci{a}} =
T^{\ci{b}}{}_{\ci{a}},\] unless $T^{\ci{a} \ci{b}}$ happens to be
symmetric. However, as long as we keep track of the order of
indices, we can repeatedly raise or lower indices according to
this convention, and such raisings and lowerings commute. Further,
if we raise and then lower the same index, or \emph{vice versa},
we return to the original tensor, because $\eta^{\ci{a}\ci{b}}$
has been defined as the inverse of $\eta_{\ci{a}\ci{b}}$. The
notation for the inverse of $\eta_{\ci{a}\ci{b}}$ is consistent
with these conventions, in that $\eta^{\ci{a b}} = \eta^{\ci{ac}}
\eta^{\ci{bd}} \eta_{\ci{cd}}$. Finally we point out that the
symmetry of the inner product means that, for example $v_{\ci{a}}
= v^{\ci{b}} \eta_{\ci{b a}} = v^{\ci{b}} \eta_{\ci{a b}}$.

\subsection{Connected components}
The orthogonal group $O(p,q)$ is not connected. It has at least
two connected components, since the determinant gives an onto map
$\det: O(p,q) \To \{\pm 1\}$. The special orthogonal group
$SO(p,q)$ is the subgroup of $O(p,q)$ consisting of the
automorphisms of determinant one. When both $p$ and $q$ are at
least $1$, the special orthogonal group $SO(p,q)$ is not connected
either \cite[Proposition 1.124]{kna:lgbi}. We take $SO_0(p,q)$ to
be the connected component of the identity, which is a closed
subgroup of $SO(p,q)$, and so itself a Lie group. We will at times
simply write $SO$, to indicate the connected component of an
arbitrary orthogonal group.

Throughout this work, we will single out the group $\LL$ for
special consideration, for two reasons. Firstly, it is the
physically relevant group in general relativity. Secondly, it is
fortuitously amenable to analysis, and much can be said about
spinor structures for this group, in particular because we can
give an explicit description of its simply connected covering
space $\SL$, in \S\ref{ssec:covering-map}.

\subsection{Lie algebras}\label{ssec:lie-algebras}
The Lie algebras of $O(p,q)$, $SO(p,q)$ and $SO_0(p,q)$ are all
isomorphic, since the Lie algebra of a Lie group depends only on
the identity component. We denote this Lie algebra as
$\LAM{so}{p,q}$, or simply as $\LA{so}$ in the general case. It
consists of all the endomorphisms ${X^{\ci{a}}}_{\ci{b}}$ of
$\Real^n$ which are antisymmetric with respect to $\eta_{\ci{a}
\ci{b}}$, in the sense that
\[{X^{\ci{a}}}_{\ci{b}} + {X_{\ci{b}}}^{\ci{a}} = 0,\]
or
\[\eta_{\ci{a}\ci{c}}({X^{\ci{a}}}_{\ci{b}} +
{X_{\ci{b}}}^{\ci{a}}) = X_{\ci{c}\ci{b}} + X_{\ci{b}\ci{c}} = 2
X_{(\ci{b}\ci{c})} = 0.\] See \cite[\S 19.4.3]{die:ta4} for
details. We will not need to know anything further about the Lie
algebra structure of $\LA{so}$ for the purpose of this thesis.

\section{Orthonormal structures: two
viewpoints}\label{sec:orthonormal-structures}

In this section we discuss \defnemph{orthonormal structures} from
two viewpoints.

Firstly, from a classical point of view, an orthonormal structure
on a smooth manifold $M$ consists of a \emph{metric tensor} with
appropriate properties. The metric tensor is a nondegenerate
valence $\valence{0}{2}$ tensor defined on all of $M$, with a
certain \emph{signature}. A manifold equipped with such a metric
tensor is called a Riemannian or pseudo-Riemannian manifold.
Additionally we might specify an \emph{orientation} on the
manifold.

The more modern second,idea of an orthonormal structure involves
principal fibre bundles. This approach was developed originally by
E. Cartan.\footnote{See \S 20 and particularly \S 20.7 of
Dieudonn{\'e} \cite{die:ta4}, and also Cartan \cite{car:tgfcgd}.}
Starting with the $\GLn$ frame bundle, we can \emph{reduce the
structure group} in various ways. We will see that the reductions
to principal fibre bundles with structure group $O(p,q)$
correspond exactly to choices of metric tensors. A reduction of
the structure group to $GL^+(n,\Real)$, the positive determinant
matrices, is equivalent to choosing an orientation. A further
reduction to $SO(p,q)$ or $SO_0(p,q)$ is equivalent to choosing
both a metric tensor and an orientation.

We begin by giving precise definitions of all these concepts, and
then proceed to show the equivalence between the two descriptions.

\subsection{Classical description of a metric tensor}

\begin{defn}\label{defn:metric-tensor}
A \defnemph{metric tensor} on a smooth $n$ dimensional manifold
$M$ is a valence $\valence{0}{2}$ tensor $g_{\ai{a b}}$ such that
\begin{enumerate}
\item it is symmetric, so $g_{\ai{a b}} = g_{\ai{b a}}$,
\item it is nondegenerate, so $g_{\ai{a b}} y^{\ai{b}} = 0$ if and only if $y^{\ai{b}} =
0$, and
\item there are positive integers $p,q$, so $p+q=n$, and at every
point of the manifold there are vectors $y_1, \dots, y_n$ so that
\[g_{\ai{a b}} y_{\ci{i}}^{\ai{a}} y_{\ci{j}}^{\ai{b}} =
\begin{cases}
 \phantom{-}0 & \text{if $i \neq j$,} \\
 \phantom{-}1 & \text{if $1 \leq i = j \leq p$,} \\
 -1 & \text{if $p+1 \leq i = j \leq q$},
\end{cases}
\]
or equivalently
\[g_{\ai{a b}} y_i^{\ai{a}} y_j^{\ai{b}} = \eta_{ij}.\]
\end{enumerate}
Such a collection of vectors is called an orthonormal frame. Note
that an orthonormal frame is in fact a frame in our previous
sense. We say that such a metric tensor has signature $(p,q)$.
\end{defn}

A manifold along with a metric tensor is called a
pseudo-Riemannian manifold. If the signature of the metric tensor
is $(n,0)$ we say that the manifold is Riemannian, and if the
signature is $(1,n-1)$ we say that it is
Lorentzian.\footnote{There is no significant difference here
between the signatures $(1,n-1)$ and $(n-1,1)$.} The physically
significant situation, in general relativity, is a $4$ dimensional
Lorentzian manifold with signature $(1,3)$.

\begin{defn*}
An \defnemph{orientation} on a smooth $n$ dimensional manifold $M$
is an equivalence class $[\omega]$ of nowhere zero antisymmetric
valence $\valence{0}{n}$ tensors $\omega_{\ai{a}_1 \dots
\ai{a}_n}$ on $M$, where two such tensors are equivalent if one is
a positive multiple of the other.
\end{defn*}

Given an orientation $[\omega]$, we say that a frame $\bdelt{p} =
(y_1, \dots ,y_n)$ is positively oriented if
\[\omega_{\ai{a}_1 \dots \ai{a}_n} y_1^{\ai{a}_1} \dotsm y_n^{\ai{a}_n} > 0.\]

It is known from linear algebra that on $\Real^n$ the space of
local valence $\valence{0}{n}$ tensors is one dimensional, and in
particular every such tensor is a multiple of the determinant,
which we write $\epsilon_{\ci{a}_1 \dots \ci{a}_n}$. Here we think
of the determinant as acting on $n$ vectors by evaluating the
determinant of the matrix formed with these vectors as columns.
Thus \[\epsilon_{\ci{a}_1 \dots \ci{a}_n} y_1^{\ci{a}_1} \cdots
y_n^{\ci{a}_n} = \det
\begin{pmatrix}
 y_1^1 & \cdots & y_n^1 \\
 \vdots & \ddots & \vdots \\
 y_1^n & \cdots & y_n^n
\end{pmatrix}.\]
Now according to Equation \eqref{eq:frame-bundle-action}, $h \in
\GLn$ transforms the frame $(y_1, \dots ,y_n)$ to $(y_{\ci{i}}
h^{\ci{i}}{}_1,\ldots,y_{\ci{i}} h^{\ci{i}}{}_n)$, and the
determinant here gives
\[
\det \begin{pmatrix}
 y_1^1 & \cdots & y_n^1 \\
 \vdots & \ddots & \vdots \\
 y_1^n & \cdots & y_n^n
\end{pmatrix} \cdot \det \begin{pmatrix}
h^{1}{}_{1} & \cdots & h^{1}{}_{n} \\
\vdots & \ddots & \vdots \\
h^{n}{}_{1} & \cdots & h^{n}{}_{n}
\end{pmatrix}.
\]
Thus $h$ acting on $\epsilon_{\ci{a}_1 \dots \ci{a}_n}$ gives
$\det(h) \epsilon_{\ci{a}_1 \dots \ci{a}_n}$. We will use these
facts presently.

\subsection{Reduction to an orthogonal group}

Our second description of an orthonormal structure is as a
reduction of the $\GLn$ frame bundle for $M$ to an $SO_0(p,q)$
bundle over $M$. As we will see, this reduction defines a metric,
and gives an orientation to $M$. If $p,q \neq 0$, it also provides
a time orientation.

Suppose $H$ is a subgroup of $G$, and that $\xi =
\pfbundle{H}{P}{\pi_P}{M}$ is an $H$ principal fibre bundle over a
base space $M$, and $\eta = \pfbundle{G}{P'}{\pi_{P'}}{M}$ is a
$G$ principal fibre bundle over $M$.
\begin{defn*}
We say that $\xi$ is a \defnemph{reduction} of $\eta$ if there is
a principal fibre bundle morphism $r:P \To P'$ such that
$r(\bdelt{p} h)=r(\bdelt{p}) h$ for every $h \in H$.
\end{defn*}

The reduction map $r$ is injective, since $H$ acts transitively on
each fibre of $P$, and freely on $P'$.

\subsection{Equivalence of these descriptions}\label{ssec:bundle-metric-equivalence}
Showing that a metric defines a reduction of the frame bundle $F
M$ to an $O(p,q)$ bundle is relatively straightforward, and we do
this first. Simply, this bundle $O M$ is the collection of all
orthonormal frames in $F M$, and $O(p,q)$ acts on it as a subgroup
of $\GLn$ acting on $F M$. We need to check that this satisfies
the axioms for an $O(p,q)$ principal fibre bundle. Almost all the
conditions of Definition \ref{defn:principal-fibre-bundle} are
satisfied immediately. We need only check that $O(p,q)$ maps $O M$
to itself, and that it acts transitively on each fibre.

Suppose $\bdelt{p} = (y_1, \dots, y_n)$ is an orthonormal frame,
so $g_{\ai{a b}} y_i^{\ai{a}} y_j^{\ai{b}} = \eta_{\ci{i j}}$ for
each $i,j=1,\dots,n$. Then, according to the action defined in
Equation \eqref{eq:frame-bundle-action}, $\bdelt{p} h =
(y_{\ci{i}} h^{\ci{i}}{}_1,\ldots,y_{\ci{i}} h^{\ci{i}}{}_n)$, and
so if $h \in O(p,q)$,
\[g_{\ai{a b}} y_{\ci{i}}^{\ai{a}}
h^{\ci{i}}{}_k y_{\ci{j}}^{\ai{b}} h^{\ci{j}}{}_l = h^{\ci{i}}{}_k
h^{\ci{j}}{}_l \eta_{\ci{i j}} = \eta_{\ci{k l}}.
\]
Thus, as we expect, elements of $O(p,q)$ map orthonormal frames to
orthonormal frames.

Further, if $\bdelt{p'} = (x_1, \dots, x_n)$ is another
orthonormal frame at the same point, there must be some element $k
\in \GLn$ that takes $\bdelt{p}$ to $\bdelt{p'}$. However,
according to the above calculation, this element $k$ preserves the
inner product $\eta_{\ci{i j}}$, and so is in fact an element of
$O(p,q)$. This establishes that $O(p,q)$ acts transitively on the
fibres.

Next, we consider orientations, claiming that an orientation
results in a reduction to a $SO(p,q)$ bundle, by taking the
collection of all positively oriented orthonormal frames.
Following exactly the argument above, and the discussion of
determinant above, we see that any element of $SO(p,q)$ preserves
the volume form, and so takes positively oriented frames to
positively oriented frames. Going the other way, given two
positively oriented orthonormal frames, there must be an element
of $O(p,q)$ taking one to the other, and the same argument shows
that this element must have positive determinant, and so lie in
$SO(p,q)$.

Conversely, suppose $O M$ is an $O(p,q)$ bundle over $M$, which is
a reduction of the frame bundle $F M$. Suppose $r$ is the
reduction map, a principal bundle morphism $r: O M \To F M$. We
will define a metric tensor on $M$. Specifically, at each point
$m$ of $M$, chose $\bdelt{b} \in \pi_{FM}^{-1}(m)$ so that
$\bdelt{b} = r(\bdelt{f})$ for some $\bdelt{f} \in OM$. Define
$g_{\ai{a b}}$ at that point by \[g_{\ai{a b}} = [\bdelt{b},
\eta_{\ci{a b}}].\] This is well defined, since if $\bdelt{b'} =
r(\bdelt{f'})$ is another point in $\pi_{FM}^{-1}(m)$, then
$\bdelt{f'} = \bdelt{f} k$ for some $k \in SO_0(p,q)$, and so
$\bdelt{b'} = \bdelt{b} k$ also, and so \[[\bdelt{b'},\eta_{\ci{a
b}}] = [\bdelt{b}, k^{\ci{c}}{}_{\ci{a}} k^{\ci{d}}{}_{\ci{b }}
\eta_{\ci{c d}}] = [\bdelt{b}, \eta_{\ci{a b}}].\] This tensor
field is smooth, since a local smooth cross section of $O M$ gives
a local smooth cross section of $F M$ via $r$. Checking that
$g_{\ai{a b}}$ satisfies the axioms of a metric tensor in
Definition \ref{defn:metric-tensor} is very straightforward.
Symmetry and nondegeneracy follow from the same properties of
$\eta_{\ci{a b}}$, and the orthonormal basis is given by
\[y_i^{\ai{a}} = [\bdelt{b}, e_i^{\ci{a}}],\] where $e_i \in \Real^n$ is
the $i$-th standard basis vector.

Further, if $O M$ is an $SO(p,q)$ bundle, then we obtain an
orientation as well. Because elements of $SO(p,q)$ preserve the
determinant, we can define a tensor field $\omega_{\ai{a}_1 \dots
\ai{a}_n} = [\bdelt{p}, \epsilon_{\ci{a}_1 \dots \ci{a}_n}]$, for
all $\bdelt{p} \in O M$. This is everywhere nonzero, and
antisymmetric, and so gives an orientation.

This argument is related to those in \cite[\S 20.7]{die:ta4} or
\cite[\S3.3]{ish:mdgfp}, but makes use of the associated bundle
construction.

Note that if $p,q > 0$, then $SO(p,q)$ is not connected. A further
reduction of the structure group to $SO_0(p,q)$, the connected
component of the identity, is achieved by choosing a time
orientation \cite[\S2.4]{ben:isgap}.
% look here also if you want to understand time orientations for (p,q) properly.
On Lorentzian manifolds this is a nowhere zero vector field
$x^{\ai{a}}$ so $g_{\ai{a b}} x^{\ai{a}} x^{\ai{b}} > 0$
everywhere.\footnote{To be precise, it is an equivalence class of
these, where $x^{\ai{a}}$ and $y^{\ai{b}}$ are equivalent if
$g_{\ai{a b}} x^{\ai{a}} y^{\ai{b}}
> 0$ everywhere.} We will not go into the details here, because for
general $p$ and $q$ they are awkward, but henceforth always
consider $SO_0(p,q)$ reductions of the frame bundle, so that the
structure group is connected.

\subsection{The world tensor algebra for an orthonormal
bundle}\label{ssec:tensor-algebra-orthonormal-bundle} At this
point we are considering two bundles, the frame bundle, and a
reduction of the frame bundle, the orthonormal bundle. We have
previously constructed the world tensor algebra as a collection of
vector bundles associated to the frame bundle. Similarly we can
now construct vector bundles associated to the orthonormal bundle.
However, we quickly find that they are equivalent. If $V$ is a
vector space carrying a representation $\lambda$ of $SO$, such
that $\lambda$ is a tensor product of copies of the matrix
representation and its dual, then we can extend this
representation to a representation of $\GLn$, simply because the
matrix representation of $SO$ extends to the matrix representation
of $\GLn$.

\begin{prop}
The map $OM \times_{SO} V \To FM \times_{\GLn} V$ given by
$[\bdelt{p},v] \mapsto [r(\bdelt{p}),v]$ is an isomorphism of the
vector bundles.
\end{prop}
\begin{proof}
It is clear that this map is linear. Additionally, it is
surjective, because any $\bdelt{b} \in FM$ can be written as
$r(\bdelt{p})g$ for some $\bdelt{p} \in OM$, and $g \in \GLn$. It
is injective, since if
$[r(\bdelt{p}_1),v_1]=[r(\bdelt{p}_2),v_2]$, then there is a $g
\in SO$ so that $\bdelt{p}_2 = \bdelt{p}_1 g$, and so $v_1 =
\lambda(g) v_2$, and finally $[\bdelt{p}_1,v_1] =
[\bdelt{p}_1,\lambda(g)v_2] = [\bdelt{p}_2,v_2]$.
\end{proof}
This shows that we can equally well consider world tensors as
lying in a vector bundle associated to $OM$ or as lying in one
associated to $FM$. This occurs because of the apparently trivial
fact that the representations of $SO$ extend to representations of
$\GLn$. We will see however that representations of the covering
group $\tilde{SO}$ need not extend to representations of
$\GLncover{n}$. This has implications for the construction of a
spinor algebra in \S\ref{sec:spinor-algebra}.

As we have seen, the metric tensor $g_{\ai{a b}}$ has a simple
form $g_{\ai{a b}} = [r(\bdelt{b}), \eta_{\ci{a b}}]$, and so the
index manipulation rules for local $\Real^n$ tensors, as in
\S\ref{ssec:index-manipulation}, carry across immediately to the
world tensor algebra. For example, given a world vector
$x^{\ai{a}}$ at a point $m \in M$, we can always find a $\bdelt{b}
\in \pi_{O M}^{-1}(m)$, and write the world vector in the form
$x^{\ai{a}} = [r(\bdelt{b}), x^{\ci{a}}]$. In this case the
associated `lowered' tensor, $x_{\ai{a}}$ is defined by
$x_{\ai{a}} = [r(\bdelt{b}), x_{\ci{a}}] = [r(\bdelt{b}),
x^{\ci{b}} \eta_{\ci{b a}}]$. The $O(p,q)$ invariance of
$\eta_{\ci{b a}}$ and the fact that $r$ is a reduction map ensures
that this is well defined.

\subsection{The orthonormal bundle as a configuration space}
At this point we briefly describe a useful way of thinking about
orthonormal bundles. Firstly recall how $SO(n)$ can be used to
describe the possible orientations\footnote{`Orientation' is
intended here in the everyday sense, not the mathematical sense
for manifolds or vector spaces.} of an object $n$ dimensional. If
we associate arbitrarily one orientation with the identity, there
is a one to one correspondence between orientations and elements
of $SO(n)$.

Next, suppose we consider an $SO(n)$ bundle reduction of the frame
bundle $F M$ over a manifold $M$. The points of this bundle
corresponds exactly to the possible configurations of an $n$
dimensional `oriented particle' on $M$, that is, an object with a
position and an orientation. The group $SO(n)$ acts in the obvious
way as rotations.

We can similarly interpret an $SO_0(1,3)$ bundle, for example, as
the configurations of a relativistic particle.

\section{Tensor calculus}\label{sec:tensor-calculus}
In the following sections, we will demonstrate, given an
orthonormal structure, the existence of a \emph{metric} connection
on the manifold. This connection is not unique
however.\footnote{The standard theory of pseudo-Riemannian
geometry picks out a particular \emph{torsion free} metric
connection. This is called the Levi--Civita connection. Although
it is possible to understand this connection in the context of
frame bundles and connection forms thereon, this will not be
needed for our purposes.} Our construction will be somewhat
unconventional, using the principal fibre bundle approach. Any
principal fibre bundle allows a connection, and we will see that
all connections on $OM$, the total space of the orthonormal
bundle, are automatically metric connections with respect to the
metric induced by the bundle. Along the way we will give a
description of the relationship between connections and covariant
derivatives for arbitrary principal fibre bundles. This
description is not absolutely complete---we try to balance
checking every detail against useful explanation. The generality
of this section will be vital later in discussing spinor covariant
derivatives in \S\ref{sec:lifting-connection} and the Dirac
equation in \S\ref{sec:dirac-equation}.

The material in the following sections is required to reach our
aim in \S \ref{ssec:metric-connections}. However, most of Part
\ref{part:spinor-structure-classification} may be read only having
looked at \S \ref{ssec:connection-forms} and the first parts of \S
\ref{ssec:connection-forms-covariant-derivatives}, introducing
connections and parallel transport. Part
\ref{part:implications-for-physics}, however, relies more heavily
on \S \ref{ssec:covariant-derivatives} and \S
\ref{ssec:metric-connections}.

\subsection{Connection forms}\label{ssec:connection-forms}
We first recall the definition of a connection form (c.f. \cite[p.
288]{cho:amp} or \cite[\S3.5]{ish:mdgfp}). We consider a principal
fibre bundle $\xi = \pfbundle{G}{P}{\pi}{M}$. At each point
$\bdelt{p} \in P$, there is the \emph{vertical subspace} of
$T_\bdelt{p} P$, given by $V_\bdelt{p} = \ker \pi_*$. We describe
two maps identifying $P_\bdelt{p} = \pi^{-1}(\pi(\bdelt{p}))$ with
$G$, defining
\begin{align*}
 \theta_\bdelt{p} & : G \To P_\bdelt{p} && \text{by} & \theta_\bdelt{p}(g) & = \bdelt{p} g && \text{and} \\
 \psi_\bdelt{p}& : P_\bdelt{p} \To G && \text{by} & \psi_\bdelt{p}(\bdelt{p'}) & = \tau(\bdelt{p},
\bdelt{p'}).
\end{align*}
($\tau$ is the translation function, described in
\S\ref{ssec:principal-fibre-bundles}.) Now
$\theta_\bdelt{p}(\psi_\bdelt{p}(\bdelt{p'}))=\bdelt{p'}$ and
$\psi_\bdelt{p}(\theta_\bdelt{p}(g))=g$. Also $\pi \compose
\theta_\bdelt{p}(g) = \pi(\bdelt{p} g)=\bdelt{p}$ so $\pi \compose
\theta_\bdelt{p}$ is a constant function for each $\bdelt{p}$, so
$\pi_* \theta_{\bdelt{p} *} v = (\pi \compose \theta_\bdelt{p})_*
v = 0$, and thus both $\theta_{\bdelt{p} *} : \LA{G} \To
V_\bdelt{p}$ and $\psi_{\bdelt{p} *} : V_\bdelt{p} \To \LA{G}$ are
linear isomorphisms. This map $\psi_{\bdelt{p} *}$, taking the
vertical subspace at a point to the Lie algebra, will reappear
many times.
\begin{defn}
\label{defn:connection-form} A \defnemph{connection form} on $\xi$
is a linear map $\omega:TP \To \LA{G}$, that is, a $1$-form on
$P$, with values in the Lie algebra of $G$, such that
\begin{enumerate}
\item $\omega_\bdelt{p}(u)=\psi_{\bdelt{p} *}u$ for all $u \in V_\bdelt{p}$,
\item $(g^* \omega)_{\bdelt{p}} (u) = \omega_{\bdelt{p} g}(g_* u)=\Ad(g^{-1})\omega_\bdelt{p}(u)$ for all $g \in
G$.
\end{enumerate}
\end{defn}

This definition prompts a comment on the notation. We will
consistently use $g_*$ to mean the derivative of the right action
by $g$ on $P$ and $g^*$ to mean the pull-back by the right action.
It is important to remember that, regardless of this notation, $g$
acts on the right!

The first part of this definition determines how the connection
form maps the vertical vectors into the Lie algebra, and the
second part is called the `elevator property'.

We now establish the existence of a connection form on any
principal fibre bundle. This connection form is by no means
unique. In particular, the result here shows that there is always
a connection available on the frame bundle, which gives a
covariant derivative on the tangent bundle and the associated
tensor bundles. Further, given a reduction of the frame bundle
associated with a metric to the orthonormal bundle, there is a
connection on the orthonormal bundle.

\begin{prop}
\label{prop:connections-exist} There exists a connection form on
any $G$ principal fibre bundle, $\xi = \pfbundle{G}{P}{\pi}{M}$.
\end{prop}
\begin{proof}
This is an entirely standard argument. However, due to its
importance, both in providing connections as technical tools, and
underlying our interest in connections on spinor structures in \S
\ref{sec:lifting-connection}, we give a proof in \S
\ref{ssec:proof-of-proposition-connections-exist}.
\end{proof}

\begin{cor}
If $M$ is an $n$ dimensional manifold, then there exists a
connection on the frame bundle $\pfbundle{\GLn}{FM}{\quad}{M}$.
\end{cor}

\begin{cor}
If $M$ is an $n$ dimensional manifold, $G$ is a subgroup of
$\GLn$, and $\xi = \pfbundle{G}{P}{\quad}{M}$ is a $G$-reduction
of the frame bundle for $M$, then there exists a connection on
$\xi$. In particular, if $G=SO$ and $P$ is a bundle of oriented
orthonormal frames then there is a connection.
\end{cor}

We will later prove in \S\ref{ssec:metric-connections} that a
connection on an orthonormal frame bundle corresponds with the
usual idea of a metric covariant derivative.

\subsection{Parallel transport}
\label{ssec:connection-forms-covariant-derivatives}

In this section we outline the relationship between connection
forms and parallel transports, and lay the groundwork for
covariant derivatives. From a geometrical point of view, the
parallel transport provides a bridge between the notions of
connection form and covariant derivative. Compare
\cite{ish:mdgfp,mil:gc}.

\subsubsection{Horizontal lifting map}
\label{ssec:horizontal-lifting-map}

Suppose a connection form $\omega$ is defined on the total space
$P$ of a bundle $\pfbundle{G}{P}{\pi}{M}$. For each point
$\bdelt{p} \in P$, we call the kernel of $\omega$ the
\emph{horizontal subspace} $H_{\bdelt{p}}$ of $T_{\bdelt{p}} P$.
Since the image of $\omega$ is all of $\LA{G}$, via the first
property in Definition \ref{defn:connection-form}, by counting
dimensions we see that the dimension of the horizontal subspace is
exactly the dimension of the base manifold. Thus $T_{\bdelt{p}} P
= V_{\bdelt{p}} \oplus H_{\bdelt{p}}$. Further, if $u \in
H_{\bdelt{p}}$, and $u \neq 0$, then $\pi_* u \neq 0$. The
derivative $\pi_*$ restricted to $H_{\bdelt{p}}$ is thus a linear
isomorphism, and we denote the inverse map $\sigma_{\bdelt{p}} :
T_{\pi(\bdelt{p})} \To H_{\bdelt{p}}$, and call it the horizontal
lifting map.

We now prove a lemma about the horizontal lifting map.
\begin{lem}\label{lem:connection-form-from-horizontal-lifting-map}
The connection form is determined by the horizontal lifting map.
\begin{equation}
\omega_{\bdelt{p}}(u) =  \psi_{\bdelt{p} *}(u - \sigma_\bdelt{p}
\pi_* u),
\end{equation}
for all $\bdelt{p} \in P$ and $u \in T_\bdelt{p} P$.
\end{lem}
\begin{proof}
Write $u = u - \sigma_\bdelt{p} \pi_* u + \sigma_\bdelt{p} \pi_*
u$. Now $\omega_{\bdelt{p}}(\sigma_\bdelt{p} \pi_* u) = 0$, since
$\image \sigma_{\bdelt{p}} = \ker \omega_{\bdelt{p}}$. Further
$\pi_*(u - \sigma_\bdelt{p} \pi_* u) = \pi_* u - \pi_* u = 0$, so
$\omega_{\bdelt{p}}(u - \sigma_\bdelt{p} \pi_* u) =
\psi_{\bdelt{p} *}(u - \sigma_\bdelt{p} \pi_* u)$, proving the
result.
\end{proof}

\subsubsection{Parallel transport}\label{ssec:parallel-transport}
The horizontal lifting map $\sigma$ allows us to define parallel
transport. Given a vector field $v^{\ai{a}} \in \vf{U}$ on an open
set $U \subset M$, we apply $\sigma$ to lift it to a horizontal
vector field defined on $\pi^{-1}(U) \subset P$. Fixing some
$\bdelt{p} \in \pi^{-1}(U)$ gives us an initial point from which
to form an integral curve of the horizontal vector field. This
integral curve is fundamental to parallel transportation.

From a simple path (smooth, with no self-intersections) $\gamma :
\I \To M$ in $M$ we can form the tangent vector field along the
curve, and, at least near $\gamma(0)$, extend this to a vector
field defined on a neighbourhood $U$ of $\gamma(0)$. Again, the
horizontal lifting map applied to this vector field gives a
horizontal vector field on $\pi^{-1}(U)$. Suppose $\bdelt{p_0} \in
\pi^{-1}(\gamma(0))$. The integral curve of the horizontal vector
field starting at $\bdelt{p_0}$ is $\bdelt{p} : \J \To P$, for
some $\eps > 0$, with $\bdelt{p}(0) = \bdelt{p_0}$, and in fact
$\pi(\bdelt{p}(t)) = \gamma(t)$. This last fact follows because
$\pi_*$ is the inverse of $\sigma$. We observe that
\begin{eqnarray*}
\frac{d}{dt}(\pi(\bdelt{p}(t))) & = & \pi_* \frac{d}{dt}
(\bdelt{p}(t)) \\
    & = & \pi_* \sigma_{\bdelt{p}(t)} \frac{d}{dt}(\gamma(t)) \\
    & = & \frac{d}{dt}(\gamma(t)).
\end{eqnarray*}

The curve $\bdelt{p}$ is the parallel transport of $\bdelt{p_0}$
along the curve $\gamma$.

A stronger version of this idea is established by the following
Proposition, allowing parallel transports along the entire length
of an arbitrary smooth path, and ensuring that the parallel
transport depends continuously upon the initial data.

\begin{prop}\label{prop:parallel-transport}
Given a smooth path $\alpha : \I \To M$, and $\bdelt{p_0} \in
\pi^{-1}(\alpha(0))$, we can form the parallel transport of
$\bdelt{p_0}$ along $\alpha$, which is a smooth curve $\bdelt{p} :
\I \To P$ such that
\begin{enumerate}
\item the projection down to $M$ is the original curve, $\pi(\bdelt{p}(t)) =
\alpha(t)$, and
\item the derivative at any point is given by the horizontal lift
of the derivative of the original curve, $\dot{\bdelt{p}}(t) =
\sigma_{\bdelt{p}(t)} \dot{\alpha}(t)$.
\end{enumerate}
Furthermore, suppose
\begin{enumerate}
\item $\left(\alpha_s\right)_{s\in \I}$ is a smooth family of
paths, in the sense that $(s,t) \mapsto \alpha_s(t)$ is smooth,
\item $s \mapsto \bdelt{p_0}{}_s$ is a smooth curve in $P$ with
$\pi(\bdelt{p_0}{}_s) = \alpha_s(0)$, and
\item $\bdelt{p}_s$ is the parallel transport of $\bdelt{p_0}{}_s$
along $\alpha_s$.
\end{enumerate}
Then the map $(s,t) \mapsto \bdelt{p}_s(t)$ is (at worst)
continuous.
\end{prop}
\begin{proof}
The method of construction is as described above---we simply add
here that the integral curve giving the parallel transport can be
extended so as to be defined over all of the interval $\I$,
following the argument of \cite[Proposition 3.1]{kob:fdg1}, or of
\cite[\S 18.6]{die:ta4}. We omit these details here.

The second part follows immediately from the fact that solutions
of differential equations depend (at worst) continuously on a
smoothly varying initial value \cite[\S IV.4]{boo:idmrg}. In more
detail, \S II.4 of \cite{kob:fdg1} proves that $s \mapsto
\bdelt{p}_s(t)$ is smooth for each $t \in \I$, and since $t
\mapsto \bdelt{p}_s(t)$ is also smooth for each $s \in I$, by the
first part of this proposition, the map $(s,t) \mapsto
\bdelt{p}_s(t)$ is certainly continuous. It is a possible, but not
necessary here, to prove a stronger result.
\end{proof}

We will use the second part of this Proposition later, in
establishing the Existence Theorem for spinor structures.

Now that we have a notion of parallel transport for the principal
fibre bundle, parallel transport in any of the associated vector
bundles is straightforward. Simply, a vector $v^{\agi{a}} = \left[
\bdelt{p_0}, v^{\cgi{a}} \right]$ at the point $\gamma(0)$ is
parallel transported as $v^{\agi{a}}(t) = \left[ \bdelt{p}(t),
v^{\cgi{a}} \right]$. We parallel transport the reference element
of $P$, leaving fixed the vector in the representation space.

\subsubsection{Local representatives and Christoffel
symbols}\label{sssec:local-representatives} A local section of a
bundle is a map $\sigma$ from an open set $U \subset M$ to $P$,
such that $\pi \compose \sigma = \textrm{id}_U$. Given a local
section, we can form a \emph{local representative} of the
connection form, $\sigma^* \omega$. The local representative is
then a $1$-form on the base space, with values in the Lie algebra.

Knowing the local section, this process can in fact be reversed
\cite[\S3.5]{ish:mdgfp}. That is, the local representatives
determine the connection. First we need to identify the tangent
space at any point of a Lie group $G$ with the Lie algebra, by
left translation. Denote the left translation by $g$ map as $L_g$,
so $L_{g^{-1}*} : T_g G \To T_e G = \LA{G}$. Thus given $\beta \in
T_g G$, we associate the element of the Lie algebra $L_{g^{-1}*}
\beta$. Suppose $\sigma : U \To P$ is a local section of a
principal fibre bundle. Define the related local trivialisation
$\psi:U \times G \To P$ by $\psi(m,g)=\sigma(m) g$.

\begin{prop}
\label{prop:local-representative} If $(\alpha,\beta) \in T_m U
\oplus T_g G$ then
\[
 (\psi^* \omega)_{(m,g)}(\alpha,\beta) =
 \Ad(g^{-1})((\sigma^* \omega)_m(\alpha)) + L_{g^{-1}*} \beta.
\]
\end{prop}
\begin{proof}
See \cite{ish:mdgfp}.
\end{proof}

In the special case of a connection on the frame bundle a
coordinate chart on the base manifold implicitly defines a cross
section of the bundle, via the coordinate basis. The local
representative formed using this cross section may be thought of
as `the connection form in local coordinates'. The above
proposition makes this precise.

The local representative of a connection form has an unusual
appearance in abstract index notation. For each choice of a
representation $\lambda$ of $G$ on a vector space $V$, we obtain a
representation of the Lie algebra $\LA{G}$ on the same vector
space. This associates with each element of the Lie algebra a
matrix acting on $V$. If a typical element of $V$ is written as
$v^{\cgi{d}}$, a kernel letter with a \ggerman superscript index,
then for each vector in the Lie algebra we obtain a tensor
${B^{\cgi{b}}}_{\cgi{c}}$. Thus the local representative is
denoted by a kernel letter with three indices, for example as
\begin{equation}\label{eq:local-represenative-index-notation}
\sigma^* \omega \leftrightarrow
A_{\ai{a}}{}^{\cgi{b}}{}_{\cgi{c}}.
\end{equation}
We will see in \S\ref{ssec:covariant-derivatives} that local
representatives written in this form are the appropriate
generalisation of \emph{Christoffel symbols} \cite[p.
62]{one:s-rgar} \cite[\S3.1]{wal:gr} to general principal fibre
bundles and their associated vector bundles.

\subsubsection{Christoffel symbols for tensor product
representations.} If the chosen representation is in fact a tensor
product of other representations, then we obtain a representation
of the Lie algebra on the tensor product space.

If ${T^{\cgi{a}_1 \dots \cgi{a}_k}_{\cgi{b}_1 \dots \cgi{b}_l}}$
lies in the representation $\tensors{k}{l}$, then an element $g$
of $G$ acts by
\[ \left({g(T)^{\cgi{a}_1 \dots \cgi{a}_k}_{\cgi{b}_1 \dots \cgi{b}_l}}\right) =
    {(g)^{\cgi{a}_1}}_{\cgi{c}_1} \cdots {(g)^{\cgi{a}_k}}_{\cgi{c}_k} {(g^{-1})^{\cgi{d}_1}}_{\cgi{b}_1} \cdots {(g^{-1})^{\cgi{d}_l}}_{\cgi{b}_l}
    {T^{\cgi{c}_1 \dots \cgi{c}_k}_{\cgi{d}_1 \dots \cgi{d}_l}}.\]
Thus if $\kappa \in \LA{G}$, and $g:\I \To G$ is a smooth path in
$G$ so $g(0) = e$ and $\kappa = \dot{g}(0)$, then, using the
Leibniz rule, $\kappa$ acts on ${T^{\cgi{a}_1 \dots
\cgi{a}_k}_{\cgi{b}_1 \dots \cgi{b}_l}}$ by
\begin{align}\label{eq:kappa-with-lots-of-indices}
 \kappa({T^{\cgi{a}_1 \dots \cgi{a}_k}_{\cgi{b}_1 \dots \cgi{b}_l}}) & =
     \dat{t}{0}
        {(g(t))^{\cgi{a}_1}}_{\cgi{c}_1} \cdots {(g(t))^{\cgi{a}_k}}_{\cgi{c}_k} {(g(t)^{-1})^{\cgi{d}_1}}_{\cgi{b}_1} \cdots {(g(t)^{-1})^{\cgi{d}_l}}_{\cgi{b}_l}
        {T^{\cgi{c}_1 \dots \cgi{c}_k}_{\cgi{d}_1 \dots \cgi{d}_l}}
        \\ \nonumber
     & =
    \left( \sum_{i=1}^{k} {\dot{g}(0)^{\cgi{a}_i}}_{\cgi{c}_i}
        {T^{\cgi{a}_1 \dots \cgi{c}_i \dots \cgi{a}_k}_{\cgi{b}_1 \dots \cgi{b}_l}}
        - \sum_{j=1}^{l} {\dot{g}(0)^{\cgi{d}_j}}_{\cgi{b}_j} {T^{\cgi{a}_1 \dots \cgi{a}_k}_{\cgi{b}_1 \dots \cgi{d}_j \dots \cgi{b}_l}}
         \right).
%     \left( \sum_{i=1}^{k} {\dot{g}(0)^{\cgi{a}_i}}_{\cgi{c}_i} \left(\delta^{\cgi{a}_1}_{\cgi{c}_1} \cdots \widehat{\delta^{\cgi{a}_i}_{\cgi{c}_i}} \cdots \delta^{\cgi{a}_k}_{\cgi{c}_k} \delta^{\cgi{d}_1}_{\cgi{b}_1} \cdots \delta^{\cgi{d}_l}_{\cgi{b}_l} \right) \right.
%     + \\ \nonumber & \quad +
%         \left. \sum_{j=1}^{l} {\dot{g}(0)^{\cgi{d}_j}}_{\cgi{b}_j} \left(\delta^{\cgi{a}_1}_{\cgi{c}_1} \cdots \delta^{\cgi{a}_k}_{\cgi{c}_k} \delta^{\cgi{d}_1}_{\cgi{b}_1} \cdots \widehat{\delta^{\cgi{d}_j}_{\cgi{b}_j}} \cdots \delta^{\cgi{d}_l}_{\cgi{b}_l} \right)
%         \right)
%        {T^{\cgi{c}_1 \dots \cgi{c}_k}_{\cgi{d}_1 \dots \cgi{d}_l}}.
\end{align}
%Here the symbols with $\delta$ as the kernel letter denote index
%substitution, and the hatted $\delta$ in each term indicates a
%factor to be omitted. Alternatively, we can write Equation
%\eqref{eq:kappa-with-lots-of-indices} as
%\begin{equation}\tag{\ref{eq:kappa-with-lots-of-indices}}
% \kappa({T^{\cgi{a}_1 \dots \cgi{a}_k}_{\cgi{b}_1 \cdots
% \cgi{b}_l}})=
%\left( \sum_{i=1}^{k} {\dot{g}(0)^{\cgi{a}_i}}_{\cgi{c}_i}
%{T^{\cgi{a}_1 \dots \cgi{c}_i \dots \cgi{a}_k}_{\cgi{b}_1 \dots
%\cgi{b}_l}}
%     +
%         \sum_{j=1}^{l} {\dot{g}(0)^{\cgi{d}_j}}_{\cgi{b}_j} {T^{\cgi{a}_1 \dots \cgi{a}_k}_{\cgi{b}_1 \dots \cgi{d}_j \dots \cgi{b}_l}}
%         \right)
%\end{equation}
%If, for the sake of clarity, we temporarily suppress each
%$\delta$, Equation \eqref{eq:kappa-with-lots-of-indices} reads
%\begin{equation}\label{eq:kappa-with-suppressed-indices}
% \kappa({T^{\cgi{a}_1 \dots \cgi{a}_k}_{\cgi{b}_1 \cdots
% \cgi{b}_l}})=
%\left( \sum_{i=1}^{k} {\dot{g}(0)^{\cgi{a}_i}}_{\cgi{c}_i}
%     +
%         \sum_{j=1}^{l} {\dot{g}(0)^{\cgi{d}_j}}_{\cgi{b}_j}
%         \right)
%        {T^{\cgi{c}_1 \dots \cgi{c}_k}_{\cgi{d}_1 \dots \cgi{d}_l}}.
%\end{equation}
%In the following two equations we also suppress all the required
%$\delta$ factors.

Now ${\dot{g}(0)^{\cgi{a}}}_{\cgi{c}}$ is exactly
${\kappa^{\cgi{a}}}_{\cgi{c}}$. Thus writing $\kappa$, acting on
elements of this tensor product representation, in abstract index
notation, we have
\[
 \kappa^{\cgi{a}_1 \dots \cgi{a}_k}_{\cgi{c}_1 \dots \cgi{c}_k}{}^{\cgi{d}_1 \dots \cgi{d}_k}_{\cgi{b}_1 \dots \cgi{b}_k}
 = {\kappa^{\cgi{a}_1}}_{\cgi{c}_1} + \cdots + {\kappa^{\cgi{a}_k}}_{\cgi{c}_k} - {\kappa^{\cgi{d}_1}}_{\cgi{b}_1} - \cdots
 -
 {\kappa^{\cgi{d}_l}}_{\cgi{b}_l}. \]
Here in each term we have omitted a product of factors of the form
$\delta^{\cgi{a}_i}_{\cgi{c}_i}$. We do the same in the next
equation.

If we are interested in the local representative of a connection
as it acts on a particular tensor product representation,
$\sigma^* \omega$ is given by
\begin{equation}\label{eq:local-representative-tensor-representation}
\sigma^* \omega \leftrightarrow A_{\ai{a}}{}^{\cgi{a}_1 \dots
\cgi{a}_k}_{\cgi{c}_1 \dots \cgi{c}_k}{}^{\cgi{d}_1 \dots
\cgi{d}_k}_{\cgi{b}_1 \dots \cgi{b}_k}
    = {{A_{\ai{a}}}^{\cgi{a}_1}}_{\cgi{c}_1} + \cdots + {{A_{\ai{a}}}^{\cgi{a}_k}}_{\cgi{c}_k}
     - {{A_{\ai{a}}}^{\cgi{d}_1}}_{\cgi{b}_1} - \cdots - {{A_{\ai{a}}}^{\cgi{d}_l}}_{\cgi{b}_l}.
\end{equation}
This fact will be used later in \S\ref{ssec:covariant-derivatives}
to explain the Leibniz rule for covariant derivatives, and  in
\S\ref{sssec:metric-connections-are-OM-connections} proving
Proposition \ref{prop:metric-connections} about metric
connections.

\subsubsection{The difference between connections is a
tensor}\label{sssec:difference-between-connections}

The following result is interesting in itself, as it constitutes
part of the `structure theory' of connections. However, our real
interest is in using this eventually to compare different spinor
connections, in \S\ref{ssec:implications}

\begin{prop}
Suppose $\omega$ and $\omega'$ are connections on $P$. The
difference between $\omega$ and $\omega'$ defines a tensor on $M$
according to the following prescription. Let $\sigma_1 : U \To P$
and $\sigma_2 : U \To P$ be local cross sections of $P$, and let
$A_{\ai{a}}{}^{\cgi{b}}{}_{\cgi{c}}$ be the local representative
$\sigma_1^*(\omega - \omega)$ in index notation, and
$B_{\ai{a}}{}^{\cgi{b}}{}_{\cgi{c}}$ that of $\sigma_2^*(\omega -
\omega)$. Then if $g : U \To G$ is such that $\sigma_2(m) =
\sigma_1(m) g(m)$ for all $m \in U$, then
\[B_{\ai{a}}{}^{\cgi{b}}{}_{\cgi{c}}(m) =
(g(m))^{\cgi{b}}{}_{\cgi{d}} (g(m)^{-1})^{\cgi{e}}{}_{\cgi{c}}
A_{\ai{a}}{}^{\cgi{d}}{}_{\cgi{e}}(m)\] and so the prescription
\begin{align*}
A_{\ai{a}}{}^{\agi{b}}{}_{\agi{c}}(m)
    & = [\sigma_1(m), A_{\ai{a}}{}^{\cgi{b}}{}_{\cgi{c}}(m)] \\
    & = [\sigma_2(m), B_{\ai{a}}{}^{\cgi{b}}{}_{\cgi{c}}(m)]
\end{align*}  gives a well defined global tensor on
$U$.
\end{prop}
\begin{rem}
Essentially the claim here is that the local representatives
transform appropriately as we change the local cross section, and
so live in the appropriate representation, so that we can define
the global tensor as an element of the associated tensor bundle.
\end{rem}
\begin{proof}
Define $\chi:P \To P$ by $\chi(\bdelt{p}) = \bdelt{p}
g(\pi(\bdelt{p}))$. Thus $\sigma_2 = \chi \compose \sigma_1$, and
$\sigma_2^* = \sigma_1^* \compose \chi^*$. For an arbitrary $v \in
T_{\bdelt{p}} P$, choose a path $n:\I \To P$ so $v = \dot{n}(0)$,
and let $u = \dat{t}{0} \pi(n(t)) \in T_m M$, where $m =
\pi(\bdelt{p})$. Then
\begin{align*}
\chi_* v & = \dat{t}{0}n(t) g(\pi(n(t))) \\
         & = v g(m) + \bdelt{p} g_* u.
\end{align*}
Here $g_* u \in T_{g(m)} G$, and $\bdelt{p} g_* u \in
V_{\bdelt{p}}$. Thus $\omega$ and $\omega'$ agree on the second
term of the expression above, and so $\chi^*(\omega-\omega')(v) =
(\omega-\omega')(v g(m)) = \Ad(g(m)^{-1})(\omega-\omega')(v)$.
Finally then $\sigma_2^*(\omega-\omega')_m =
\Ad(g(m)^{-1})\sigma_1^*(\omega-\omega')$, and this is easily seen
to imply the result.
\end{proof}

\subsubsection{Parallel transport in a local trivialisation}
\label{sssec:transport-trivialisation} Parallel transportation can
be described more explicitly when a local trivialisation is given.
Fix a local cross section $\sigma : U \To P$ and the related local
trivialisation $\psi:U \times G \To P$ defined by
$\psi(m,g)=\sigma(m) g$. Let $m : \I \To U$ be a path, and
$\bdelt{p_0} \in \pi^{-1}(m(0))$, so that in this trivialisation
$\bdelt{p_0} = \psi(m(0),e)$. The parallel transport of
$\bdelt{p_0}$ along $m$ is the unique curve in $P$ starting at
$\bdelt{p_0}$ which has an everywhere horizontal tangent vector
and which projects down via $\pi$ to the curve $m$. Thus in the
trivialisation this curve is of the form $\bdelt{p}:\I \To U
\times G$, $t \mapsto (m(t),g(t))$, for some function $g:\I \To G$
with $g(0)=e$. The condition that the tangent vector is horizontal
is expressed by
\[ \omega\left( \frac{d}{dt} \psi(\bdelt{p}(t)) \right) = 0. \]
This is equivalent to
\begin{align*}
0=\omega\left( \psi_* \dot{\bdelt{p}}(t)\right) & = (\psi^* \omega)(\dot{\bdelt{p}}(t)) \\
            & = (\psi^* \omega)(\dot{m}(t),\dot{g}(t)) \\
            & = \Ad(g(t)^{-1})(\sigma^* \omega (\dot{m}(t))) + L_{g(t)^{-1}*} \dot{g}(t).
\end{align*}
The final step is an application of Proposition
\ref{prop:local-representative}. We conclude from this that
\[L_{g(t)^{-1}*} \dot{g}(t) = - \Ad(g(t)^{-1})(\sigma^* \omega (\dot{m}(t))). \]
Further, we can write this in abstract index notation, writing
$x^{\ai{a}}$ for the tangent vector field $\dot{m}(t)$, using
$\sigma^* \omega \leftrightarrow
{{A_{\ai{a}}}^{\cgi{b}}}_{\cgi{c}}$ and explicitly applying
$\Ad(g(t)^{-1})$, to obtain
\begin{align}
 {\left(L_{g(t)^{-1}*} \dot{g}(t)\right)^{\cgi{b}}}_{\cgi{c}} & =
    - {\left(g(t)^{-1}\right)^{\cgi{b}}}_{\cgi{d}}
      {{A_{\ai{a}}}^{\cgi{d}}}_{\cgi{f}}
      {\left(g(t)\right)^{\cgi{f}}}_{\cgi{c}}
      x^{\ai{a}}, \nonumber \\
\intertext{and more simply at $t=0$}
 {\left(\dot{g}(0)\right)^{\cgi{b}}}_{\cgi{c}} & =
    - {{A_{\ai{a}}}^{\cgi{b}}}_{\cgi{c}}
      x^{\ai{a}} \label{eq:gdot-indices}
\end{align}
We will use these expressions in the next section.

\noop{ If we are instead interested in a tensor product
representation, we use Equation
\eqref{eq:local-representative-tensor-representation} instead of
Equation \eqref{eq:local-represenative-index-notation}, and obtain
\begin{equation}\label{eq:p-dot-many-indices-formula}
{{\left(\dot{\bdelt{p}}(0)\right)^{\ai{b}}}^{
    \cgi{a}_1 \dots \cgi{a}_k}_{\cgi{c}_1 \cdots
    \cgi{c}_k}}^{\cgi{d}_1 \dots \cgi{d}_k}_{\cgi{b}_1 \cdots
    \cgi{b}_k} =
    \left(x^{\ai{b}},\left(- \sum_{i=1}^k
      {{A_{\ai{a}}}^{\cgi{a}_i}}_{\cgi{c}_i}
     - \sum_{j=1}^1 {{A_{\ai{a}}}^{\cgi{d}_j}}_{\cgi{b}_j}\right)
      x^{\ai{a}}\right).
\end{equation}
(Note that again in this equation in each term of the summations
there is understood a product of $\delta$ symbols of the form
$\delta^{\cgi{a}_i}_{\cgi{c}_i}$ and
$\delta^{\cgi{d}_j}_{\cgi{b}_j}$. These perform index
substitutions on corresponding pairs of omitted indices, as in
Equation \eqref{eq:kappa-with-suppressed-indices}.) }

\subsection{Covariant
derivatives}\label{ssec:covariant-derivatives} The chief
difficulty in defining the derivative of one vector field with
respect to another is that although the vector spaces at each
point of the base manifold are isomorphic, they are not
canonically so, and therefore we have no intrinsic way of
comparing vectors at two different points of the manifold. More
concretely, vectors based at different points are elements of
different vector spaces, and so we have no way to apply the usual
vector space operations to them. Without this, we cannot form the
difference quotient familiar from the usual definition of
derivative. Parallel transportation bridges this difficulty.

Given a local cross section $y^{\agi{b}}$ defined on $U \subset M$
of an associated vector bundle, a connection $\omega$ on the
principal fibre bundle, and a tangent vector field $x^{\ai{a}}$
also defined on $U$, we define the covariant derivative of
$y^{\agi{b}}$ in the direction $x^{\ai{b}}$, written $x^{\ai{a}}
\nabla_{\ai{a}} y^{\agi{b}}$, as follows. Fix a point $m_0 \in U$.
Let $m:\J \To U$, for some $\eps>0$, be the integral curve of
$x^{\ai{a}}$ starting at $m_0$. Parallel transportation of
$y^{\agi{b}}(m_0)$ along $m$ defines a curve $t \mapsto
y^{\agi{b}}(t)$ in the associated vector bundle, such that the
vector $y^{\agi{b}}(t)$ is based at the point $m(t)$. Notice that
we distinguish between $y^{\agi{b}}(t)$ and $y^{\agi{b}}(m(t))$.
The first is the parallel transport by $t$ of $y^{\agi{b}}(m_0)$,
and the second is the value of $y^{\agi{b}}$ at the point $m(t)$.
We can thus compare $y^{\agi{b}}(t)$ and $y^{\agi{b}}(m(t))$
because they are vectors at the same point. We define
\begin{equation}
\label{eq:covariant-derivative}
 x^{\ai{a}} \nabla_{\ai{a}} y^{\agi{b}} = \lim_{t\To 0} \frac{y^{\agi{b}}(m(t))
- y^{\agi{b}}(t)}{t}.
\end{equation}
Note that this limit is in the topology on the fibre bundle, as
the vectors $y^{\agi{b}}(m(t))$ and $y^{\agi{b}}(t)$ do not lie at
a fixed point. An alternative definition of parallel transport is
available that uses only the topology of the fibre at a point, but
it is more cumbersome in other places, and finally makes little
difference. Analogously, if ${T^{\agi{b}_1 \dots
\agi{b}_k}}_{\agi{c}_1 \dots \agi{c}_k}$ is a local cross section
of a tensor bundle, we define the covariant derivative in the same
way, so
\begin{equation}
 x^{\ai{a}} \nabla_{\ai{a}} {T^{\agi{b}_1 \dots \agi{b}_k}}_{\agi{c}_1 \dots \agi{c}_k} =
    \lim_{t\To 0} \frac{{T^{\agi{b}_1 \dots \agi{b}_k}}_{\agi{c}_1 \dots \agi{c}_k}(m(t)) -
    {T^{\agi{b}_1 \dots \agi{b}_k}}_{\agi{c}_1 \dots \agi{c}_k}(t)}{t}.
\end{equation}

This description is sufficient to define a covariant derivative,
but we will need to develop the details further for the purposes
of later theorems. Although we are about to perform calculations
in a specific local trivialisation, the prescription given here is
well defined. The symbol $\nabla_{\ai{a}}$ itself is not a tensor,
but $\nabla_{\ai{a}} y^{\agi{b}}$ is,\footnote{We must keep in
mind that the indices here correspond to two different principal
fibre bundles, one the frame bundle, and so we need to use the
idea of a product bundle in \S\ref{ssec:product-bundles}.} because
it is clear from Equation \eqref{eq:covariant-derivative} that
$x^{\ai{a}} \nabla_{\ai{a}} y^{\agi{b}}$ is a tensor for every
vector field $x^{\ai{a}}$. \noop{\footnote{For this purpose we
need to remember the parallel transported $y^{\agi{b}}(t)$ is
actually a tensor at $m(t)$.}}

In order to evaluate the covariant derivative, we choose a local
cross section of the principal fibre bundle $\sigma: U \To P$,
with $m \in U$. As usual, this gives a local trivialisation
$\psi(m,g)=\sigma(m) g$. The vector field $y^{\agi{b}}$ can be
expressed in terms of this trivialisation in the form
\[ y^{\agi{b}}(m) = \left[(m,e), y^{\cgi{b}}(m)\right],\]
where $y^{\cgi{b}}$ takes values in the fixed underlying vector
space of the representation. It is important to remember here the
notational distinction between $y^{\agi{b}}$, which is a section
of the vector bundle, and $y^{\cgi{b}}$, which is a map from $U$
to a fixed vector space. Parallel transportation of $(m_0,e)$
along the curve $m$ gives the curve $t \mapsto (m(t),g(t))$ in $P$
described above in \S\ref{sssec:transport-trivialisation}. Thus
$y^{\agi{b}}(t) = \left[(m(t),g(t)),y^{\cgi{b}}(m_0)\right]$, and
so we can calculate the derivative defined in Equation
\eqref{eq:covariant-derivative} as
\begin{align} \tag{\ref{eq:covariant-derivative}}
 x^{\ai{a}} \nabla_{\ai{a}} y^{\agi{b}}(m_0) & =  \lim_{t\To 0} \frac{y^{\agi{b}}(m(t)) - y^{\agi{b}}(t)}{t}  \nonumber \\
            & =  \lim_{t\To 0} \frac{ \left[
              (m(t),e),y^{\cgi{b}}(m(t))\right] - \left[
             (m(t),g(t)),y^{\cgi{b}}(m_0)\right] }{t} \nonumber \\
            & =  \lim_{t\To 0} \left[
               (m(t),e),\frac{ y^{\cgi{b}}(m(t))-{(g(t))^{\cgi{b}}}_{\cgi{c}} (y^{\cgi{c}}(m_0))}{t}\right]   \nonumber
\end{align}
\begin{align}
 x^{\ai{a}} \nabla_{\ai{a}} y^{\agi{b}}(m_0)
            & =  \left[
               (m_0,e),\dat{t}{0} \left(y^{\cgi{b}}(m(t))-{(g(t))^{\cgi{b}}}_{\cgi{c}} (y^{\cgi{c}}(m_0))\right)\right]
                    \label{eq:covariant-derivative-with-g}\\
            & =  \left[
               (m_0,e),x^{\ai{a}} (\dd y^{\cgi{b}})_{\ai{a}} (m_0)-{(\dot{g}(0))^{\cgi{b}}}_{\cgi{c}} (y^{\cgi{c}}(m_0))\right] \nonumber \\
            & =  \left[
              (m_0,e),x^{\ai{a}} \left\{(\dd y^{\cgi{b}})_{\ai{a}} (m_0) + {{A_{\ai{a}}}^{\cgi{b}}}_{\cgi{c}}
                y^{\cgi{c}}(m_0)\right\}\right] \nonumber
\end{align}
In the last line here we have utilised Equation
\eqref{eq:gdot-indices}. Using the ideas of
\S\ref{ssec:product-bundles} we can `cancel' the $x^{\ai{a}}$. To
do this we need to consider not just a local trivialisation of
$P$, but also a local trivialisation of the frame bundle $FM$, so
that $(\dd y^{\cgi{b}})_{\ai{a}} (m_0) = \left[(m_0,e' \in \GLn),
(\dd y^{\cgi{b}})_{\ci{a}} (m_0)\right]$. Then
\begin{equation}\label{eq:covariant-derivative2}
 \nabla_{\ai{a}} y^{\agi{b}}(m_0)
             =  \left[
              (m_0,e \times e'),(\dd y^{\cgi{b}})_{\ci{a}} (m_0) + {{A_{\ci{a}}}^{\cgi{b}}}_{\cgi{c}}
                y^{\cgi{c}}(m_0)\right].
\end{equation} %-% explain better!

It becomes clear at this point how the local representatives of
the connection form are related to the more familiar Christoffel
symbols of Riemannian geometry. They specify the difference
between covariant differentiation and partial differentiation in a
particular local trivialisation. Similarly,
\begin{multline}\label{eq:covariant-derivative-for-tensors}
 \nabla_{\ai{a}} {T^{\agi{b}_1 \dots \agi{b}_k}}_{\agi{c}_1 \dots \agi{c}_k}(m_0)
    = \biggl[
    (m_0,e),
    (\dd {T^{\cgi{b}_1 \dots \cgi{b}_k}}_{\cgi{c}_1 \dots
    \cgi{c}_k})_{\ci{a}}
    (m_0) + \\
     + \sum_{i=1}^{k} {{A_{\ci{a}}}^{\cgi{b}_i}}_{\cgi{d}_i} {T^{\cgi{b}_1 \dots \cgi{d}_i \dots \cgi{b}_k}}_{\cgi{c}_1 \dots
     \cgi{c}_k}(m_0)
      - \sum_{j=1}^{l} {{A_{\ci{a}}}^{\cgi{f}_j}}_{\cgi{c}_j} {T^{\cgi{b}_1 \dots \cgi{b}_k}}_{\cgi{c}_1 \dots \cgi{f}_j \dots \cgi{c}_k}(m_0)
    \biggr],
\end{multline}
using Equation
\eqref{eq:local-representative-tensor-representation} instead of
Equation \eqref{eq:local-represenative-index-notation}. It is
clear from this expression that covariant differentiation
satisfies the Leibniz rule. This follows because the term
involving the derivative of the components, $(\dd {T^{\cgi{b}_1
\dots \cgi{b}_k}}_{\cgi{c}_1 \dots \cgi{c}_k})_{\ci{a}}$,
satisfies the Leibniz rule, and we can rearrange the terms
involving Christoffel symbols appropriately.
%-%keep going a bit... say what some of this means :-)

\subsection{Metric connections} \label{ssec:metric-connections}
We now complete the demonstration of the equivalence between the
two viewpoints of orthonormal structures. This section shows that
connections on an orthonormal bundle correspond to metric
covariant derivatives.

Suppose $g_{\ai{b c}}$ is a metric of signature $(r,s)$ and
further that $\nabla_{\ai{a}}$ is a covariant derivative,
associated with a connection form $\omega$ on the frame bundle
$FM$. Let $OM$ be the $SO_0(r,s)$ principal fibre bundle of
orthonormal frames for $g_{\ai{b c}}$. Since $OM$ is a reduction
of $FM$, in a natural sense $\omega$ can be restricted to a form
on $OM$. This restriction will not generally be a connection form
on $OM$, since its values lie in the Lie algebra
$\LAM{gl}{n,\Real}$. We say that $\omega$ restricts to a
connection form on $OM$ if its range lies within the Lie
subalgebra $\LAM{so}{r,s}$.

We say that $\nabla_{\ai{a}}$ is \emph{metric} with respect to
$g_{\ai{b c}}$ if \[\nabla_{\ai{a}} g_{\ai{b c}} = 0.\] We can
also call the connection form $\omega$ itself metric, if its
associated covariant derivative is metric. This section gives the
proof of the following proposition.

\begin{prop}
\label{prop:metric-connections} The following two conditions are
equivalent:
\begin{enumerate}
\item The covariant derivative $\nabla_{\ai{a}}$ is metric.
\item The connection form $\omega$ restricts to a connection form on
the orthonormal frame bundle $OM$.
\end{enumerate}
\end{prop}

This problem will be addressed in two steps, in the following
sections.

A corollary of Proposition \ref{prop:metric-connections} is that
the connection forms on the orthonormal frame bundle $OM$ provided
by Proposition \ref{prop:connections-exist} give metric covariant
derivatives. That is, given a metric $g_{\ai{b c}}$, there is
always a compatible covariant derivative $\nabla_{\ai{a}}$ so that
$\nabla_{\ai{a}} g_{\ai{b c}} = 0$.

This fact prompts a final note on the abstract index notation. The
conventions for raising and lowering indices are compatible with
the metric covariant derivative, in that if we have a valence
$\valence{0}{1}$ tensor $y_{\ai{b}}$, and the corresponding
$\valence{1}{0}$ tensor $y^{\ai{b}} = g^{\ai{b c}} y_{\ai{c}}$,
then \[ \nabla_{\ai{a}} y^{\ai{b}} = \nabla_{\ai{a}} g^{\ai{b c}}
y_{\ai{c}} = g^{\ai{b c}} \nabla_{\ai{a}} y_{\ai{c}}. \] Here we
have used the fact that $\nabla_{\ai{a}}$ is a metric covariant
derivative, and the Leibniz rule.

\subsubsection{An $OM$ connection is metric}

To show that any connection on $OM$ is metric with respect to the
metric induced by the bundle, we will step back slightly, and
describe how this metric is parallel transported by the
connection. Specifically, if the metric at one point is parallel
transported to another point, it is found to be equal to the
metric defined at that point. Using the definition of the
covariant derivative in terms of parallel transports, this then
ensures that the covariant derivative of the metric is zero, that
is, $\nabla_{\ai{a}} g_{\ai{b c}} = 0$

We actually prove this result in a more general setting. Suppose
$\pfbundle{G}{P}{\pi}{M}$ is a principal fibre bundle, and
$\lambda : G \To \Aut(V)$ is a representation of the group on $V$.
An invariant vector $v^{\cgi{b}}$ in $V$ for this representation
is a vector such that
\[\lambda(g) \left( v^{\cgi{b}} \right) = v^{\cgi{b}} \] for all $g
\in G$. The metric tensor $\eta_{\ci{b c}}$ is an invariant tensor
for the orthonormal group, since the relevant representation acts
as \[ \lambda(g) \left( \eta_{\ci{b} \ci{c}} \right) =
g^{\ci{d}}_{\ci{b}} g^{\ci{e}}_{\ci{c}} \eta_{\ci{d} \ci{e}} =
\eta_{\ci{b} \ci{c}}.
\]

Any such invariant defines an element of the corresponding vector
bundle $P \times_\lambda V$ at each point $m \in M$, by
$v^{\agi{b}}=\left[\bdelt{b}, v^{\cgi{b}}\right]$, for an
arbitrary $\bdelt{b} \in \pi^{-1}(m)$. This is well defined, since
for some other $\bdelt{b'} \in \pi^{-1}(m)$, $\bdelt{b'} =
\bdelt{b} g$ for some $g \in G$, and \[\left[\bdelt{b} g,
v^{\cgi{b}}\right] = \left[\bdelt{b}, \lambda(g)
(v^{\cgi{b}})\right] = \left[\bdelt{b}, v^{\cgi{b}}\right].\] The
element $v^{\agi{b}}$ of the vector bundle is also called an
invariant vector, or tensor, if appropriate. Further, defining
$v^{\agi{b}}$ in this fashion at each point gives an invariant
vector field.

\begin{prop}\label{prop:invariant-vector-fields-parallel}
Suppose $v^{\agi{b}}$ is an invariant vector field for the
principal fibre bundle $\pfbundle{G}{P}{\pi}{M}$. If $\omega$ is a
connection on $P$, and $\nabla_{\ai{a}}$ is the associated
covariant derivative, then \[\nabla_{\ai{a}} v^{\agi{b}} = 0.\]
\end{prop}
\begin{proof}
Let $\gamma : \I \To M$ be a curve in $M$, and let $m_0 =
\gamma(0)$, $m_1 = \gamma(1)$. Suppose $\bdelt{b} \in
\pi^{-1}(m_0)$, and parallel transport along $\gamma$ carries
$\bdelt{b}$ to $\bdelt{b'} \in \pi^{-1}(m_1)$. Thus parallel
transport carries $v^{\agi{b}}(m_0) = \left[ \bdelt{b} ,
v^{\cgi{b}} \right]$ to $\left[ \bdelt{b'} , v^{\cgi{b}} \right]$,
which is exactly $v^{\agi{b}}(m_1)$, since $v^{\agi{b}}$ in an
invariant vector field.

Thus parallel transport along any curve carries $v^{\agi{b}}$ to
itself, and so, from the definition of the covariant derivative in
terms of parallel transportation in
\S\ref{ssec:covariant-derivatives},
\[\nabla_{\ai{a}} v^{\agi{b}} = 0.\qedhere\]
\end{proof}

This general result now specialises easily to prove the desired
result. It will also prove an important result of the
$\smash{\widetilde{SO}}_0(1,3)$ spinor calculus, Proposition
\ref{prop:varepsilon-covariantly-parallel}.

\begin{cor}
The metric tensor $g_{\ai{b c}}$ is the invariant tensor field
defined by the invariant tensor $\eta_{\ci{b} \ci{c}}$ for the
orthonormal group. Thus if $\nabla_{\ai{a}}$ is the covariant
derivative defined by a connection on the orthonormal frame
bundle,
\[ \nabla_{\ai{a}} g_{\ai{b c}} = 0.\]
\end{cor}

\subsubsection{Metric connections are $OM$
connections}\label{sssec:metric-connections-are-OM-connections}
For the converse, we need only show that a metric connection takes
values solely in the Lie algebra $\LA{so}$. If this is true, the
properties of the connection on the frame bundle ensure that the
restriction to the orthonormal bundle also satisfies the
connection form axioms of Definition \ref{defn:connection-form}.

Since $\eta_{\ci{b} \ci{c}}$ is an invariant tensor, $g_{\ai{b
c}}(m) = \left[(m,e),\eta_{\ci{b} \ci{c}}\right]$ in any local
cross section. Thus the derivative is given by Equation
\eqref{eq:covariant-derivative-for-tensors} as
\[\nabla_{\ai{a}} g_{\ai{b c}}(m) = \left[(m,e), (\dd \eta_{\ci{b}
\ci{c}})_{\ci{a}}(m) - {{A_{\ci{a}}}^{\ci{d}}}_{\ci{b}}
\eta_{\ci{d} \ci{c}} - {{A_{\ci{a}}}^{\ci{d}}}_{\ci{c}}
\eta_{\ci{b} \ci{d}} \right].\] Since $\eta_{\ci{b} \ci{c}}$ is a
certain fixed tensor, the first term, involving its exterior
derivative, vanishes. Further, in the last two terms we use the
index lowering convention for $\eta_{\ci{b} \ci{c}}$, to obtain
\[\nabla_{\ai{a}} g_{\ai{b c}}(m)
 = - \left[(m,e), A_{\ci{a} \ci{c b}}
                            + A_{\ci{a} \ci{b c}} \right].\]

Since this expression vanishes, we find the simple condition
$A_{\ci{a} \ci{c b}} + A_{\ci{a} \ci{b c}} = 0$ governing the
local representatives of the connection. This implies that the
connection always takes values in the Lie algebra of $SO$ as
described in \S\ref{ssec:lie-algebras}, since the values in the
full $\GLn$ Lie algebra are always antisymmetric with respect to
the invariant tensor $\eta_{\ci{b} \ci{c}}$.

\newpage
\part{Spinor Structure Classification}
\label{part:spinor-structure-classification} We now begin our
treatment of spinor structures. The idea is to take a principal
fibre bundle, and replace the structure group with its simply
connected covering group in an appropriate fashion. The precise
definition is given in \S\ref{sec:spinor-structure}. To start, we
need to introduce the fundamentals of covering space theory, which
underlie all the results in this part of the thesis. The necessary
material is summarised in \S\ref{sec:covering-spaces}.

In \S\ref{sec:spinor-structure} we state and prove the Existence
and Classification Theorems for spinor structures in a general
setting, and compare these results with previously published work.
\longversion{Further, in \S\ref{sec:metric-independence}, we
analyse the spinor structures of reduced bundles.}

In \S\ref{sec:classifying-as-bundles} we discuss classifying
inequivalent spinor structures in terms of the underlying
principal fibre bundle. With the available methods it is only
possible to do this completely in special cases, but we show that
these include the physically significant situation. We also
conjecture an extension of the result presented here.

Finally, in \S\ref{sec:lifting-connection} and
\S\ref{sec:inequivalent-connections} we give a thorough discussion
of connections on spinor structures. With the aid of our `bundle
classification' of spinor structures, we show how connections on
inequivalent spinor structures can be compared. This leads
naturally into Part \ref{part:implications-for-physics}, as it
allows us to explain how the classification of spinor structures
is relevant to the physics of the Dirac equation.

\section{A preamble on covering spaces}\label{sec:covering-spaces}

Much of the theory of spinor structures that we develop will rely
upon covering space theory. In fact, the geometric definition of a
spinor structure which we will give relies intimately upon the
notion of a covering space. Thus, in this section, we give the
relevant definitions, as well as a suitable version of the
fundamental Covering Space Classification Theorem. This result
forms the basis of the results of \S \ref{sec:spinor-structure}.

\subsection{Definitions}

The two basic definitions are of continuous covering maps and
smooth covering maps.

\begin{figure}[!htbp]
\includegraphics[width=0.3\textwidth]{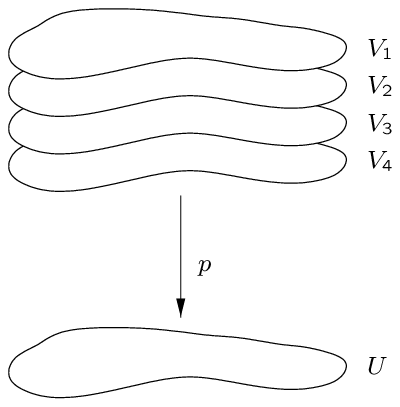} \caption{Covering maps are
`locally trivial'.\label{fig:covering-spaces}}
\end{figure}

\begin{defn}
A \defnemph{continuous covering map} $p: Y \To X$ is a continuous
map from a connected topological space $Y$ to a connected
topological space $X$ such that each $x$ in $X$ has a
neighbourhood $U \subset X$ so that $p^{-1}(U)$ is a disjoint
union of sets $\bigcup_{\beta \in \mathcal{B}} V_\beta$, so that
the restriction $\restrict{p}{V_\beta}$ is a homeomorphism for
each $\beta \in \mathcal{B}$. (See Figure
\ref{fig:covering-spaces}.) We call $Y$ the covering space.
\end{defn}

Notice that we consider only connected covering spaces.

\begin{defn}
A \defnemph{smooth covering map} $p: Y \To X$ is a continuous
covering map so that the restrictions $\restrict{p}{V_\beta}$ are
all diffeomorphisms.
\end{defn}

\begin{prop}\label{prop:continuous-smooth-coverings}
If $p : Y \To X$ is a continuous covering map, and $X$ is a smooth
manifold, then there is a unique differentiable structure for $Y$
so $p$ is a smooth covering map.
\end{prop}
\begin{proof}
We construct this smooth structure as follows. Let
$\left(V_\alpha\right)_{\alpha \in \mathcal{A}}$ be an open
covering of $Y$ by sets so that $p$ maps $V_\alpha$
homeomorphically onto its image, and $p(V_\alpha)$ is a chart for
$X$, with coordinate map $\psi_\alpha : u(V_\alpha) \To W_\alpha
\subset \Real^n$ for each $\alpha \in \mathcal{A}$. Such an open
covering certainly exists. Define $\varphi_\alpha : V_\alpha \To
W_\alpha$ by $\varphi_\alpha(y) = \psi_\alpha(u(y))$ for each
$\alpha \in \mathcal{A}$. This map is a homeomorphism, because it
is a composition of homeomorphisms. Further, the `transition maps'
$\varphi_\alpha \compose
\restrict{\varphi_{\alpha'}^{-1}}{W_\alpha \cap W_{\alpha'}}$ are
all diffeomorphisms, because $\varphi_\alpha \compose
\varphi_{\alpha'}^{-1} = \psi_\alpha \compose
\psi_{\alpha'}^{-1}$. Thus the collection
$\left(V_\alpha,\varphi_\alpha\right)_{\alpha \in \mathcal{A}}$
defines an atlas for $Y$, and it is clear that $p$ is a smooth
map, and further a smooth covering map, with respect to this
differentiable structure.

Uniqueness is trivial, since for $p$ to be a diffeomorphism, all
of the charts described above must be in the atlas for $Y$. The
differentiable structure is uniquely determined by any atlas,
establishing the result.
\end{proof}

Thus there is no essential difference between continuous and
smooth covering maps.

\begin{defn}\label{defn:covering-space-equivalence}
We say two continuous (respectively, smooth) covering maps $p: Y
\To X$ and $p' : Y' \To X$ are \defnemph{equivalent} if there is a
homeomorphism (resp. diffeomorphism) $a:Y \To Y'$ so $p' \compose
a = p$.
\end{defn}

\subsection{Paths and loops}

We next introduce the notions of paths and loops in a manifold. A
path is a map $\I \To M$, and a loop is a map $\I \To M$ taking
$0$ and $1$ to the same point of $M$. In this section, we will
distinguish between continuous and smooth paths or loops, but we
will also see that for the purposes of later sections this
distinction is not important.

We say that two paths $\alpha, \beta$ such that $\alpha(0) =
\beta(0)$ and $\alpha(1) = \beta(1)$ are continuously homotopic if
there is a continuous map $H : \I \times \I \To M$ so that
\begin{align*}
    H(0,t) & = \alpha(t)\\
    H(1,t) & = \beta(t)\\
    H(s,0) & = \alpha(0) = \beta(0)\\
    H(s,1) & = \alpha(1) = \beta(1).
\end{align*}
Two smooth paths are smoothly homotopic  \cite[\S 4]{mil:tdv} if
there is such a smooth map $\I \times \I \To M$. Again, we will
see that this distinction is unimportant for our purposes, and so
in later sections we always mean `smoothly homotopic' by
`homotopic'. Continuous homotopy gives an equivalence relation. It
is clear that the relation is reflexive and symmetric. Continuous
homotopies can be patched together, showing that continuous
homotopy is a transitive relation. We thus denote the equivalence
class of a path $\alpha$ under continuous homotopy by $[\alpha]$.

Continuous paths can be concatenated. Given $\alpha, \beta : \I
\To M$, such that $\alpha(0) = \beta(1)$, the path $\alpha \cat
\beta : \I \To M$ is defined by
\[
    (\alpha \cat \beta)(t)=\left\{
        \begin{array}{ll}
            \beta(2t)  & \textrm{if $t \in \left[0,\frac{1}{2}\right]$} \\
            \alpha(2t-1) & \textrm{if $t \in \left[\frac{1}{2},1\right]$}
        \end{array}
        \right.
        .
\]
Smooth paths cannot necessarily be concatenated, as the resulting
path may not be smooth at $t=\frac{1}{2}$.

Concatenation is neither commutative nor associative. Up to
homotopy, however, it \emph{is} associative. That is, $[(\alpha
\cat \beta) \cat \gamma] = [\alpha \cat (\beta \cat \gamma)]$ for
all paths $\alpha, \beta, \gamma$ such that these concatenations
are defined. This relation is trivially proved by providing the
appropriate homotopy. It is easy to see that $[\alpha \cat \beta]$
depends only on the equivalence classes $[\alpha]$ and $[\beta]$,
so we can use the notation $[\alpha] \cat [\beta]$ for $[\alpha
\cat \beta]$.

The claim that the distinction between the continuous and smooth
cases is unimportant follows from two facts.

\begin{prop}\label{prop:continuous-smooth-homotopy}
Firstly, every continuous path in a smooth manifold is homotopic
to a smooth path. Secondly, if two smooth paths are continuously
homotopic, they are smoothly homotopic.
\end{prop}
\begin{proof}
See Theorem 7 and the following discussion in Chapter 2 of
\cite{pon:smaht}, and Theorem 8 of the same. Related results are
given in \cite[\S16.26]{die:ta3}.
\end{proof}

Given that henceforth we will work only in the smooth setting, it
may seem redundant to have mentioned the continuous case at all.
This infelicity is forced upon us by the fact that covering space
theory is most natural in the continuous setting, and the theorems
that we will rely on are proved there. On the other hand, much of
the work described here, particularly the proofs in
\S\ref{sec:spinor-structure} of the Existence and Classification
Theorems for spinor structures, \longversion{and
\S\ref{sec:metric-independence},} relies intimately on smooth
connections to provide accessible and geometric arguments. At the
price of dealing here with both the continuous and the smooth
case, we may later combine the power of both covering space theory
and the theory of smooth connections. Additionally, of course, we
want to work with smooth manifolds, so that we can do calculus.

With these results in hand, we can improve upon the theory of
smooth paths and smooth homotopies. Firstly, we can define the
equivalence relation of smooth homotopy. Again, it is clear that
the relation is reflexive and symmetric. Now, if
$\alpha,\beta,\gamma$ are three smooth paths so $\alpha$ is
smoothly homotopic to $\beta$, and $\beta$ is smoothly homotopic
to $\gamma$, then $\alpha$ must be \emph{continuously} homotopic
is $\gamma$. Using the result that continuously homotopic smooth
paths are smoothly homotopic, we see that smooth homotopy is also
transitive. Again, we denote the smooth homotopy equivalence class
of $\alpha$ by $[\alpha]$. This overlap of notation is consistent.
That is, the smooth paths in the smooth homotopy equivalence class
of $\alpha$ are exactly the smooth paths in the continuous
homotopy equivalence class of $\alpha$.

Secondly, although smooth paths $\alpha,\beta$ with $\alpha(0) =
\beta(1)$ cannot necessarily be concatenated, up to homotopy they
can be.\footnote{This result, and the previous, that smooth
homotopy is a transitive relation, can be proved more concretely,
without the use of Proposition
\ref{prop:continuous-smooth-homotopy}. See for example \cite[\S
4]{mil:tdv}. Define \[\lambda(t) =
\frac{\mu(t-\frac{1}{3})}{\mu(t-\frac{1}{3})+\mu(\frac{2}{3}-t)},\]
where $\mu(t) = 0$ for $t\leq 0$, and $\mu(t) = e^{-\frac{1}{t}}$
for $t>0$. Then $\lambda : \I \To \I$ is smooth (but not
analytic), and $\lambda([0,\frac{1}{3}]) = 0$ and
$\lambda([\frac{2}{3}]) = 1$. Using this, $\alpha \compose
\lambda$ is smoothly homotopic to $\alpha$, and for any smooth
paths $\alpha, \beta$ such that $\alpha(0) = \beta(1)$, $(\alpha
\compose \lambda) \cat (\beta \compose \lambda)$ is a smooth path.
A similar argument using $\lambda$ shows that smooth homotopy is
transitive.} This is because $\alpha$ and $\beta$ can be
concatenated to form a continuous path $\alpha \cat \beta$, and
this continuous path is homotopic to a smooth path. Thus we can
define $[\alpha] \cat [\beta]$ by $[\alpha \cat \beta]$. It is
straightforward to see that $[\alpha] \cat [\beta]$ depends only
on the equivalence classes $[\alpha]$ and $[\beta]$. Again, up to
homotopy, concatenation is associative.

Concatenation always has a inverse, up to homotopy. If $\alpha$ is
a path, we will write $\alpha^{-1}$ for the reverse path, defined
by $\alpha^{-1}(t) = \alpha(1-t)$. Then $[\alpha^{-1} \cat \alpha]
= [\alpha(0)] = [\alpha \cat \alpha^{-1}]$, where $[\alpha(0)]$
denotes the homotopy class of the constant path at $\alpha(0)$.

\subsection{Fundamental groups}

We now introduce the fundamental group of a manifold. This
construction requires a fixed base point in the manifold. Suppose
$M$ is a smooth manifold, and $m_0 \in M$ is a base point. Define
$\Pi M$ to be the set of all smooth paths in $M$ starting at
$m_0$. Define $\Omega M$ to be the set of all smooth loops
$\alpha$ in $M$ based at $m_0$ so $\alpha(0) = \alpha(1) = m_0$,
and $\Omega^c M$ to be the set of all continuous loops in $M$.
Define $\pi_1(M,m_0)$ to be the set of smooth homotopy equivalence
classes in $\Omega M$, and give it a group structure by
concatenation. Similarly define $\pi_1^c(M,m_0)$ in the continuous
case. In both cases the identity is given by the constant path at
$m_0$. We now reach the result which will allow us for the most
part to dispense with the continuous case.

\begin{prop}\label{prop:smooth-fundamental-group-isomorphism}
The map of $\pi_1(M,m_0)$ into $\pi_1^c(M,m_0)$, taking the smooth
homotopy equivalence class $[\alpha]$ to the continuous homotopy
equivalence class $[\alpha]$ is an isomorphism.
\end{prop}
\begin{proof}
This follows immediately from Proposition
\ref{prop:continuous-smooth-homotopy}. Firstly it is surjective,
because any path in a smooth manifold is homotopic to a smooth
path. Secondly, it is injective, since if two smooth paths are
continously homotopic, they are smoothly homotopic.
\end{proof}

Henceforth we will not distinguish the continuous and smooth
versions of the fundamental group. In particular, every element of
the fundamental group has a smooth representative, and any two
such representatives have a smooth homotopy between them. This
will simplify our proofs, and will be vital in allowing certain
constructions to work at all. With this knowledge in hand, we
exclusively consider smooth paths, loops and homotopies, unless
stated otherwise.

A map $\psi : X \To Y$, taking $x_0$ to $y_0$ induces a
homomorphism of the fundamental groups, from $\pi_1(X,x_0)$ to
$\pi_1(Y,y_0)$. This is given by $\psi_* : [\alpha] \mapsto [\psi
\compose \alpha]$. A moment's consideration confirms this is a
homomorphism and well defined on $\pi_1(X,x_0)$.

\subsection{Classification of covering spaces}

With the definitions of covering spaces and fundamental groups in
place, we now state the main theorem for this section. It will be
used in several places in the ensuing work.

\begin{classification-of-covering-spaces-theorem}\label{prop:classification-covers}
Let $P$ be a smooth connected manifold, with base point
$\bdelt{p_0}$.

For any covering space $Q$ of $P$, with covering map $u: Q \To P$
and base point $\bdelt{q_0} \in u^{-1}(\bdelt{p_0}) \subset Q$,
the induced map $u_* :\pi_1(Q,\bdelt{q_0}) \To
\pi_1(P,\bdelt{p_0})$ is injective.

For each subgroup $K \subgroup \pi_1(P,\bdelt{p_0})$, there exists
a connected smooth covering space $Q$ of $P$, with smooth covering
map $u : Q \To P$, and a base point $\bdelt{q_0} \in
u^{-1}(\bdelt{p_0}) \subset Q$ such that the image of $u_*
:\pi_1(Q,\bdelt{q_0}) \To \pi_1(P,\bdelt{p_0})$ is exactly $K$.

Two coverings spaces $Q_1$ and $Q_2$, with covering maps $u_1 :Q_1
\To P$ and $u_2:Q_2 \To P$ and base points $\bdelt{q_1} \in
u_1^{-1}(\bdelt{p_0})$ and $\bdelt{q_2} \in u_2^{-1}(\bdelt{p_0})$
respectively, are equivalent as in Definition
\ref{defn:covering-space-equivalence} if and only if $u_{1
*}(\pi_1(Q_1,\bdelt{q_1}))$ and $u_{2 *}(\pi_1(Q_2,\bdelt{q_2}))$
are conjugate subgroups in $\pi_1(P,\bdelt{p})$.
\end{classification-of-covering-spaces-theorem}

\noindent \textsl{A preparatory remark.} For the most part,
smoothness is not particularly important in this theorem. The
hypothesis that $P$ is a smooth manifold enables us to dispose
easily of several of the necessary conditions for constructing
covering spaces which occur in the continuous setting. The
existence of smooth covering spaces follows very simply from the
existence of continuous covering spaces.

\begin{proof}
A complete proof of this theorem, as stated, cannot be found in
any one place. Furthermore, for later work we will need some of
the details of the constructions involved. For this reason, we
present here an outline of the proof, citing appropriate
references for each intermediate result, and in places extending
standard results to fit the particular circumstances of this
theorem.

The first part of the theorem, that the covering map induces an
injective map of the fundamental groups, is very straightforward,
using the lifting properties of covering maps. A proof is given in
\cite[\S 13]{ful:atfc}, and \cite[\S16.28.4]{die:ta3}.

Next, we consider the implications of the smoothness of $P$. Since
$P$ is a manifold, it is locally path connected and locally simply
connected, on account of each point of $P$ having a neighbourhood
homeomorphic to an open ball in $\Real^n$. Further, connectedness
implies that $P$ is path connected. This is because local path
connectedness means that the path connected components of $P$ are
open and closed, and so equal to connected components of $P$. See
also \cite[\S 3-4]{mun:tfc}.

The second part of the theorem, on existence of coverings, is
proved in the continuous setting in \cite[\S 8-14]{mun:tfc}. It
depends upon $P$ being path connected, locally path connected, and
locally (or semilocally) simply connected. As we have seen all
these conditions are automatically true for smooth manifolds. To
improve that result for this theorem, we need only show that this
covering can be given a smooth structure so that the covering map
becomes a smooth covering map, and this has already been achieved
above, in Proposition \ref{prop:continuous-smooth-coverings}. The
statement about the fundamental groups remains true in the smooth
setting, on account of Proposition
\ref{prop:smooth-fundamental-group-isomorphism}.

Finally, the last part, giving conditions for equivalence of
covering spaces, is proved in the continuous case in \cite[\S
8-14]{mun:tfc}. To improve this for the current theorem, we need
to show that if $u_1 : Q_1 \To P$ and $u_2 : Q_2 \To P$ are
continuously equivalent covering maps, then they are smoothly
equivalent covering maps, with respect to the differentiable
structures defined above. This follows immediately from the
definitions, and the fact that the continuous equivalence is given
by a homeomorphism $a : Q_1 \To Q_2$ such that $u_2 \compose a  =
u_1$, which is then also a diffeomorphism.
\end{proof}

\noindent \textsl{A concluding remark.} Later results will require
some of the details of the construction of covering spaces. To
that end, we describe this construction, and define the covering
map. We will not explicitly describe the topology on the covering
map. This is given in the references above, but we do not need the
details beyond knowing that the covering map is in fact a covering
map.

For a subgroup $K \subgroup \pi_1(P,\bdelt{p_0})$, the associated
covering space $Q$, as a set, is the collection of equivalence
classes of paths in $P$, starting at $\bdelt{p_0}$, and ending
anywhere in $P$, with two such paths $\alpha$ and $\beta$
considered equivalent if $\alpha(1)=\beta(1)$ and the homotopy
class ${[\alpha]^{-1}} \cat {[\beta]}$ is in $K$. We will write
$\sh{\alpha}$ for the equivalence class of $\alpha$. In
particular, if $[\alpha] = [\beta]$, then $\sh{\alpha} =
\sh{\beta}$. Moreover, if $[\gamma] \in K$, then $[\alpha \cat
\gamma]^{-1} \cat [\alpha] = [\gamma^{-1}] \in K$, so
$\sh{({\alpha} \cat {\gamma})} = \sh{\alpha}$. The covering map
$u$ maps such a element of $Q$ to its endpoint. Thus $u([\alpha])
= \alpha(1)$. This is clearly well defined.

\begin{cor}
Every smooth connected manifold has a universal covering manifold,
that is, a simply connected smooth covering space. Further, this
is essentially unique.
\end{cor}
\begin{proof}
Take the trivial subgroup $\langle e \rangle$ in
$\pi_1(P,\bdelt{p_0})$, and form the associated covering space
$Q$. Since the covering map $u$ induces an injective map
$\pi_1(Q,\bdelt{q_0}) \To \langle e \rangle$,
$\pi_1(Q,\bdelt{q_0})$ is itself trivial, and so $Q$ is simply
connected. If $u' : Q' \To P$ is any other covering map with $Q'$
simply connected with base point $\bdelt{q'_0}$, then $u'_*:
\pi_1(Q',\bdelt{q'_0}) \To \pi_1(P,\bdelt{p_0})$ has a trivial
image, and so the covering map $u' : Q' \To P$ is equivalent to
the one we have constructed, $u:Q \To P$.
\end{proof}

\subsection{Covering spaces of Lie
groups}\label{ssec:covering-lie-groups} Given a connected Lie
group $G$ we can form its universal covering manifold $\tilde{G}$,
with covering map $\rho : \tilde{G} \To G$. We always consider the
identity $e$ to be the base point of a group. Fix some $\tilde{e}
\in \rho^{-1}(e)$, the inverse image of the identity in $G$.

\begin{prop}
This manifold $\tilde{G}$ has a unique group structure with
identity $\tilde{e}$ so that $\rho$ becomes a homomorphism.
\end{prop}
\begin{proof}
This is proved in \cite[\S16.30]{die:ta3}. \footnote{An
alternative, less abstract sketch proof is as follows. The fact
that $\rho$ is locally a diffeomorphism near $\tilde{e}$ ensures
that there is a unique group structure on a neighbourhood of
$\tilde{e}$. By path connectedness, and the Lebesgue number lemma
\cite[\S 3-7]{mun:tfc}, every element of the group is a finite
product of elements of this neighbourhood. This extends the local
group structure to a group structure for the entire manifold. We
then have to check that it is well defined. For the sake of
brevity, we will not do the details here.}
\end{proof}

We will henceforth always mean the \emph{group} when we write
$\tilde{G}$, and call it the \emph{universal covering group}.

According to the construction given in the Classification of
Covering Spaces Theorem, the set underlying $\tilde{G}$ is the set
of homotopy classes of paths in $G$ starting at $e$ and ending
somewhere in the group $G$. The covering map $\rho:\tilde{G} \To
G$ then takes such a class of paths to the common endpoint.

\begin{example}
Familiar Lie groups with well known covering groups are $S^1$,
covered by $\Real$, where the covering map is $x \mapsto e^{2 \pi
i x}$, and $SO(3)$, covered by $SU(2)$. In Part
\ref{part:implications-for-physics} we will be particularly
concerned with the double covering of $\LL$ by $\SL$. This is the
physically relevant group in relativity theory.
\end{example}

\section{Spinor structures}
\label{sec:spinor-structure}

It is at this point, when we come to define a \emph{spinor
structure}, that the effort required to reformulate geometrically
the ideas of metrics and compatible covariant derivatives in terms
of orthonormal bundles and connections thereon comes to fruition.
The spinor structure will be explicitly constructed from the
orthonormal bundle. The alternative approach to spinors, which is
more common, is interested only in the algebraic side, and mostly
proceeds from the axioms for a spinor algebra \cite{pen:ss-t1}.
(Compare \S \ref{sec:spinor-algebra}.) The comparison of
constructive and axiomatic viewpoints in \cite[pp.
211--212]{pen:ss-t1} is especially worthwhile.

For the following definition, take $G$ to be a connected but not
simply connected Lie group, and $\tilde{G}$ to be its universal
covering group. The covering map will be denoted $\rho: \tilde{G}
\To G$.

\begin{defn*}
Given a $G$ principal fibre bundle $\pfbundle{G}{P}{\pi_P}{M}$, a
\defnemph{spinor structure} is a $\tilde{G}$ principal fibre bundle
$\pfbundle{\tilde{G}}{Q}{\pi_Q}{M}$, along with a principal fibre
bundle morphism relative to $\rho$, that is, a map $u: Q \To P$,
so that $u(\bdelt{q} \tilde{g}) = u(\bdelt{q}) \rho(\tilde{g})$,
for all $\bdelt{q} \in Q$ and $\tilde{g} \in \tilde{G}$. We call
$u$ the \emph{spinor map}.
\end{defn*}

This definition implies in particular that the projection maps are
related according to $$\pi_Q = \pi_P \compose u.$$

Accordingly, given a pseudo-Riemannian manifold, and suitable
orientations, we have seen that there is a corresponding
$SO_0(p,q)$ principal fibre bundle, which we have called the
orthonormal bundle. A spinor structure for such a
pseudo-Riemannian manifold is then just a spinor structure for
this bundle. Having recast pseudo-Riemannian geometry in terms of
principal fibre bundles, the theory of spinor structures for
pseudo-Riemannian manifolds can be subsumed into the general
discussion that we give here. We will see also that the
correspondence between covariant derivatives and connections on an
orthonormal bundle fits into this theory. In
\S\ref{sec:lifting-connection} we show how to generate connections
on a spinor structure from connections on the original bundle.

In the special case of a $(1+3)$ dimensional Lorentz structure,
where $G = \LL$ and $\Lambda M$ is an orthonormal frame bundle, a
spinor structure is an $\SL$ principal fibre bundle $\Sigma M$,
along with a map $u: \Lambda M \To \Sigma M$, so $u(\bdelt{q} g) =
u(\bdelt{q}) \rho(g)$ for all $g \in \SL$, where $\rho$ is the two
fold covering map described in detail in
\S\ref{ssec:covering-map}.

We will next state the main results on the existence and
uniqueness of spinor structures. We will later be particularly
interested in the case of Lorentz bundles and $\SL$ bundles.
However the discussion will apply to the more general situation.
Investigating the general case allows us later to discuss the
degree to which the choice of metric on a manifold affects the
existence and classification of the spinor structures, in
\S\ref{sec:metric-independence}.

To begin, we need the following fundamental lemma relating spinor
structures and covering maps.

\begin{lem}
\label{lem:principal-morphism-covering} If $Q$ is a spinor
structure for the bundle $P$, the principal fibre bundle morphism
$u: Q \To P$ is a covering map.
\end{lem}
\begin{proof}
Consider a local cross section of $Q$, defined on an open subset
$U \subset M$, $\sigma : U \To Q$. The composition $u \compose
\sigma$ then defines a local cross section of $P$. We can use
these cross sections to define local trivialisations of both
bundles, by Lemma \ref{lem:sections-trivialisations}.
\begin{eqnarray*}
 \psi :    & U \times \tilde{G} & \To      \pi_Q^{-1}(U) \\
           & (m,\tilde{g})      & \mapsto  \sigma(m) \tilde{g} \\
 \varphi : & U \times G & \To      \pi_P^{-1}(U) \\
           & (m,g)      & \mapsto  u(\sigma(m)) g
\end{eqnarray*}
Both of these maps are diffeomorphisms, and in fact principal
morphisms. We can compose these maps with $u$, to obtain
 $$ \varphi^{-1} \compose u \compose \psi : U \times \tilde{G} \To U
 \times G .$$
However this map acts very simply, as follows,
\begin{eqnarray*}
 \left(\varphi^{-1} \compose u \compose \psi\right)(m,\tilde{g})
    & = & (\varphi^{-1} \compose u)(\sigma(m) \tilde{g}) \\
    & = & \varphi^{-1}(u(\sigma(m)) \rho(\tilde{g})) \\
    & = & (m, \rho(\tilde{g})).
\end{eqnarray*}
Thus $\varphi^{-1} \compose u \compose \psi = \operatorname{id}_M
\times \rho$, and as $\psi$ and $\varphi$ are diffeomorphisms, we
can write the covering map as $\restrict{u}{\pi_Q^{-1}(U)} =
\varphi \compose (\operatorname{id}_M \times \rho) \compose
\psi^{-1}$. This expresses $u$ locally as a trivial map in the
sense of covering spaces, and so $u$ is a covering map.
\end{proof}

This enables us to apply the powerful Classification of Covering
Spaces Theorem to the task at hand. It also indicates the dual
appearance of covering space theory in the description of a spinor
structure. To look for a principal fibre bundle whose structure
group has been `unwrapped' to the simply connected covering group,
we must `unwrap' the bundle itself. This is not always possibly,
and we will see that the desired covering bundle is not itself
simply connected, and so need not be unique when one does exist.

An immediate and simple result of Lemma
\ref{lem:principal-morphism-covering} and the Classification of
Covering Spaces Theorem is the following.
\begin{prop}
If the fundamental group of $P$ is trivial then there is no spinor
structure.
\end{prop}
\begin{proof}
Since $P$ is simply connected, every covering space is equivalent
to $P$ itself, and so $P$ has no connected covering spaces larger
than itself, and thus no spinor structure is possible.
\end{proof}

Next, we need to say exactly what we mean by `classification' of
spinor structures, by defining what it means to say that two are
\emph{equivalent}.

\begin{defn}\label{defn:spinor-structure-equivalence}
Two spinor structures
\[\pfbundle{\tilde{G}}{Q}{\pi_Q}{M} \quad \text{and} \quad
\pfbundle{\tilde{G}}{Q'}{\pi_{Q'}}{M}\] with spinor maps $u:Q \To
P$ and $u' :Q' \To P$ respectively are said to be
\defnemph{equivalent} is there is a principal fibre bundle morphism
$a: Q \To Q'$ such that $u = u' \compose a$.
\end{defn}

The main results of this section are summarised by the following
theorems.

We begin by defining the map $i : G \To P$ by $i(g) = \bdelt{p_0}
g$. This induces a homomorphism $i_* : \pi_1(G) \To \pi_1(P)$.
\begin{existence-theorem}
A principal fibre bundle $\pfbundle{G}{P}{\pi_P}{M}$ has a spinor
structure if and only if the fundamental group of the bundle
$\pi_1(P)$ can be written as a direct product of subgroups $K$ and
$I$, \[\pi_1(P) = K \times I,\] such that $K$ and $I$ have trivial
intersection, and $\pi_{P *}$ maps $K$ isomorphically to
$\pi_1(M)$ and $i_*$ maps $\pi_1(G)$ isomorphically to
$I$.\footnote{We could state this condition more concisely, but
more abstractly, as `there is a short exact sequence
\[
\xymatrix{
    0 \ar[r] & \pi_1(G) \ar[r]^{i_*} & \pi_1(P) \ar[r]^{\pi_{P *}}
    & \pi_1(M) \ar[r] & 0}
\]
and this sequence is split'. We will not be thinking in these
terms however.}
\end{existence-theorem}

Note that if $P$ is trivial, so $P = M \times G$, then there is an
obvious spinor structure, given by $Q = M \times \tilde{G}$, and
$u:Q \To P$ according to $u(m,\tilde{g}) = (m,\rho(\tilde{g}))$.
In this case the theory of fundamental groups shows that $\pi_1(P)
= \pi_1(M) \times \pi_1(G)$. We can think of the existence theorem
as the statement that even if $P$ is not trival, to have a spinor
structure `its fundamental group must look as if $P$ is trivial'.

\begin{classification-theorem}
In the case that the conditions of the Existence Theorem obtain,
the inequivalent spinor structures are in one to one
correspondence with the homomorphisms from $\pi_1(M) \To
\pi_1(G)$.
\end{classification-theorem}

This is a `relative' classification. Given a particular spinor
structure, each of the other spinor structures corresponds to a
particular nontrivial homomorphism $\pi_1(M) \To \pi_1(G)$.

To reach these results, we will first establish the necessary
conditions for the existence of a spinor structure. This is
achieved in \S\ref{sssec:necessary-conditions}. That these
conditions are sufficient will follow, in
\S\ref{sssec:sufficiency}, and subsequently we will describe the
classification of spinor structures in
\S\ref{sssec:classification}.

\subsection{Necessary conditions}
\label{sssec:necessary-conditions}

Suppose now that there exists a spinor structure $Q$ for $P$, with
spinor map $u:Q \To P$. The four main results that follow from
this are Propositions \ref{prop:piQ-isomorphism},
\ref{prop:piP-isomorphim}, \ref{prop:i-injective} and
\ref{prop:sigma-isomorphism}. Together, these establishe the
necessity of the conditions in the Existence Theorem.

\begin{prop}\label{prop:piQ-isomorphism}
The map \[\pi_{Q *} : \pi_1(Q) \To \pi_1(M)\] is an isomorphism.
\end{prop}
\begin{proof}
For the purposes of this proof, we will fix a connection on $Q$.
Such a connection always exists by the results in
\S\ref{ssec:connection-forms}.

Proving that $\pi_{Q *}$ is surjective is relatively easy, so we
first do that.

Suppose $\alpha$ is any smooth loop in $M$ based at $m_0$. Define
$\tilde{\alpha}_\bdelt{q_0}$ to the parallel transport of
$\bdelt{q_0} \in \pi_Q^{-1}(m_0)$ along $\alpha$. This curve will
generally not be a loop. However,
$\pi_Q(\tilde{\alpha}_\bdelt{q_0}(1)) = m_0$, and since the fibres
of $Q$ are path connected, we can find a path $\delta : \I \To
\tilde{G}$ so $\delta(0) = \tilde{e}$ and $\delta(1) =
\tau(\tilde{\alpha}_\bdelt{q_0}(1), \bdelt{q_0})$. Now consider
the path $\tilde{\alpha}_\bdelt{q_0} \delta$, which is in fact a
loop since $(\tilde{\alpha}_\bdelt{q_0} \delta)(0) = \bdelt{q_0}$
and $(\tilde{\alpha}_\bdelt{q_0} \delta)(1) =
\tilde{\alpha}_\bdelt{q_0}(1)
\tau(\tilde{\alpha}_\bdelt{q_0}(1),\bdelt{q_0}) = \bdelt{q_0}$.
Further $\pi_Q(\tilde{\alpha}_\bdelt{q_0} \delta) =
\pi_Q(\tilde{\alpha}_\bdelt{q_0}) = \alpha$, and so $\pi_{Q
*}[\tilde{\alpha}_\bdelt{q_0} \delta] = [\alpha]$. Thus $\pi_{Q
*}$ is surjective.

We now turn to the more technical problem of demonstrating that
$\pi_{Q *}$ is injective. The underlying result, however, has
already been established, the idea here being to use a connection
to `lift' a homotopy in $M$ to a map into $Q$, and then using the
simply connectedness of fibres to modify this into the appropriate
homotopy. Suppose $[\alpha]$ and $[\beta]$ are elements of
$\pi_1(Q)$, and $\pi_{Q*}([\alpha]) = \pi_{Q*}([\beta])$. Then
there are smooth loops $\gamma_0$ and $\gamma_1$ in $M$, so
$[\gamma_0] = \pi_{Q*}([\alpha])$ and $[\gamma_1] =
\pi_{Q*}([\beta])$, and, further, there are smooth loops $\alpha'$
and $\beta'$ in $Q$ so $[\alpha] = [\alpha']$, $[\beta]=[\beta']$
and $\pi_Q(\alpha') = \gamma_0$, and $\pi_Q(\beta') = \gamma_1$.
Thus there is a smooth homotopy from $\gamma_0$ to $\gamma_1$.
Call this homotopy $\gamma$, so $\gamma(0,t) = \gamma_0(t)$, and
$\gamma(1,t) = \gamma_1(t)$. We will write $\gamma_s$ for the
function $t \mapsto \gamma(s,t)$. According to the second part of
Proposition \ref{prop:parallel-transport}, we can parallel
transport $\bdelt{q_0}$ along $\gamma_s$, to obtain a smooth curve
$\tilde{\gamma_s}_{\bdelt{q_0}}$, so that
$\pi_Q(\tilde{\gamma_s}_{\bdelt{q_0}}(t)) = \gamma(s,t)$, and the
map $H : (s,t) \mapsto \tilde{\gamma_s}_{\bdelt{q_0}}(t)$ is
continuous. We will next modify $H$ to form a homotopy between
$\alpha'$ and $\beta'$.

The particular properties of $H$ that we require are
\begin{align*}
H(s,0) & = \bdelt{q_0}, \\
\pi_Q(H(0,t)) & = \gamma_0(t) = \pi_Q(\alpha'(t)),\\
\pi_Q(H(1,t)) & = \gamma_1(t) = \pi_Q(\beta'(t)), \textrm{ and}\\
\pi_Q(H(s,1)) & = m_0 = \pi_Q(\bdelt{q_0}).
\end{align*}
Define $\partial$ to be the boundary of $\I \times \I$, that is
\[\partial = \left(\{0,1\} \times \I\right) \cup \left(\I \times
\{0,1\}\right).\] Define $\zeta:\partial \To Q$ according to
$\zeta(s,0)=\zeta(s,1) = \bdelt{q_0}$ for all $s \in \I$, and
$\zeta(0,t) = \alpha'(t)$, $\zeta(1,t) = \beta'(t)$. According to
this definition, and the above properties of $H$, $\pi_Q \compose
\zeta = \pi_Q \circ \restrict{H}{\partial}$ on $\partial$, and so
$\zeta = (\restrict{H}{\partial})\tilde{g}$, for some function
$\tilde{g} :
\partial \To \tilde{G}$. Since $\tilde{G}$ is simply connected, we
can extend $\tilde{g}$ to a continuous function $\tilde{g} : \I
\times \I \To \tilde{G}$. Now define $K : \I \times \I \To Q$ by
$K = H \tilde{g}$. Thus on $\partial$, $K$ and $\zeta$ agree, and
so $K$ is a continuous homotopy between $\alpha'$ and $\beta'$.
Finally, this implies that there is a continuous homotopy between
$\alpha$ and $\beta$, and so by Proposition
\ref{prop:continuous-smooth-homotopy}, there is a smooth homotopy
between $\alpha$ and $\beta$. This establishes the injectivity of
$\pi_{Q *}$, and so proves that it is an isomorphism.
\end{proof}

\begin{prop}\label{prop:piP-isomorphim}
Let $K = u_*(\pi_1(Q)) \subgroup \pi_1(P)$. Then the restriction
of $\pi_{P *}$ to $K$, mapping $K$ to $\pi_1(M)$, is an
isomorphism.
\end{prop}
\begin{proof}
Firstly, the map $u_* : \pi_1(Q) \To \pi_1(P)$ is injective,
according to the Covering Space Classification Theorem. We now
consider the following commuting diagram,
\[
 \xymatrix@R+3pt{
    \pi_1(Q) \ar[dd]^{\pi_{Q *}} \ar[dr]^{u_*} & \\
        & K \subgroup \pi_1(P) \ar[dl]^{\pi_{P *}}       \\
    \pi_1(M)
 }
\]
and the restriction of $\pi_{P *}$ to $K$, $\restrict{(\pi_{P
*})}{K}$. Since $u_*$ is injective and $\pi_{Q *}$ is an
isomorphism, by Proposition \ref{prop:piQ-isomorphism},
$\restrict{(\pi_{P *})}{K}$ is injective. Further,
$\restrict{(\pi_{P *})}{K} u_* = \pi_{Q *}$, so $\restrict{(\pi_{P
*})}{K}$ must be surjective, and thus $\restrict{(\pi_{P *})}{K}$
is an isomorphism.
\end{proof}

An important property of the map $i : G \To P$ is that $i_*$ maps
$\pi_1(G)$ into the centre of $\pi_1(P)$. This is made clear by
the following Lemma.

\begin{lem}\label{lem:homotopy-centre}
Suppose $g \in \Pi G$, and $\alpha \in \Pi P$. Then
\begin{equation*}
 \left[ \alpha g \right] =
 \left[ {(\alpha g(1))} \cat {i(g)}\right].
\end{equation*}
If $g \in \Omega G$, and $\alpha \in \Omega P$, then
\begin{equation*}
\left[ {i(g)} \cat {\alpha} \right ]
    = \left[ \alpha g \right] =
\left[ {\alpha} \cat {i(g)}\right].
\end{equation*}
Thus $i_* : \pi_1(G) \To Z(\pi_1(P))$.
\end{lem}
\begin{proof}
See \S\ref{ssec:proof-lem:homotopy-centre}.
\end{proof}

It should also be pointed out that $\pi_1(G)$ is always itself
commutative when $G$ is a Lie group, as discussed in Lemma
\ref{lem:pi1G-commutative}.

\begin{prop}\label{prop:i-injective}
The map $i_*:\pi_1(G) \To \pi_1(P)$ is injective, and so if we
define $I = i_*(\pi_1(G)) \subgroup \pi_1(P)$, then $i_* :
\pi_1(G) \To I$ is an isomorphism.
\end{prop}
\begin{proof}
Suppose $[g] \in \pi_1(G)$, and $i_*[g] = [\alpha]$. Suppose
$\alpha$ is homotopically trivial. Then, according to the Path
Lifting Lemma \cite[\S 8-4]{mun:tfc} we can lift $\alpha$ via the
covering map $u$ to a path $\tilde{\alpha}$ in $Q$, and according
to the Homotopy Lifting Lemma \cite[\S 8-4]{mun:tfc}, it is a loop
homotopic to the constant loop. This loop lies within a single
fibre, and, since the fibres are homeomorphic to $\tilde{G}$, they
are simply connected, and so $\tilde{\alpha}$ is homotopic to the
constant loop by a homotopy that stays within the fibre
$\pi_Q^{-1}(m_0)$. Applying $u$ to this homotopy gives a homotopy
of $\alpha$ to the constant loop by a homotopy that stays within
the fibre $\pi_P^{-1}(m_0)$, and thus $g$ is homotopic to the
constant loop in $G$. Thus $[g]=[e]$, and so $i_*$ is injective.
\end{proof}

\begin{prop}\label{prop:sigma-isomorphism}The groups $K$ and $I$ have trivial
intersection in $\pi_1(P)$, and the internal direct product $K
\times I$ is exactly $\pi_1(P)$.
\end{prop}
\begin{proof}
The proof is in two steps.

Say $[\alpha] \in K$, $[g] \in \pi_1(G)$, and $[\alpha] = i_*[g]$.
Then, applying $\pi_{P *}$ to both sides,
\[\pi_{P *}[\alpha] = \pi_{P *} i_*[g] = [e],\]
since $i_*[g]$ has a representative lying within a single fibre.
Now, since $\pi_{P *}$ restricted to $K$ is an isomorphism,
$[\alpha] = [e]$ also, and since $i_*$ is injective by Proposition
\ref{prop:i-injective}, $[g]=[e]$ as well. Thus the two groups
have a trivial intersection.

Next, take any $[\alpha] \in \pi_1(P)$. We define $[\hat{\alpha}]
\in K \subgroup \pi_1(P)$ as follows. Firstly let $\beta = \pi_P
\compose \alpha : \I \To M$. Then, as in the discussion of
Proposition \ref{prop:piQ-isomorphism}, let $\beta' =
\tilde{\beta}_{\bdelt{q_0}} : \I \To Q$ be the parallel transport
of $\bdelt{q_0}$ along $\beta$. Further, chose a path $\delta : \I
\to \tilde{G}$ so $\delta(0) = \tilde{e}$ and $\beta'' = \beta'
\delta$ is a loop in $Q$. Now $\pi_Q \compose \beta'' = \pi_Q
\compose \beta' = \pi_P \compose \alpha$. Define $\hat{\alpha}$ =
$u \compose \beta''$. We see from this, and Proposition
\ref{prop:piQ-isomorphism}, that $[\hat{\alpha}] = u_* \pi_{Q
*}{}^{-1} \pi_{P *} [\alpha]$, and moreover that $\pi_P \compose
\hat{\alpha} = \pi_P \compose \alpha$.

Thus $\alpha = \hat{\alpha} g$ for some loop $g : \I \To G$. Then
\begin{align*}
 [\alpha] & = [\hat{\alpha} g] \\
          & = {[\hat{\alpha}]} \cat {i_*[g]},
\end{align*}
applying Lemma \ref{lem:homotopy-centre}. Thus the internal direct
product $K \times I$ generates all of $\pi_1(P)$.
\end{proof}

\subsection{Sufficient conditions}
\label{sssec:sufficiency} Now we suppose the conditions stated in
the Existence Theorem. That is, suppose that $P$ is the total
space of a $G$ principal fibre bundle over $M$, and that there is
a subgroup $K$ of $\pi_1(P)$ isomorphic to $\pi_1(M)$ via $\pi_{P
*}$ such that $\pi_1(P) = K \times i_*(\pi_1(G))$. We will show
that these conditions are sufficient for the existence of a spinor
structure $Q$.

Let $Q$ be the covering space of $P$ associated with the subgroup
$K$, according to the Covering Space Classification Theorem, and
$u$ be the corresponding covering map. There is a base point
$\bdelt{q_0} \in u^{-1}(\bdelt{p_0}) \subset Q$ so that $u_*$ maps
$\pi_1(Q,\bdelt{q_0}) \To K \subgroup \pi_1(P,\bdelt{p_0})$ and is
injective. Define the projection map $\pi_Q : Q \To M$ by $\pi_Q =
\pi_P \compose u$.

We now define a $\tilde{G}$ action on $Q$. We will then show that
with respect to this action $Q$ becomes a $\tilde{G}$ principal
fibre bundle, and $u$ a principal morphism relative to $\rho :
\tilde{G} \To G$.

\begin{figure}[!hbp]
\centering
\includegraphics[width=0.5\textwidth]{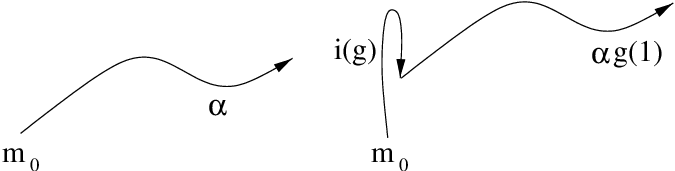}
\caption{The paths corresponding to $\bdelt{q}$ and $\bdelt{q}
\tilde{g}$.\label{fig:group-action-paths}}
\end{figure}

Fix $\bdelt{q} \in Q$ and $\tilde{g} \in \tilde{G}$. According to
the construction of the covering space, outlined in the remark
following the proof of the Covering Space Classification theorem,
$\bdelt{q}$ is an equivalence class of paths in $P$, written
$\sh{\alpha}$, for some $\alpha : \I \To P$, with $\alpha(0) =
\bdelt{p_0}$. Two such paths are equivalent, $\sh{\alpha} =
\sh{\beta}$, if $\alpha(1)=\beta(1)$ and $[\alpha^{-1} \cat \beta]
\in K$. Since $\tilde{G}$ is the universal covering group of $G$,
$\tilde{g}$ can be thought of as a homotopy class of paths in $G$
starting at the identity, as in \S\ref{ssec:covering-lie-groups}.
Choose a path from this homotopy class, and denote it $g : \I \To
G$, so $g(0) = e$. The $G$ action on $P$ allows us to define a
path $\alpha g :\I \To P$ by $(\alpha g)(t) =\alpha(t) g(t)$.
Define the action of $\tilde{g}$ on $\bdelt{q}$ by $\bdelt{q}
\tilde{g} = \sh{(\alpha g)}$. According to the first part of Lemma
\ref{lem:homotopy-centre}, $\left[ \alpha g \right] = \left[
{(\alpha g(1))} \cat {i(g)}\right]$, and so we can alternatively
write
\begin{equation}\label{eq:tildeG-action}
\bdelt{q} \tilde{g} = \sh{({(\alpha g(1))} \cat {i(g)})}.
\end{equation}
These paths are illustrated in Figure
\ref{fig:group-action-paths}.
This is clearly independent of the particular path $g$ we have
chosen, because homotopic paths in $P$ are equivalent as points in
$Q$. To check that this definition is also independent of the
representative of $\sh{\alpha}$, we suppose $\sh{\alpha} =
\sh{\beta}$, so $[\beta^{-1} \cat \alpha] \in K$. Then $[(\beta
g)^{-1} \cat (\alpha g)] = [\beta^{-1} \cat i(g^{-1}) \cat i(g)
\cat \alpha] \in K$, using Lemma \ref{lem:homotopy-centre}, and so
$\sh{(\alpha g)} = \sh{(\beta g)}$.

We have now defined the projection map $\pi_Q$ and the $\tilde{G}$
action on $Q$. Our claim is that these provide a spinor structure
for $P$. Thus the remainder of the proof of the Existence Theorem
is contained in the following two results. Proposition
\ref{prop:Q-isa-pfb} checks the consistency of $\pi_Q$ and the
$\tilde{G}$ action, in the sense that together they satisfy the
axioms for a principal fibre bundle, in Definition
\ref{defn:principal-fibre-bundle}. Lemma
\ref{lem:u-isa-principal-morphism} then proves that $u$, the
covering map from $Q$ to $P$, is in fact a principal morphism
relative to $\rho$, respecting the principal fibre bundle
structures of $Q$ and $P$.

\begin{prop}
\label{prop:Q-isa-pfb} The above construction of
$\pfbundle{\tilde{G}}{Q}{\pi_Q}{M}$ is in fact a principal fibre
bundle. Specifically, the $\tilde{G}$ action on $Q$ must be
\begin{description}
\item [free] in the sense that if
$\bdelt{q} \tilde{g} = \bdelt{q}$ for any $\bdelt{q} \in Q$, then
$\tilde{g} = e$, and,
\item [transitive on fibres] so if $\bdelt{q}, \bdelt{q'} \in
Q$ are such that $\pi_Q(\bdelt{q}) = \pi_Q(\bdelt{q'})$, then
there is some $\tilde{g} \in \tilde{G}$ so that $\bdelt{q'} =
\bdelt{q} \tilde{g}$.
\end{description}
Further, there must be local trivialisations of $Q$ compatible
with the $\tilde{G}$ action.
\end{prop}
\begin{proof}
The proof is in three parts. All are straightforward, but somewhat
involved, especially the second.

\emph{The action is free.} Suppose $\tilde{g} \in \tilde{G}$ is
such that $\bdelt{q} \tilde{g} = \bdelt{q}$ for some $\bdelt{q} =
\sh{\alpha} \in Q$. Take a path in $G$ representing $\tilde{g}$,
say $g : \I \To G$, so $g(1) = \rho(\tilde{g})$. As in Equation
\eqref{eq:tildeG-action}, $\bdelt{q} \tilde{g} = \sh{((\alpha
g(1)) \cat i(g))}$. Then $\bdelt{q} \tilde{g} = \bdelt{q}$ implies
\begin{align*}
{[\alpha]^{-1}} \cat {[{(\alpha g(1))} \cat {i(g)}]} & = [i(g)] \\
    & \in K,
\end{align*}
and so $[i(g)] = [e]$, by the hypothesis that $\pi_1(P) = K \times
i(\pi_1(G))$. Thus $g$ is homotopically trivial, and so $\tilde{g}
= e$.

\emph{The action is transitive.} Suppose we have two elements
$\bdelt{q}, \bdelt{q'}$ of $Q$ within the same fibre, such that
$\bdelt{q} = \sh{\alpha}$ and $\bdelt{q'} = \sh{\beta}$ for two
paths $\alpha, \beta : \I \To P$. Since $\bdelt{q}$ and
$\bdelt{q'}$ are in the same fibre, $\alpha(1), \beta(1) \in
{\pi_P}^{-1}(m)$ for some $m \in M$. Consider ${[\beta]^{-1}} \cat
{[{\alpha g} \cat {\gamma}]}$ for some $[\gamma] \in K \subset
\pi_1(P)$, and $g: \I \To G$ so $g(0) = e$ and $\alpha(1) g(1) =
\beta(1)$. Such a $g$ exists since $G$ is path connected, and $G$
acts transitively on the fibres of $P$. Further, $g$ represents
some $\tilde{g} \in \tilde{G}$. We calculate
\begin{align*}
[\beta]^{-1} \cat [\alpha g \cat \gamma]
    & = [\beta^{-1} \cat \alpha g(1) \cat i(g) \cat \gamma] \\
    & = [\beta^{-1} \cat \alpha g(1) \cat i(h^{-1}) \cat i(h \cat g) \cat \gamma] \\
    & = [\beta^{-1} \cat \alpha g(1) \cat i(h^{-1}) \cat \gamma \cat i(h \cat
    g)].
\end{align*}
Here $h :\I \To G$ is any path in $G$ so $h(0)=g(1)$ and $h(1) =
e$, and $h^{-1}$ denotes the reversed path $h^{-1}(t) = h(1-t)$,
not the inverse path, and we have used Lemma
\ref{lem:homotopy-centre} in the last line. Note that by varying
$g$, subject still to the conditions $g(0) = e$ and $\alpha(1)
g(1) = \beta(1)$, we can make  $h \cat g$ homotopic to any
arbitrary loop $j$ in $G$. This is achieved by setting $g = h^{-1}
\cat j$, so $[h \cat g] = [j]$. Thus the first two paths, $\gamma$
and $i(h \cat g)$ can be chosen to generate any element of
$\pi_1(P)$, since $\pi_1(P) = K \times i(\pi_1(G))$. In
particular, we can chose $g$ and $\gamma$ so that
\[[\gamma \cat i(h \cat g)] = [\beta^{-1} \cat \alpha g(1) \cat
i(h^{-1})]^{-1},\] so
\[[\beta]^{-1} \cat [\alpha g \cat \gamma] = [e] \in K.\] With this choice,
\[\bdelt{q'} = \sh{\beta} = \sh{\alpha g \cat \gamma} = \sh{\alpha
g} = \bdelt{q} \tilde{g}.\]

This proves that the $\tilde{G}$ action is transitive on the
fibres, as required.

\emph{There are local trivialisations compatible with the
$\tilde{G}$ action.} Since $P$ is a principal fibre bundle, for
any point $m_0 \in M$ there is an open set $V$ with $m_0 \in V
\subset M$ and a local section $\sigma : V \To P$, in accordance
with Lemma \ref{lem:sections-trivialisations}. Find a simply
connected open set $U \subset V$, and restrict $\sigma$ to $U$. By
the monodromy principle
%-%really necessary?
\cite[\S16.28.8]{die:ta3} there is a lifting of $\sigma$ to a map
$\tilde{\sigma} : U \To Q$ via the covering map $u$. This is then
a local section of $Q$, and applying Lemma
\ref{lem:sections-trivialisations} a second time we find a local
trivialisation.
\end{proof}

The final step in establishing that our construction generates a
spinor structure is now easy.
\begin{lem}
\label{lem:u-isa-principal-morphism} The map $u: Q \To P$ is a
principal morphism relative to $\rho$.
\end{lem}
\begin{proof}
In the notation above $\rho$ acts on $\tilde{G}$ by taking
$\tilde{g}$ to $g(1)$ . Thus
\begin{align*}
 u(\bdelt{q} \tilde{g}) & = u(\sh{(\alpha g)}) \\
                       & = \alpha(1) g(1) \\
                       & = u(\bdelt{q})
                       \rho(\tilde{g}).&&\qedhere
\end{align*}
\end{proof}

Following from these results, we obtain the following statement
about the spinor structure equivalence, which will be vital in
proving the classification in \S\ref{sssec:classification}.

\begin{prop}\label{prop:spinor-structure-equivalence}
Two spinor structures $u : Q \To P$ and $u' : Q' \To P$ are
equivalent (in the sense of Definition
\ref{defn:spinor-structure-equivalence}) if and only if they are
equivalent as covering maps (Definition
\ref{defn:covering-space-equivalence}).
\end{prop}
\begin{proof}
If $u : Q \To P$ and $u' : Q' \To P$ are equivalent as spinor
structures then there is a principal bundle morphism $a : Q \To
Q'$ so $u = u' \compose a$. This $a$ is then \mbox{\emph{a
fortiori}} a diffeomorphism, and so $u$ and $u'$ are immediately
seen to be equivalent as covering maps.

Conversely, suppose $u: Q \To P$ and $u' : Q' \To P$ are
equivalent as covering spaces, so there is a diffeomorphism $a : Q
\To Q'$ such that $u = u' \compose a$. Since $u$ and $u'$ are
spinor maps, we can easily calculate
\begin{align*}
 u'(a(\bdelt{q} \tilde{g})) & = u(\bdelt{q} \tilde{g}) \\
        & = u(\bdelt{q}) \rho(\tilde{g}) \\
        & = u'(a(\bdelt{q})) \rho(\tilde{g}) \\
        & = u'(a(\bdelt{q}) \tilde{g}).
\end{align*}
The equality between the first and last expressions then implies
that $a(\bdelt{q} \tilde{g}) = a(\bdelt{q}) \tilde{g} \tilde{k}$,
for some $\tilde{k} \in \ker\rho \subset \tilde{G}$. Further,
since $u'$ is a covering map, if we fix $\bdelt{q}$, $\tilde{k}$
depends continuously on $\tilde{g}$. Since $\ker\rho$ is discrete,
$\tilde{k}$ is constant, and since if $\tilde{g} = e$, $\tilde{k}
= e$, we must have $a(\bdelt{q} \tilde{g}) = a(\bdelt{q})
\tilde{g}$ for all $\tilde{g} \in \tilde{G}$. That is, $a$ is
additionally a principal bundle morphism, and so $u$ and $u'$ are
equivalent as spinor structures.
\end{proof}

Finally, this result guarantees that every spinor structure (up to
equivalence, of course) is obtained via the construction of this
section. The argument is as follows. Suppose $u': Q' \To P$ is a
spinor structure. According to Lemma
\ref{lem:principal-morphism-covering}, $u'$ is a covering map.
Now, up to equivalence, a covering map $u:Q' \To P$ is determined
by $u'_*(\pi_1(Q'))$, according to the Classification of Covering
Space Theorem. The previous section, on necessary conditions,
ensures that $u'_*(\pi_1(Q'))$ satisfies the hypotheses required
for the construction of the spinor structure $u : Q \To P$. Since
$u_*(\pi_1(Q')) = u'_*(\pi_1(Q'))$, $u$ and $u'$ are equivalent as
covering maps, and so, by this latest result, equivalent as spinor
structures. This underpins the proof of the Classification
Theorem, given in the next section.

\subsection{Classification of inequivalent spinor structures}
\label{sssec:classification}

The next step of the analysis describes the uniqueness or
otherwise of spinor structures, in the case that one exists at
all. Thus in this section will we give the proof of the
Classification Theorem.

The condition for the existence of a spinor structure requires
that we can write $\pi_1(P)$ in a particular way, as a direct
product of groups isomorphic to $\pi_1(M)$ and $\pi_1(G)$.
Moreover, the $\pi_1(G)$ factor is determined by the image of $i_*
: \pi_1(G) \To \pi_1(P)$. We therefore have some freedom in
choosing the first factor, in that we can choose any subgroup of
$\pi_1(P)$ isomorphic to $\pi_1(M)$ via $\pi_{P *}$, as long as
the internal direct product of this subgroup with the fixed
$\pi_1(G)$ subgroup is all of $\pi_1(P)$, as in the statement of
the Existence Theorem.

Figure \ref{fig:pi1Pdecomposition} indicates this freedom, with
the diagrams a) and b) depicting two choices of a subgroup
isomorphic to $\pi_1(M)$, $K$ and $L$. Perhaps an analogy could be
made with the choice of horizontal subspace made in defining a
connection. In that case the vertical subspace, tangent to a
fibre, is fixed, just as here the $\pi_1(G)$ factor is fixed as
the image of $i$.
\begin{figure}[!hbp]
{\setlength{\unitlength}{3947sp}%
\begingroup\makeatletter\ifx\SetFigFont\undefined
% extract first six characters in \fmtname
\def\x#1#2#3#4#5#6#7\relax{\def\x{#1#2#3#4#5#6}}%
\expandafter\x\fmtname xxxxxx\relax \def\y{splain}%
\ifx\x\y   % LaTeX or SliTeX?
\gdef\SetFigFont#1#2#3{%
  \ifnum #1<17\tiny\else \ifnum #1<20\small\else
  \ifnum #1<24\normalsize\else \ifnum #1<29\large\else
  \ifnum #1<34\Large\else \ifnum #1<41\LARGE\else
     \huge\fi\fi\fi\fi\fi\fi
  \csname #3\endcsname}%
\else \gdef\SetFigFont#1#2#3{\begingroup
  \count@#1\relax \ifnum 25<\count@\count@25\fi
  \def\x{\endgroup\@setsize\SetFigFont{#2pt}}%
  \expandafter\x
    \csname \romannumeral\the\count@ pt\expandafter\endcsname
    \csname @\romannumeral\the\count@ pt\endcsname
  \csname #3\endcsname}%
\fi \fi\endgroup
\begin{picture}(5197,1507)(54,-721)
\thicklines \put(4126,764){\line( 0,-1){1200}}
\put(3076,-136){\line( 4, 1){2103.529}} \thinlines \put(
76,764){\line( 1, 0){2100}} \put(2176,764){\line( 0,-1){1200}}
\put(2176,-436){\line(-1, 0){2100}} \put( 76,-436){\line( 0,
1){1200}} \thicklines \put(1126,764){\line( 0,-1){1200}} \put(
76,164){\line( 1, 0){2100}}
\put(4201,614){\makebox(0,0)[lb]{\smash{\SetFigFont{8}{9.6}{rm}$\pi_1(G)$}}}
\thinlines \put(3076,764){\line( 1, 0){2100}}
\put(5176,764){\line( 0,-1){1200}} \put(5176,-436){\line(-1,
0){2100}} \put(3076,-436){\line( 0, 1){1200}}
\put(5251,614){\makebox(0,0)[lb]{\smash{\SetFigFont{8}{9.6}{rm}$\pi_1(P)$}}}
\put(4076,-721){\makebox(0,0)[lb]{\smash{\SetFigFont{12}{14.4}{rm}b)}}}
\put(4726,164){\makebox(0,0)[lb]{\smash{\SetFigFont{8}{9.6}{rm}$L$}}}
\put(1726,
14){\makebox(0,0)[lb]{\smash{\SetFigFont{8}{9.6}{rm}$K$}}}
\put(1201,614){\makebox(0,0)[lb]{\smash{\SetFigFont{8}{9.6}{rm}$\pi_1(G)$}}}
\put(2251,614){\makebox(0,0)[lb]{\smash{\SetFigFont{8}{9.6}{rm}$\pi_1(P)$}}}
\put(1061,-721){\makebox(0,0)[lb]{\smash{\SetFigFont{12}{14.4}{rm}a)}}}
\end{picture}
} \caption{Possible decompositions
of $\pi_1(P)$.\label{fig:pi1Pdecomposition}}
\end{figure}

Each such choice results in a spinor structure for $P$, according
to the above construction. We have seen previously that these
choices exhaust all the possible spinor structures. That these
choices all result in inequivalent structures is straightforward,
using the result furnished by Proposition
\ref{prop:spinor-structure-equivalence}.

\begin{prop}
Any two different choices of the subgroup isomorphic to $\pi_1(M)$
result in inequivalent spinor structures.
\end{prop}
\begin{proof}
Suppose $K$ and $L$ are subgroups of $\pi_1(P)$, each isomorphic
to $\pi_1(M)$ via $\pi_{P *}$, such that we can construct spinor
structures in accordance with \S\ref{sssec:sufficiency}. Say these
are $Q$ and $Q'$ with spinor maps $u:Q \To P$ and $u':Q' \To P$.

According to Proposition \ref{prop:spinor-structure-equivalence},
these spinor structures will be equivalent if the spinor maps $u$
and $u'$ are equivalent as covering maps. The classification of
covering maps given in \S\ref{sec:covering-spaces} states that two
such covering maps are equivalent if and only if the groups $K$
and $L$ are conjugate.

Thus suppose $K$ and $L$ are conjugate, so there is an $x \in
\pi_1(P)$ so that $L = x K x^{-1}$. Now $\pi_1(P)$ can be written
as the product $K \times i_*(\pi_1(G))$, so $x = k g$, for some $k
\in K$, and $g$ in the image under $i_*$ of $\pi_1(G)$. Moreover,
$g$ lies in the centre of $\pi_1(P)$, by Lemma
\ref{lem:homotopy-centre}. Thus $L = k g K g^{-1} k^{-1} = k K
k^{-1} = K$. This establishes the desired result.
\end{proof}

At this point we have established that the inequivalent spinor
structures are in one to one correspondence with the subgroups $K$
of $\pi_1(P)$ such that $\pi_1(P) = K \times i_*(\pi_1(G))$ and
$\pi_{P*}:K \To \pi_1(M)$ is an isomorphism. The following lemma
gives a simplification of this classification, once a particular
subgroup has been singled out.

\begin{lem}
\label{lem:counting-subgroups-homomorphisms} Suppose $\pi_1(P)$
can be written $\pi_1(P) = K \times i_*(\pi_1(G))$, where $K$ is
isomorphic to $\pi_1(M)$ via $\pi_{P *}$. Subgroups $L$ of
$\pi_1(P)$ isomorphic to $\pi_1(M)$ via $\pi_{P *}$ such that $L
\times i_*(\pi_1(G)) = \pi_1(P)$ are in one to one correspondence
with homomorphisms $\varphi: \pi_1(M) \To \pi_1(G)$.
\end{lem}
\begin{proof}
Suppose $\varphi: \pi_1(M) \To \pi_1(G)$ is a homomorphism. Define
$L \subgroup \pi_1(P)$ by $L = \setc{k \cat i_*
\varphi(\pi_{P*}k)}{k \in K}$. Now take $l = k \cat i_*
\varphi(\pi_{P*}k) \in L$ and suppose $l \in i_*(\pi_1(G))$ also.
Now because $\pi_1(P) = K \times i_*(\pi_1(G))$ gives a unique
decomposition, $k = e$, and so $l = e$. This establishes that $L$
and $i_*(\pi_1(G))$ have a trivial intersection. Next, for any
$[\alpha] \in \pi_1(P)$, there is some $k \in K$, $h \in \pi_1(G)$
so
\begin{align*}
[\alpha] & = k \cat i_* h \\
         & = k \cat i_* \varphi(\pi_{P*}k) \cat i_*
             \varphi(\pi_{P*}k^{-1}) \cat i_* h \\
         & = l \cat i_* h',
\end{align*}
where $l = k \cat i_* \varphi(\pi_{P*}k) \in L$, and $h' =
\varphi(\pi_{P*}k^{-1}) \cat h \in \pi_1(G)$. Thus the internal
direct product of $L$ and $i_*(\pi_1(G))$ is all of $\pi_1(P)$, as
required.

Conversely, define an isomorphism $\chi: K \To L \subgroup K
\times i_*(\pi_1(G)$ by \[\chi = (\restrict{\pi_{P *}}{L})^{-1}
\compose (\restrict{\pi_{P *}}{K}).\] Then we must have $\chi(k) =
\omega(k) \cat i_* \psi(k)$ for some maps (not necessarily, at
this stage, homomorphisms) $\omega : K \To K$, and $\psi:K \To
\pi_1(G)$. Now $\pi_{P *} \chi(k) = \pi_{P*} \omega(k)$, so
$\pi_{P *}k = \pi_{P *} \omega(k)$, and since $\pi_{P *}$
restricted to $K$ is an isomorphism, $\omega(k) = k$ for all $k
\in K$. Using this simplification, we write $\chi(k_1 \cat k_2)$
in two ways.
\begin{align*}
k_1 \cat k_2 \cat i_* \psi(k_1 \cat k_2) & = \chi(k_1 \cat k_2) \\
    & = k_1 \cat i_* \psi(k_1) \cat k_2 \cat i_* \psi(k_2) \\
    & = k_1 \cat k_2 \cat i_* \psi(k_1) \cat i_* \psi(k_2) && \text{by Lemma \ref{lem:homotopy-centre}.}
\end{align*}
Thus by the uniqueness of the $\pi_1(P) = K \times i_*(\pi_1(G))$
decomposition and the injectivity of $i_*$, we conclude that
$\psi$ is a homomorphism. Finally, define $\varphi:\pi_1(M) \To
\pi_1(G)$ by \[\varphi = \psi \compose (\restrict{\pi_{P
*}}{K})^{-1},\] and note that now $\chi(k) = k \cat i_*
\varphi(\pi_{P *} k)$, and so $\varphi$ is exactly the required
homomorphism, relating $K$ and $L$ as in the first part of the
proof.
\end{proof}

The above discussion completes the proof of the Classification
Theorem.

\subsection{Comparison with results in the literature}

Spinor structures are described in the literature for $SO(n)$ or
$SO_0(1,n-1)$ structure groups. The results of
\S\ref{sec:fundamental-groups-SO} show that for $n\geq 3$, the
fundamental groups of $SO(n)$ and $SO_0(1,n-1)$ are isomorphic to
$\Integer_2$. This implies that the simply connected covering
groups are two fold covers. Most results on existence of spinor
structures which have been proved previously are only relevant in
this context, and so do not allow for structure groups
$SO_0(p,q)$, with both $p$ and $q$ greater than or equal to $2$,
where the fundamental group is $\Integer_2 \times \Integer_2$.
(See \S\ref{sec:fundamental-groups-SO}.) Moreover these results
only treat spinor structures for reductions of a frame bundle. In
this sense our results above generalise these results.

The usual result stated for the existence of spinor structures is
as follows.
\begin{prop}
Suppose $\pfbundle{G}{P}{\pi}{M}$ is a principal fibre bundle
which is a reduction of the frame bundle $FM$ of $M$. Suppose the
structure group $G$ is connected and has a two fold simply
connected covering group. Then $P$ admits a spinor structure if
and only if the second Stiefel--Whitney class $w_2$ of $M$ is
zero.
\end{prop}

The second Stiefel--Whitney class is defined in \cite[II
\S1]{law:sg} and in \cite[\S1.5]{pen:ss-t1}. It is related to the
tangent bundle of the manifold, restricting the relevance of this
result to the case where $P$ is a reduction of the frame bundle.

An article by J. Milnor \cite{mil:ssm} which gives one of the
earlier definitions of spinor structures (we use a slight
generalisation of this here) also mentions this result for
$G=SO(n)$. In turn, we are referred for the proof to
\cite{bor:cchs2}, which is fairly impenetrable, and in fact only
gives an outline of the result, saying that the detail is ``a
standard argument''. A discussion of this result for $G=SO(4)$ and
$G=\LL$ with $M$ compact is given in \cite[\S 10]{lic:ts-t}. The
result for $G=\LL$ is mentioned in \cite{ger:sss-tgr1} and
\cite[p. 155]{pen:ss-t}, and discussed in \cite[\S
1.5]{pen:ss-t1}.

A sketch proof of this theorem is given in \cite[II \S1]{law:sg},
based on a Serre spectral sequence argument, for $G=SO(n)$. The
condition stated here, in terms of the Stiefel--Whitney class, is
of quite a different nature from that in our Existence Theorem, in
terms of the fundamental group of the principal fibre bundle. We
will not give a separate proof that they equivalent, but at this
juncture point out that the Existence Theorem covers the general
case for any group $G$, whereas the theorem stated here in terms
of the second Stiefel--Whitney class does not have a
straightforward generalisation. The Serre spectral sequence
argument can still be performed\footnote{The argument is very
briefly as follows. (We are generalising the argument in \cite[II
\S1]{law:sg}. Refer there for details of the notation.) Spinor
structures are in one to one correspondence with elements of
$H^1(P, \pi_1(G))$ such that the restriction to a fibre is
nonzero. Associated to the fibration $\pfbundle{G}{P}{\pi}{M}$
there is an exact sequence $0 \To H^1(M, \pi_1(G))
\xrightarrow{\pi^*} H^1(P, \pi_1(G)) \xrightarrow{i^*}
H^1(G,\pi_1(G)) \xrightarrow{w_E} H^2(M, \pi_1(G))$, which we
obtain from the Serre spectral sequence. Thus existence of a
spinor structure is equivalent to $\image i^* \neq \{0\}$, which
is in turn equivalent to $\ker (w_E) \neq \{0\}$. This is the
generalisation of the condition that the second Stiefel--Whitney
class vanishes.}, but the result does not have such a simple
interpretation if $P$ is not a reduction of the frame bundle or
$\pi_1(G) \neq \Integer_2$.

In summary, the result stated above has appeared in several
similar forms widely throughout the literature. Nevertheless, it
seems no thorough proof has been published, whether employing
methods as elementary as appear here, or techniques such as
spectral sequence arguments.

Essentially the same result as we have given, showing existence
depends on the fundamental group of $P$, is mentioned in
\cite{ger:sss-tgr1}\footnote{This article makes a promising
mention of \cite{wal:iat} in regards a proof of this theorem. This
reference turns out to be simply an introductory text explaining
no more than the meaning of the terms of the theorem.} and
\cite[\S 13.2]{wal:gr} in the case that $G = \LL$. In both cases
the result is stated imprecisely, and no proofs are given. It
seems likely that at least in the general situation described
here, the proof has not appeared in the literature.

A number of other existence results for the $SO_0(1,3)$ case were
given in \cite{ger:sss-tgr2}. These results connect quite varied
properties of the manifold with the existence of a spinor
structure. As examples, there is a result depending on the index
of topological $2$ spheres in the manifolds,\footnote{See also
Proposition 1.12 in \S II of \cite{law:sg}.} another result
depending on the algebraic type of the Weyl tensor, and yet
another ensuring that every globally hyperbolic space-time has a
spinor structure. We refer the reader to this article for the
definitions of all these concepts! The author makes a compelling
case that all physically reasonable space-times have a spinor
structure.

Our classification result is a simple generalisation of the result
given in the literature, for the same situation as in the theorem
above.

\begin{prop}
The inequivalent spinor structures are in one to one
correspondence with elements of $H^1(M,\Integer_2)$.
\end{prop}

This result is mentioned in \cite{mil:ssm}, and a brief discussion
given in \cite[II Theorem 1.7]{law:sg}. In the special case that
$M$ is $4$ dimensional and $G=\LL$ there is an incomplete, but
reasonably elementary, proof in \cite{ish:sffds-t}. (We extend the
idea behind this proof, and this proof, in \S
\ref{sec:classifying-as-bundles}.) Another proof appears in
\cite[\S 4]{hug:itt}, and there is a discussion in \cite[\S
13.2]{wal:gr} (with an error, in footnote 11 on p. 369).

The following lemma shows that our result generalises this.
\begin{lem}\label{lem:homomorphisms-cohomology-group}
The homomorphisms $\pi_1(M) \To \pi_1(G)$ correspond naturally to
the elements of the cohomology group $H^1(M,\pi_1(G))$.
\end{lem}
\begin{proof}
For any topological space $M$, the first homology group with
integer coefficients is isomorphic to the commutative factor group
of the fundamental group. That is, if $N$ denotes the commutator
subgroup of $\pi_1(M)$,
\[H_1(M,\Integer) = \pi_1(M)/N.\]
This called the Hurewicz isomorphism and is a standard result from
algebraic topology. See \cite[II \S6]{hus:ht} for the proof. In
particular, since $\pi_1(G)$ is commutative, by Lemma
\ref{lem:pi1G-commutative}, for any homomorphism $\varphi:\pi_1(M)
\To \pi_1(G)$, the commutator subgroup $N$ is contained in the
kernel, and so $\varphi$ descends to a map
$\varphi:H_1(M,\Integer) \To \pi_1(G)$. Clearly any such map
extends to a homomorphism $\pi_1(M) \To \pi_1(G)$.

This has established that the homomorphisms $\pi_1(M) \To
\pi_1(G)$ correspond naturally to the homomorphisms
$H_1(M,\Integer) \To \pi_1(G)$. Finally, because $\pi_1(G)$ is
commutative we can use the Universal Coefficient Theorem \cite[Ch.
5, \S 5]{spa:at}, relating homology and cohomology, to show that
\[H^1(M,\pi_1(G)) \cong \Hom(H_1(M,\Integer),\pi_1(G)) \oplus
\Ext(H_0(M,\Integer),\pi_1(G)).\] Here
$\Hom(H_1(M,\Integer),\pi_1(G))$ is precisely the group of
homomorphisms $H_1(M,\Integer) \To \pi_1(G)$, and we do not define
in detail $\Ext(H_0(M,\Integer),\pi_1(G))$, pointing out that as
$H_0(M,\Integer) = \Integer$ by \cite[Ch. 5, \S 5]{spa:at} it is
always trivial. Putting this together, we see that the spinor
structures are classified by
\[\Hom(H_1(M,\Integer),\pi_1(G)) \cong H^1(M,\pi_1(G)).\qedhere\]
\end{proof}

In the particular case where $\tilde{G}$ is a double cover of $G$,
$\pi_1(G) = \Integer_2$, and the spinor structures correspond to
the elements of $H^1(M,\Integer_2)$.

%\longversion{

\section{Metric independence of spinor
structures}\label{sec:metric-independence} To begin this section
we will restrict our attention to the Lorentz group, and spinor
structures for Lorentz structures. In this context, the result of
this section will be to prove, in a precise sense, that the
existence and classification of spinor structures is in fact
entirely independent of the particular Lorentz structure we began
with! That is, the Existence Theorem and the Classification
Theorem, whose hypotheses are requirements on the topology of the
Lorentz structure, can be reformulated so that they only refer to
the topology of the base manifold. See also \cite[II \S5]{law:sg}
for a related discussion.

To understand this, we need to consider the bundle of oriented
frames $F^+M$ on the base manifold $M$. This is a $GL^+(4,\Real)$
principal fibre bundle. The notation $GL^+(4,\Real)$ indicates the
group of orientation preserving, or, equivalently, positive
determinant, linear automorphisms of $\Real^4$. This group is
connected, and could alternatively be described as the connected
component of the identity in $GL(4,\Real)$. Recall that the
definition of an orthonormal structure above is as an $\LL$
reduction of this frame bundle. The general linear group
$GL^+(4,\Real)$ is not simply connected, and in fact the inclusion
of $SO(4)$ into $GL^+(4,\Real)$ induces an isomorphism $
\pi_1(SO(4)) \To \pi_1(GL^+(4,\Real))$. (See
\S\ref{sec:maximal-compact-subgroups} for the details, and a more
general result.) Thus $\pi_1(GL^+(4,\Real)) = \Integer_2$, and so
$GL^+(4,\Real)$ has a double covering group, which we will denote
by $\GLncover{4}$. We might refer to it as the `metalinear' group
(just as the metaplectic group is a cover of the symplectic
group). This group is not a particularly easy group to work with,
as it is not an algebraic group (that is, it cannot be expressed
as a matrix group). To see this, we can prove that $\GLncover{4}$
has no finite dimensional representations other than those which
descend to representations of $GL^+(4,\Real)$, and so no faithful
finite dimensional representations. See \cite[II \S5]{law:sg} for
details.

In fact the inclusion of $\LL$ into $GL^+(4,\Real)$ also induces
an isomorphism of fundamental groups. We see this by considering
the following commuting diagram of inclusion maps,
\begin{equation*}
\xymatrix{
    SO(3) \ar@{^{(}->}[r] \ar@{^{(}->}[d] & SO(4) \ar@{^{(}->}[d] \\
    \LL   \ar@{^{(}->}[r]        & GL^+(4,\Real)
}
\end{equation*}
and the diagram of induced maps between fundamental groups,
\begin{equation*}
\xymatrix{
    \pi_1(SO(3)) \ar[r] \ar[d] & \pi_1(SO(4)) \ar[d] \\
    \pi_1(\LL)   \ar[r]        & \pi_1(GL^+(4,\Real))
}
\end{equation*}
The inclusions $SO(3) \hookrightarrow SO(4)$, $SO(3)
\hookrightarrow \LL$ and $SO(4) \hookrightarrow GL^+(4,\Real)$
each give an isomorphism of fundamental groups by
\S\ref{ssec:fundamental-group-SO(n)},
\S\ref{ssec:fundamental-group-SO(p,q)} and
\S\ref{sec:maximal-compact-subgroups} respectively. Thus we can
conclude that the induced map $\pi_1(\LL) \To
\pi_1(GL^+(4,\Real))$ must also be an isomorphism.

For the purposes of stating the next results, we will consider a
general case corresponding to this situation. Suppose $G$ is a Lie
group, with covering group $\tilde{G}$, and $H$ is a Lie subgroup
of $G$, such that the inclusion $\iota: H \To G$ induces an
isomorphism of fundamental groups $\iota_* : \pi_1(H) \To
\pi_1(G)$. In particular, one can prove that this is always the
case when the maximal compact subgroup (see
\S\ref{sec:maximal-compact-subgroups}) of $G$ is contained in $H$.
%lemma that if H is compact (does it need to be compact?
%   monodromy might do it), $\tilde{H}$ lifts 'isomorphically'
%   to a subgroup of $\tilde{G}$?

\begin{lem}
\label{lem:exact-sequence} For any principal fibre bundle
$\pfbundle{G}{P}{\pi_P}{M}$, there is an exact sequence, part of
which is
\begin{equation}
\label{eq:exact-sequence}
 \pi_2(M) \xrightarrow{h_*} \pi_1(G) \xrightarrow{i_*}
 \pi_1(P) \xrightarrow{\pi_{P *}} \pi_1(M) \rightarrow 0
\end{equation}
Here $i_* : \pi_1(G) \To \pi_1(P)$ is the map induced from the
action of $G$ on a fibre, as above.
\end{lem}
\begin{proof}
The maps $i_*$ and $\pi_{P *}$ have been considered previously. It
is obvious that the sequence is exact at $\pi_1(P)$. Exactness at
$\pi_1(M)$ states simply that $\pi_{P *}$ is onto. This is clear,
since any path in $M$ can be lifted arbitrarily to give a path in
$P$, and the lift of a loop in $M$ can be extended within the
initial fibre to form a closed loop. This loop then maps down via
$\pi_P$ to give the original loop in $M$.

Next we turn to the map $h_*$. The construction of this map in a
similar context is mentioned in \cite{cla:mchccimtr}. There are
theorems proved in a very general setting giving exact sequences
for homotopy groups of spaces with fibrations \cite{hus:ht},
\cite{ste:tfb}. To use such a theorem here we would have to
introduce relative homotopy groups \cite{ste:tfb}, which would
take us rather far afield. However, in this particular situation,
where we are content to assume that our spaces are smooth and
paracompact, we can give a simple and geometric argument.
Interestingly, the proof here will introduce a connection, but as
it will turn out this particular choice will not affect the final
construction. Providing our own argument here rather than the
general one mentioned above simplifies the proof of Lemma
\ref{lem:commuting-diagram} below.

We first give some notation for parallel transportation. For this
purpose we will fix a particular connection on the principal fibre
bundle. Given a path $\alpha$ in $M$, with initial point $m_0$, we
can parallel transport $\bdelt{p_0}$ along $\alpha$, to obtain a
point in the bundle in the fibre of $\alpha(1)$. Denote this point
by $j(\alpha)$, so that $j$ becomes a map $j:\Pi M \To P$.

Parallel transportation along a loop in $M$ is of interest because
it returns $\bdelt{p_0}$ to the initial fibre. Thus $j$ restricted
to loops becomes a map $\Omega M \To \pi^{-1}(m_0)$. For any
$\bdelt{p} \in \pi^{-1}(m_0)$, there is a unique $g \in G$ such
that $\bdelt{p} g = \bdelt{p_0}$. This $g$ is value of the
translation function $\tau(\bdelt{p},\bdelt{p_0})$. Define a new
function $h$ which, given a loop in $M$, produces this $g$. Thus
$h:\Omega M \To G$. Moreover, $h$ acting on the constant loop
gives the identity element of $G$, and so is base point
preserving. In fact, $h$ is actually a homomorphism, because of
the reparametrisation properties of parallel transport, but we
shall not need this fact. More importantly, $h$ is continuous. Not
having specified the topology for $\Omega M$, we cannot make this
precise, but it is clear that Proposition
\ref{prop:parallel-transport} ensures that $h$ is relatively well
behaved.

Since $h$ is a base point preserving map, it induces a map of the
homotopy classes, $h_* : \pi_1(\Omega M) \To \pi_1(G)$. The
fundamental group of the loop space of $M$ is just the second
homotopy group of $M$, $\pi_2(M)$, and so this $h_*$ is of the
form indicated in the statement of this Lemma. It is not too hard
to prove that $h_*$ is in fact independent of the particular
choice of connection in the definition of $h$. However the
argument is lengthy and unnecessary here.

%we have relegated it to \S\ref{ssec:h-technical-lemmas} as Lemma
%\ref{lem:h-connection-independent}. The independence will prove
%important, however, in our discussion of Lemma
%\ref{lem:commuting-diagram}, where we chose a particular pair of
%related connections.

The remaining part of the series is
\[
\pi_2(M) \xrightarrow{h_*} \pi_1(G) \xrightarrow{i_*}
 \pi_1(P).
\]
Thus we want to prove that $\image h_* = \ker i_*$.

Suppose $[g] \in \pi_1(G)$ is in $\image h_*$, so $[g] = h_*
[\alpha]$ for some $\alpha \in \Omega \Omega M$. Thus for each $t
\in \I$, $g(t) = h(\alpha_t)$. Define $\beta : \I \To P$ by
\[\beta(t) = i(h(\alpha_t)) = \bdelt{p_0} h(\alpha_t),\] so $i_*
[g] = [\beta]$. We now want to prove that $i_*[g] = [e]$, that is,
that $\beta$ is homotopic to the constant map in $P$.

For each $s \in \I$, define $\alpha_{t,s} \in \Pi M$ as the path
$\alpha_t$ traversing only the interval $[0,s]$. Thus
$\alpha_{t,s}(r) = \alpha_t(r s)$ and in particular
$\alpha_{t,s}(0)=m_0$, and $\alpha_{t,s}(1) = \alpha_t(s)$. We now
define a homotopy $H :\I \times \I \To P$ according to \[H(s,t) =
j(\alpha_{t,s}).\] A calculation shows that this is a homotopy
from $\beta$ to the constant path at $\bdelt{p_0}$.
\begin{eqnarray*}
H(0,t)  =  j(\alpha_{t,0}) &=& \bdelt{p_0} \\
H(1,t)  =  j(\alpha_{t,1}) &=& j(\alpha_t) = \bdelt{p_0}
h(\alpha_t) = \beta(t).
\end{eqnarray*}
Also, $H$ is a homotopy fixing endpoints, that is,
$H(s,0)=H(s,1)=\bdelt{p_0}$. This follows from the fact that
$\alpha_0$ and $\alpha_1$ are both the constant path in $P$. Thus
$[\beta] = [e]$, and so $\image h_* \subset \ker i_*$.

Next we want to prove that $\ker i_* \subset \image h_*$, and so
we suppose that $g \in \Omega G$, and $i_*[g] = [e]$. Now, $i_*[g]
= [i \compose g]$, and $i(g(t)) = \bdelt{p_0} g(t)$. Therefore,
from the hypothesis there must exist some homotopy $H : \I \times
\I \To P$ so that
\begin{eqnarray*}
H(1,t) & = & \bdelt{p_0} g(t) \\
H(0,t) & = & \bdelt{p_0}  \\
H(s,0) & = & \bdelt{p_0}  \\
H(s,1) & = & \bdelt{p_0}.
\end{eqnarray*}
We will next use this homotopy to define an element $\alpha$ of
$\Omega \Omega M$ so that $h_* [\alpha] = [g]$. Let $\alpha_t(s) =
\pi(H(s,t))$. Thus, for each $t \in \I$, $\alpha_t$ is a path in
$M$. Again, define $\alpha_{t,s} \in \Pi M$ as the path $\alpha_t$
restricted to the interval $[0,s]$. We can use the connection to
perform parallel transportations along these paths, resulting in a
map $\I \times \I \To P$, given by $(t,s) \mapsto
j(\alpha_{t,s})$. This is a continuous function, by Proposition
\ref{prop:parallel-transport}. However, there is no reason for
$j(\alpha_{t,s})$ to be equal to $H(s,t)$. On the other hand, it
must be in the same fibre as $H(s,t)$, since the parallel
transport projects down to the original curve. Thus for each $t,s
\in \I$, there is some $k(t,s) \in G$ so $j(\alpha_{t,s}) = H(s,t)
k(t,s)$. Since $j(\alpha_{t,s})$ and $H(s,t)$ are continuous, $k$
is a continuous function also. Now, $j(\alpha_t) = j(\alpha_{t,1})
= H(1,t) k(t, 1) = \bdelt{p_0} g(t) k(t,1)$. Thus $k$ in fact
defines a homotopy between $\bdelt{p_0} g(t)$ and $j(\alpha_t)$.
Moreover, this homotopy stays within the fibre of $\bdelt{p_0}$,
and so gives a homotopy of the loop $h(\alpha_t)$ and $g(t)$. This
proves that $h_*[\alpha] = [g]$, and so $h_*$ maps onto the kernel
of $i_*$, completing the result.
%-% this needs a picture!
\end{proof}

Now, suppose we have a reduction of the bundle $P$ to a $H$
principal fibre bundle $\pfbundle{H}{R}{\pi_R}{M}$.  Thus there is
a map $\kappa : R \To P$ such that $\kappa(\bdelt{r} h) =
\kappa(\bdelt{r}) h$ for all $\bdelt{r} \in R$ and $h \in H$. As
usual, from $\kappa$ we obtain a map $\kappa_* : \pi_1(R) \To
\pi_1(P)$. Denote the base points as $\bdelt{r_0} \in R$ and
$\kappa(\bdelt{r_0}) = \bdelt{p_0} \in P$. Suppose also, as above,
that the inclusion $\iota:H \To G$ induces an isomorphism of
fundamental groups of the structure groups. We can write two exact
sequences as in Equation \eqref{eq:exact-sequence}, and link them
together with the maps $\iota_*$ and $\kappa_*$. To avoid
confusion we will define two maps
\begin{eqnarray*}
    i_P : G & \To & P \\
          g & \mapsto & \bdelt{p_0} g \\
    i_R : H & \To & R \\
          h & \mapsto & \bdelt{r_0} h,
\end{eqnarray*}
and the induced maps $i_{P *} :\pi_1(G) \To \pi_1(P)$ and $i_{R *}
:\pi_1(H) \To \pi_1(R)$. Moreover, to define the maps $h_{H *}$
and $h_{G *}$ we use a related pair of connections. Choose first
an arbitrary connection on $R$. As it turns out, there is a unique
extension of this connection to a connection of $P$. This
(unexciting) argument is given in
\S\ref{ssec:extending-connections}. Use these connections to
define $h_H:\Omega M \To \pi_1(H)$ and $h_G:\Omega M \To
\pi_1(G)$, and thence $h_{H*}$ and $h_{G*}$. Collecting all these
maps, we obtain the following diagram.

\begin{figure}[!ht]
\begin{equation*}
 \xymatrix{
    \pi_2(M) \ar[r]^{h_{H *}} \ar[d] & \pi_1(H) \ar[r]^{i_{R *}} \ar[d]^{\iota_*} & \pi_1(R) \ar[r]^{\pi_{R *}} \ar[d]^{\kappa_*} & \pi_1(M) \ar[r] \ar[d] & 0 \ar[d] \\
    \pi_2(M) \ar[r]^{h_{G *}}        & \pi_1(G) \ar[r]^{i_{P *}}                  & \pi_1(P) \ar[r]^{\pi_{P *}}                   & \pi_1(M) \ar[r]        & 0
}
\end{equation*}
\caption{\label{fig:commuting-diagram}}
\end{figure}

Each of the three unlabelled vertical maps in Figure
\ref{fig:commuting-diagram} is simply the identity map.

\begin{lem}
\label{lem:commuting-diagram} The diagram in Figure
\ref{fig:commuting-diagram} commutes, and the map \[\kappa_* :
\pi_1(R) \To \pi_1(P)\] is an isomorphism.
\end{lem}
\begin{proof}
To prove that a diagram of this form commutes, we need only check
that each square commutes.

To prove the first square commutes we use the result of
\S\ref{ssec:extending-connections} that the two parallel
transports are the same, and so $\iota(h_H(\alpha)) = h_G(\alpha)$
for every $\alpha \in \Omega M$. Thus $\iota_* \compose h_{H *} =
h_{G *}$.

The second and third squares commute, using the identities
\begin{eqnarray*}
\kappa \compose i_{R}(h) & = & \kappa(\bdelt{r_0} h) \\
                      & = & \kappa(\bdelt{r_0}) h \\
                      & = & i_{P} \compose \iota (h)
\end{eqnarray*}
and
\[\pi_P \compose \kappa = \pi_R.\]
The fourth square commutes trivially.

Now that we have established that the diagram commutes, we can
apply a powerful technique from homological algebra, the five
lemma. (For a proof, see \cite[Ch. I, \S4]{eil:fat}.) The five
lemma states that if we have two exact sequences, linked by four
isomorphisms in a commuting diagram as above, then the central
vertical map is also an isomorphism. Thus $\kappa_*$ is an
isomorphism.
\end{proof}

Using the above argument in this context was suggested by
\cite{cla:mchccimtr}, but a proof has not previously appeared.

We can now begin reformulating the existence and classification
results for spinor structures for the $H$ bundle in terms of the
topology of the $G$ bundle. Since $\kappa_*$ is an isomorphism
$\pi_1(R) \To \pi_1(P)$, $\pi_1(R)$ can be written in product form
$\pi_1(M) \times \pi_1(H)$ if and only if $\pi_1(P)$ can be
written in the form $\pi_1(M) \times \pi_1(G)$. However, this
itself is not sufficient to prove that $R$ has a spinor structure
if and only if $P$ has a spinor structure. For this we need to
consider a subdiagram of the commuting diagram in Figure
\ref{fig:commuting-diagram}, namely
\begin{equation*}
\xymatrix{
    \pi_1(H) \ar[r]^{i_{R *}} \ar[d]^{\iota_*} & \pi_1(R) \ar[r]^{\pi_{R *}} \ar[d]^{\kappa_*} & \pi_1(M) \ar[d] \\
    \pi_1(G) \ar[r]^{i_{P *}}                  & \pi_1(P) \ar[r]^{\pi_{P *}}                   & \pi_1(M)
}
\end{equation*}
From this we easily obtain the following result.

\begin{prop}\label{prop:spinor-structure-reductions}
If $K \subgroup \pi_1(R)$ is such that
\begin{enumerate}
\item $\restrict{\pi_{R*}}{K}$ is an isomorphism,
\item $K \cap i_{R*}(\pi_1(H)) = \{e\}$, and
\item $\pi_1(R) = K \times i_{R*}(\pi_1(H))$
\end{enumerate}
then $\kappa_* (K) \subgroup \pi_1(P)$ is such that
\begin{enumerate}
\item $\restrict{\pi_{P*}}{\kappa_* (K)}$ is an isomorphism,
\item $\kappa_* (K) \cap i_{P*}(\pi_1(G)) = \{e\}$, and
\item $\pi_1(P) = \kappa_* (K) \times i_{P*}(\pi_1(G))$
\end{enumerate}
and conversely. Thus, by the Existence Theorem,  $R$ has a spinor
structure if and only if $P$ has a spinor structure.
\end{prop}

We will next continue this line of analysis, showing that the
classification of spinor structures is similarly unaffected by
such a reduction of the structure group. Again, suppose $P$ is a
$G$ bundle, $R$ an $H$ bundle which is a reduction of $P$ via the
map $\kappa:R \To P$, and the inclusion $\iota:H \To G$ induces an
isomorphism of the fundamental groups.

\begin{prop}
There is a one to one correspondence between spinor structures for
$R$ and for $P$. If $S$ is a spinor structure for $R$, and $Q$ is
the corresponding spinor structure for $P$, then $S$ is a
reduction of $Q$.
\end{prop}
\begin{proof}
We have seen in the proof of the Classification Theorem that the
spinor structures are in one to one correspondence with subgroups
of the fundamental group of the bundle with satisfy the hypotheses
of the Existence Theorem. Proposition
\ref{prop:spinor-structure-reductions} shows that such subgroups
for $\pi_1(R)$ and $\pi_1(P)$ are in one to one correspondence via
$\kappa_*$. Thus, fix $K \subgroup \pi_1(R)$ and $\kappa_* (K)
\subgroup \pi_1(P)$, and form the associated spinor structures $S$
and $Q$.

To construct the reduction map $\tilde{\kappa} : S \To Q$, recall
that in the construction of the spinor structures, each point of
$S$ is an equivalence class of paths in $R$, and each point of $Q$
is an equivalence class of paths in $Q$. Thus the typical point of
$S$ is $\sh{\alpha}$ where $\alpha : \I \To R$, $\alpha(0) =
\bdelt{r_0}$, and $\sh{\alpha} = \sh{\beta}$ if and only if
$\alpha(1) = \beta(1)$ and $[\alpha^{-1} \cat \beta] \in K$.
Similarly, the typical point of $Q$ is $\na{\gamma}$, where
$\gamma : \I \To P$, $\gamma(0) = \bdelt{p_0}$, and $\na{\gamma} =
\na{\delta}$ if and only if $\gamma(1) = \delta(1)$ and
$[\gamma^{-1} \cat \delta] \in \kappa_*(K)$. Define
$\tilde{\kappa}$ in the natural way, as
\[\tilde{\kappa}(\sh{\alpha}) = \na{(\kappa\compose\alpha)}.\]
This is well defined, since if $\sh{\alpha} = \sh{\beta}$,
$[(\kappa\compose\alpha)^{-1} \cat (\kappa \compose \beta)] =
\kappa_* [\alpha^{-1} \cat \beta] \in \kappa_*(K)$.

Further, it is a reduction map. If $\tilde{h} \in \tilde{H}$, then
$\tilde{h}$ is an equivalence class of homotopic paths in $H$. Say
$h: \I \To H$ is a representative, so $\tilde{h} = [h]$. Define
$\tilde{\iota} : \tilde{H} \To \tilde{G}$ by
$\tilde{\iota}(\tilde{h}) = [\iota(h)]$. This is well defined,
since if $[h] = [h']$, then $[\iota(h)] = [\iota(h')]$, and the
elements of $\tilde{G}$ are homotopy classes of paths in $G$. Now,
\begin{align*}
\tilde{\kappa}(\sh{\alpha} \tilde{h})
    & = \tilde{\kappa}(\sh{\alpha h}) \\
    & = \na{(\kappa \compose (\alpha h))} \\
    & = \na{((\kappa \compose \alpha) h)} \\
    & = \tilde{\kappa}(\sh{\alpha}) \tilde{h},
\end{align*} as required.
\end{proof}

With these two results in hand we have a complete description of
the spinor structures of a reduced bundle, as long as the reduced
structure group is `large enough', in the sense that the inclusion
map induces an isomorphism of the fundamental groups. In this
case, we see that there is essentially no interplay between the
reduction and the process of forming a spinor structure.

In the specific case of a reduction of the frame bundle, we have
seen that the inclusion of $\LL$ into $GL^+(4,\Real)$ induces an
isomorphism of the fundamental groups, and so we have the
following.
\begin{cor}
A spinor structure exists for the Lorentz structure if and only if
the oriented frame bundle $F^+M$ has a $\GLncover{4}$ spinor
structure. In this case, every such spinor structure is a
reduction of a spinor structure for $F^+ M$.
\end{cor}
An important implication of this result is that the existence of a
spinor structure for a Lorentz structure is determined solely by
the orientation and topology of the base manifold. This is because
the oriented frame bundle is defined without reference to the
metric. We can easily extend this corollary to the case $G=SO(n)$
or $G=SO_0(1,n)$ for any $n \geq 3$ via the results on \S
\ref{sec:fundamental-groups-SO}.

In the light of this result, one might wonder why spinor
structures for $\LL$ reductions of the frame bundle are
interesting, given that they are all reductions of a
$\GLncover{4}$ spinor structure. This is because while
$\GLncover{4}$ is not algebraic, and has no finite dimensional
representations which do not descend to representations of
$GL^+(4,\Real)$, the group $\SL$ \emph{does} have additional
finite dimensional representations relative to $\LL$, as we shall
see. Thus only once we have made a particular choice of reduction
can we use this representation theory to construct the `spinor
algebra', as in \S\ref{sec:spinor-algebra}, which is used to give
a new formulation of the Dirac equation in
\S\ref{sec:dirac-equation}.

%}

\section{Lifting a connection to the spinor structure}
\label{sec:lifting-connection} We now prove that a connection
$\omega$ on a bundle \mbox{$\pfbundle{G}{P}{\quad}{M}$} can always
be lifted to a connection on a spinor structure
\mbox{$\pfbundle{\tilde{G}}{Q}{\quad}{M}$}.  It might seem
unlikely that this could be possible -- after all, $\omega$ takes
values in the Lie algebra of $G$, whereas a connection on $Q$ must
take values in the Lie algebra of $\tilde{G}$. However, the
covering map $\rho:\tilde{G} \To G$ provides an isomorphism of
these Lie algebras, since it is locally a diffeomorphism, by its
derivative at the identity of $\tilde{G}$, denoted $\rho_{*
e}:\LA{\tilde{G}} \To \LA{G}$.

We define the connection on the spinor bundle by $\hat{\omega} =
(\rho_{* e})^{-1} u^* \omega$. That is, we simply pull-back the
connection form via the spinor map, and identify the Lie algebras.
We next prove a proposition to the effect that this defines a
valid connection form on the spinor bundle. In fact, the following
proposition gives a stronger result. If we consider arbitrary Lie
algebra valued forms on $P$, then this construction only results
in a valid connection form if the form on $P$ is actually a
connection form.

\begin{prop}\label{prop:lifting-connection}
Suppose $\omega$ is a $\LA{G}$ valued $1$-form on $P$. Define
$\hat{\omega} = (\rho_{* e})^{-1} u^* \omega$. Then $\hat{\omega}$
is a connection on $Q$ if and only if $\omega$ is a connection on
$P$.
\end{prop}
\begin{rem}
One half of this proposition, that if $\omega$ is a connection
then $\hat{\omega}$, as defined, is a connection, is essentially
equivalent to Proposition 6.1 in \S 6 of Chapter II in
\cite{kob:fdg1}. The other half will be used to prove Proposition
\ref{prop:push-down-spinor-connection}.
\end{rem}
\begin{proof}
The proof is relatively straightforward, although requiring
several technical calculations. It will be useful to define a
partial inverse function to $\rho$ for this proof. Since $\rho$ is
a covering map, there is a neighbourhood of the identity in
$\tilde{G}$, say $U$, so that $\restrict{\rho}{U}$ is one to one.
We will abbreviate $\left(\restrict{\rho}{U}\right)^{-1}$ to
simply $\rho^{-1}$. Notice $(\rho_{* e})^{-1} = (\rho^{-1})_{*
e}$.

Firstly we need to check that vertical vectors are mapped
appropriately into the Lie algebra. Firstly define functions
$\psi_\bdelt{p} : P_{\bdelt{p}} \To G$ and
$\hat{\psi}_{\bdelt{q}}:Q_{\bdelt{q}} \To \tilde{G}$ by
\[\psi_{\bdelt{p}}(\bdelt{p'})=\tau(\bdelt{p},\bdelt{p'}) \textrm{ and }
\hat{\psi}_{\bdelt{q}}(\bdelt{q'})=\tau(\bdelt{q},\bdelt{q'}).\]
Then, in accordance with Definition \ref{defn:connection-form},
the condition on vertical vectors is that
\begin{eqnarray*}
\omega_{\bdelt{p}}(x) & = & \psi_{\bdelt{p} *}x \quad \forall x
\in T P \textrm{ such that } \pi_* x = 0, \textrm{ and} \\
\hat{\omega}_{\bdelt{q}}(y) & = & \hat{\psi}_{\bdelt{q} *}y \quad
\forall y \in T Q \textrm{ such that } \hat{\pi}_* y = 0.
\end{eqnarray*}
We will show that these conditions are equivalent.

We easily see that for $y \in T Q$, $\hat{\pi}_* y = 0$ if and
only if $\pi_* u_* y = 0$, since $\pi \compose u = \hat{\pi}$.
Moreover, every $x \in T P$ such that $\pi_* x = 0$ is of the form
$x = u_* y$, with $y \in T Q$ such that $\hat{\pi}_* y = 0$. (That
is, $u_*$ maps $V_\bdelt{q}$ onto $V_{u(\bdelt{q})}$.)
% if we're being careful a proof here would be appropriate.

Next since $\bdelt{q} \tau(\bdelt{q}, \bdelt{q'}) = \bdelt{q'}$,
we can apply $u$ to both sides and use the fact that $u$ is a
principal bundle morphism to obtain $u(\bdelt{q})
\rho(\tau(\bdelt{q}, \bdelt{q'})) = u(\bdelt{q'})$, and so
$\rho(\tau(\bdelt{q}, \bdelt{q'})) = \tau(u(\bdelt{q}),
u(\bdelt{q'}))$. When $\tau(\bdelt{q}, \bdelt{q})) \in U$, we have
$\tau(\bdelt{q}, \bdelt{q'}) = \rho^{-1} (\tau(u(\bdelt{q}),
u(\bdelt{q'})))$. Thus $(\rho^{-1} \compose \psi_{u(\bdelt{q})}
\compose u)(\bdelt{q'}) = \tau(\bdelt{q}, \bdelt{q'}) =
\hat{\psi}_{\bdelt{q}}(\bdelt{q'})$, and
\begin{equation}\label{eq:psi-equation1}
\rho_{* e}^{-1} \psi_{u(\bdelt{q}) *} u_* =
\hat{\psi_{\bdelt{q}}}{}_*.
\end{equation}
Equivalently,
\begin{equation}\label{eq:psi-equation2}
\psi_{u(\bdelt{q}) *} u_* = \rho_{* e} \hat{\psi_{\bdelt{q}}}{}_*.
\end{equation}

Now suppose that $\omega_{\bdelt{p}}(x) = \psi_{\bdelt{p} *}x$ for
all vertical vectors $x \in T P$. Then
\begin{eqnarray*}
\hat{\omega}_{\bdelt{q}}(y) & = & \rho_{* e}^{-1}
\omega_{u(\bdelt{q})} (u_* y) \\
    & = & \rho_{* e}^{-1} \psi_{u(\bdelt{q}) *} u_* y \\
    & = & \hat{\psi}_{\bdelt{q}} y,
\end{eqnarray*}
applying Equation \eqref{eq:psi-equation1}. This holds for every
vertical vector $y \in T Q$. Conversely, suppose that
$\hat{\omega}_{\bdelt{q}}(y) = \hat{\psi}_{\bdelt{q} *} y$ for all
vertical vectors $y \in T Q$. Then
\begin{eqnarray*}
\rho_{* e} \hat{\psi}_{\bdelt{q} *} y & = &
        \rho_{* e} \hat{\omega}_{\bdelt{q}}(y) \\
    & = & (u^* \omega)_{\bdelt{q}}(y) \\
    & = &  \omega_{u(\bdelt{q})}(u_* y).
\end{eqnarray*}
Applying Equation \eqref{eq:psi-equation2}, we obtain
\begin{align*}
\psi_{u(\bdelt{q}) *} u_* y & = \omega_{u(\bdelt{q})}(u_* y) &&
\forall y \in T Q \textrm{ such that } \hat{\pi}_* y = 0, \\
\intertext{and so} \psi_{u(\bdelt{q}) *} x &=
\omega_{u(\bdelt{q})}(x) && \forall x \in T P \textrm{ such that }
\pi_* x = 0.
\end{align*}
This completes this section of the proof.

Secondly, to confirm that the `elevator properties',
\begin{eqnarray*}
\tilde{g}^* \hat{\omega} & = & \Ad_{\tilde{G}}(\tilde{g}^{-1})
\hat{\omega} \qquad \forall \tilde{g}
\in \tilde{G}, \textrm{ and} \\
g^* \omega & = & \Ad_G(g^{-1}) \omega \qquad \forall g \in G
\end{eqnarray*}
are equivalent, we need to prove the following simple commutation
relations.
\begin{subequations}
\begin{eqnarray}
\label{eq:cr-1}
u^* \Ad_G(g^{-1}) & = & \Ad_G(g^{-1}) u^* \\
\label{eq:cr-2}
g^* \rho_{* e}^{-1} & = & \rho_{* e}^{-1} g^* \\
\label{eq:cr-3}
\tilde{g}^* u^* & = & u^* \rho(\tilde{g})^* \\
\label{eq:cr-4} \Ad_{\tilde{G}}(\tilde{g}^{-1}) \rho_{* e}^{-1} &
= & \rho_{* e}^{-1} \Ad_G(\rho(\tilde{g})^{-1}).
\end{eqnarray}
\end{subequations}
Equations \eqref{eq:cr-1} and \eqref{eq:cr-2} are obvious, because
the adjoint map and $\rho_{* e}^{-1}$ act on values in a Lie
algebra. Similarly Equation \eqref{eq:cr-3} follows from the fact
that $u$ is a principal fibre bundle morphism. Finally, to
establish Equation \eqref{eq:cr-4}, we calculate, for $g' \in
\rho(U)$,
\begin{align*}
    I_{\rho(\tilde{g}^{-1})}(g') & = \rho(\tilde{g}^{-1}) g' \rho(\tilde{g}) \\
        & = \rho(\tilde{g}^{-1}) \rho(\rho^{-1}(g')) \rho(\tilde{g}) \\
        & = (\rho \compose I_{\tilde{g}^{-1}} \compose \rho^{-1})(g').
\end{align*}
Thus for $\tilde{g} \in \tilde{G}$, using $(\rho_{* e})^{-1} =
(\rho^{-1})_{* e}$, we have
\begin{align*}
\Ad_G(\rho(\tilde{g}^{-1})) & = (I_{\rho(\tilde{g}^{-1})})_{*e} \\
    & = (\rho \compose I_{\tilde{g}^{-1}} \compose \rho^{-1})_{*e} \\
    & = \rho_{*e} \Ad_{\tilde{G}}(\tilde{g}^{-1})\rho_{*e}^{-1}.
\end{align*}
We finish the proof as follows. Suppose firstly that the elevator
property holds for $\omega$. Then for every $\tilde{g} \in
\tilde{G}$,
\begin{align*}
\tilde{g}^* \hat{\omega} & = \tilde{g}^* \rho_{* e}^{-1} u^* \omega                 && \\
                 & = \rho_{* e}^{-1} \tilde{g}^* u^* \omega                 && \text{by \eqref{eq:cr-2}} \\
                 & = \rho_{* e}^{-1} u^* (\rho(\tilde{g}))^* \omega         && \text{by \eqref{eq:cr-3}}\\
                 & = \rho_{* e}^{-1} u^* \Ad_G(\rho(\tilde{g}^{-1})) \omega   && \text{by the elevator property for $\omega$}\\
                 & = \rho_{* e}^{-1} \Ad_G(\rho(\tilde{g}^{-1})) u^* \omega   && \text{by \eqref{eq:cr-1}}\\
                 & = \Ad_{\tilde{G}}(\tilde{g}^{-1}) \rho_{* e}^{-1} u^* \omega         && \text{by \eqref{eq:cr-4}}\\
                 & = \Ad_{\tilde{G}}(\tilde{g}^{-1}) \hat{\omega}.                     &&
\end{align*}
Thus the elevator property holds for $\hat{\omega}$. Conversely,
suppose the elevator property holds for $\hat{\omega}$. Then for
every $\tilde{g} \in \tilde{G}$,
\begin{align*}
\Ad_{\tilde{G}}(\tilde{g}^{-1}) \hat{\omega} & = \tilde{g}^* \hat{\omega} \\
                         & = \tilde{g}^* \rho_{* e}^{-1} u^* \omega \\
                         & = \rho_{* e}^{-1} \tilde{g}^* u^* \omega && \text{by \eqref{eq:cr-2}} \\
                         & = \rho_{* e}^{-1} u^* \rho(\tilde{g})^* \omega.
                          && \text{by \eqref{eq:cr-3}}
\end{align*}
Next, expressing the $\hat{\omega}$ on the left hand side in terms
of $\omega$, and applying $\rho_{* e}$ to both sides, we find
\begin{align*}
\rho_{* e} \Ad_{\tilde{G}}(\tilde{g}^{-1}) \rho_{* e}^{-1} u^*
\omega
    & = u^* \rho(\tilde{g})^* \omega \\
 \Ad_G(\rho(\tilde{g})^{-1}) u^* \omega & = u^* \rho(\tilde{g})^* \omega && \text{by \eqref{eq:cr-4}} \\
 u^* \Ad_G(\rho(\tilde{g})^{-1}) \omega & = u^* \rho(\tilde{g})^* \omega && \text{by \eqref{eq:cr-1}}
\end{align*}
Thus for every $y \in T Q$,
\[\Ad_G(\rho(\tilde{g})^{-1}) \omega(u_* y) = \rho(\tilde{g})^* \omega(u_* y).\]
Now every $g \in G$ can be written as $g=\rho(\tilde{g})$ for some
$\tilde{g} \in \tilde{G}$, and every $x \in T P$ can be written as
$x = u_* y$ for some $y \in T Q$, and so we reach our desired
result
\[\Ad_G({g}^{-1}) \omega = {g}^* \omega \qquad \forall g \in G.\]
This completes the proof of the proposition.
\end{proof}

In particular, when we consider the case of a spinor structure for
an orthonormal frame bundle, this proposition can be used to pick
out a special connection on the spinor bundle. In particular the
Levi--Civita connection, the unique torsion free connection on the
orthonormal bundle, can be lifted by this procedure. In the
$(1+3)$ dimensional Lorentzian case, the spinor connection is thus
an $\LAM{sl}{2,\Complex}$ valued $1$-form on the spinor bundle.

Conversely, we can prove that every spinor connection is obtained
in precisely this way. This appears to be a new result. Having
already established the relevant necessary and sufficient
condition for a form being a connection form in Proposition
\ref{prop:lifting-connection}, the proof is not too difficult. The
idea is to use the fact that the spinor map is a covering map, and
hence locally invertible, and push the spinor connection form down
from the $\tilde{G}$ bundle to the $G$ bundle. There is a
potential obstacle in that the spinor map is many to one, and so
does not necessarily give a well defined form on the $G$ bundle.
This is overcome by means of the following proposition.

\begin{prop}\label{prop:kernel-in-centre}
Let $\rho : \tilde{G} \To G$ be the covering homomorphism from a
simply connected Lie group $\tilde{G}$ to a Lie group $G$. Then
the subgroup $\rho^{-1}(e)$ is contained in the centre of
$\tilde{G}$.
\end{prop}
\begin{proof}
See \cite[\S16.30.2.1]{die:ta3}.
\end{proof}

\begin{cor}
The subgroup $\rho^{-1}(e)$ is contained in the kernel of the
adjoint representation of $\tilde{G}$.
\end{cor}

\begin{prop}\label{prop:push-down-spinor-connection}
Any connection $\hat{\omega}$ on the spinor bundle $Q$ defines a
unique form $\omega$ on $P$. The connections are related by
$\hat{\omega} = (\rho_{* e})^{-1} u^* \omega$ as in the
construction of Proposition \ref{prop:lifting-connection}, and so
$\omega$ is a connection on $P$.
\end{prop}
\begin{proof}
We propose to define $\omega$ by $\omega = \rho_{* e} (u^{-1})^*
\hat{\omega}$. The problem with this is that $u^{-1}$ is not
uniquely defined, so we must check that regardless of which
inverse we use the same answer is reached. For this purpose, say
$\bdelt{p} \in P$, and $\bdelt{q} \in Q$ is an inverse image, so
$u(\bdelt{q}) = \bdelt{p}$. Now all the inverse images are of the
form $\bdelt{q} \tilde{g}$, where $\tilde{g} \in \ker(\rho)$.
Suppose further that $v \in T_\bdelt{p} P$, and the inverse of $u$
taking $\bdelt{p}$ to $\bdelt{q}$ takes $v$ to $y \in T_\bdelt{q}
Q$. Then the inverse taking $\bdelt{p}$ to $\bdelt{q} \tilde{g}$
acts as $u^{-1}_* v = \tilde{g}_* v \in T_{\bdelt{q} g} Q$. Now
\begin{align*}
\hat{\omega}_{\bdelt{q} \tilde{g}}(\tilde{g}_* v)
    & = \tilde{g}^*(\hat{\omega}_{\bdelt{q}})(v) & \\
    & = \Ad_{\tilde{G}}(\tilde{g}^{-1}) \hat{\omega}_\bdelt{q}(v) && \text{by the elevator property}\\
    & = \hat{\omega}_\bdelt{q}(v) && \text{since $\tilde{g} \in \ker(\Ad_{\tilde{G}})$}.
\end{align*}
This has established that $(u^{-1})^* \hat{\omega}$ is independent
of the inverse used in the calculation, and so the proposed
definition is well defined. Finally, it is clear that
$\hat{\omega} = (\rho_{* e})^{-1} u^* \omega$, so Proposition
\ref{prop:lifting-connection} applies. This proves that $\omega$
is a connection.
\end{proof}

Finally, we describe the relationship between the parallel
transports using $\omega$ and $\hat{\omega}$. This is particularly
straightforward.

\begin{prop}\label{prop:spinor-connection-parallel-transport}
Let $\alpha : \I \To M$ be a path in $M$. Let $\bdelt{q} \in
\pi_Q^{-1}(\alpha(0))$, and let $\bdelt{p} = u(\bdelt{q})$. Then
the parallel transport paths $\tilde{\alpha}_\bdelt{q} : \I \To Q$
and $\tilde{\alpha}_\bdelt{p} : \I \To P$, obtained using
$\hat{\omega}$ and $\omega$ respectively, are related by
\[ \tilde{\alpha}_{\bdelt{p}} = u(\tilde{\alpha}_{\bdelt{q}}). \]
\end{prop}
\begin{proof}
Clearly
$u(\tilde{\alpha}_\bdelt{q})(0)=\bdelt{p}=\tilde{\alpha}_\bdelt{p}(0)$.
Further, \[\pi_{P *}(\dot{\tilde{\alpha}}_\bdelt{p}(t)) =
\dot{\alpha}(t) \quad \text{and} \quad \pi_{P *} \frac{d}{dt}
u(\tilde{\alpha}_\bdelt{q}(t)) = \pi_{Q
*}(\dot{\tilde{\alpha}}_\bdelt{q}(t)) = \dot{\alpha}(t).\] Next,
according to the definition of parallel transport as an integral
curve of a horizontal vector field, in
\S\ref{ssec:parallel-transport},
\[\omega(\dot{\tilde{\alpha}}_\bdelt{p}(t)) = 0 \quad \text{and}
\quad \hat{\omega}(\dot{\tilde{\alpha}}_\bdelt{q}(t)) = 0.\] We
then calculate
\begin{align*}
\omega(\frac{d}{dt} u(\tilde{\alpha}_\bdelt{q}(t))) & = \omega(u_*
\dot{\tilde{\alpha}}_\bdelt{q}(t)) \\
    & = u^* \omega (\dot{\tilde{\alpha}}_\bdelt{q}(t)) \\
    & = \rho_{* e} \rho_{* e}^{-1} u^* \omega (\dot{\tilde{\alpha}}_\bdelt{q}(t)) \\
    & = \rho_{* e} \hat{\omega}(\dot{\tilde{\alpha}}_\bdelt{q}(t)) \\
    & = 0.
\end{align*}
Finally, rearranging the result of Lemma
\ref{lem:connection-form-from-horizontal-lifting-map} shows that a
vector $v \in TP$ is determined by $\omega(v)$ and $\pi_{P *}v$,
and so
\[\frac{d}{dt} u(\tilde{\alpha}_\bdelt{q}(t)) =
\frac{d}{dt}\tilde{\alpha}_\bdelt{p}(t).\] Thus the integral
curves $\tilde{\alpha}_\bdelt{p}$ and
$u(\tilde{\alpha}_\bdelt{q})$ are equal.
\end{proof}

\section{Classifying spinor structures as bundles}\label{sec:classifying-as-bundles}
We next consider the problem of classifying spinor structures as
bundles. As we will discover, inequivalent spinor structures may
or may not be equivalent as bundles. There is a rich
classification theory of principal fibre bundles which we can
bring to bear on the spinor structures problem.

\begin{defn*}
Two $\tilde{G}$ principal fibre bundles
$\pfbundle{\tilde{G}}{Q}{\pi_Q}{M}$ and
$\pfbundle{\tilde{G}}{Q'}{\pi_{Q'}}{M}$ are \defnemph{equivalent
as bundles} if there is a principal fibre bundle morphism $a:Q \To
Q'$ so $\pi_{Q'} \compose a = \pi_Q$.
\end{defn*}

Clearly if two spinor structures are equivalent as in Definition
\ref{defn:spinor-structure-equivalence}, they are equivalent as
bundles. The converse is not true.

The classification of principal fibre bundles is achieved by the
following proposition. (Here the higher homotopy groups $\pi_i$
are defined inductively, so $\pi_i(M) = \pi_{i-1}(\Omega M)$ for
$i>1$, where $\Omega M$ is the loop space of $M$, discussed
earlier, with an appropriate topology.)

\begin{prop}
Let $M$ be a smooth manifold, and
$\pfbundle{G}{\mathcal{P}}{\quad}{N}$ be a principal fibre bundle
such that $\pi_i(\mathcal{P}) = 0$ for $i \leq \dim(M)$. There is
a bijection between $[M, N]$, the collection of homotopy classes
of maps $M \To N$, and $\mathbf{k}(M)$, the collection of
equivalence classes of principal $G$-bundles over $M$.
\end{prop}
\begin{proof}
We say that $\mathcal{P}$ is $\dim(M)$-universal. The bijection is
given by $[f:M \To N] \mapsto f^*(\mathcal{P})$, where
$f^*(\mathcal{P})$ denotes the `pull-back' bundle, defined in
\cite[2.5.3]{hus:fb} or \cite[\S3.1]{ish:mdgfp}. A simple proof
that this map is well defined, that is, that homotopic maps give
isomorphic bundles, appears in \cite{mor:p-bbhi}. The proposition
itself is a deep result of the algebraic topology of bundles, and
is discussed in \cite[Ch. 5]{bor:scefpehglc} and
\cite[\S3.1]{ish:mdgfp} and proved in \cite[4.13.1]{hus:fb}.
\end{proof}

We also have
\begin{prop}\label{prop:reducing-structure-group}
Let $M$ be a paracompact manifold and let $G$ be a Lie group, and
$H$ a closed subgroup, so that $G/H$ is homeomorphic to $\Real^n$
for some $n$. Then the equivalence classes of principal
$G$-bundles over $M$ are in one to one correspondence with the
equivalence classes of principal $H$-bundles over $M$. Thus
\[\mathbf{k}_G(M) = \mathbf{k}_H(M).\]
\end{prop}
\begin{proof}
Since $M$ is paracompact in particular it has a countable basis
for its topology. With this fact, this Proposition is a slight
weakening of a theorem proved in \cite[\S 12.8]{ste:tfb}. Another
result which implies this theorem, but less obviously, is given in
\cite[\S 6.2.3 and \S 6.3.2]{hus:fb}.
\end{proof}

\subsection{Classifying spinor bundles in general relativity}
With these results in hand, we can now deal with what proves to be
a relatively simply case. We will see that all spinor structures
for the structure group $G = \LL$ over a noncompact $4$-manifold
$M$ are trivial as bundles. This is not a new result. The
difference between the available spinor structures is solely in
the spinor map itself. We will see that in this situation each of
the different spinor structures can be obtained from any one
spinor structure by modifying the spinor map. Essentially, if we
consider the classification of spinor structures in terms of
homomorphisms $\pi_1(M) \To \pi_1(G)$, we will see that each such
map can be \emph{realised} by a smooth map $M \To G$, and the
spinor structure corresponding to this homomorphism is constructed
by `multiplying' the spinor map by this realisation. This argument
appears to be an improvement over previous results along these
lines, and the details appear later. This very direct construction
of the spinor structures prompts the question---`what can we do
when not all the spinor structures are trivial?' We will suggest a
possible resolution of this problem.

We first tackle the problem of classifying all spinor structures
over a noncompact $4$-manifold $M$, when $G = \LL$ . The interest
in noncompact manifolds is justified by \cite[Proposition
6.4.2]{haw:lsss-t}, which shows that all compact Lorentz manifolds
have closed timelike curves, which are generally rejected on the
physical grounds of violating causality. The argument here is
adapted from that given in \cite{ish:sffds-t}. The group $\SL$ has
maximal compact subgroup $SU(2)$, and the quotient $\SL/SU(2)$ is
homeomorphic to $\Real^3$, by Proposition
\ref{prop:maximal-compact-subgroups}. Thus by Proposition
\ref{prop:reducing-structure-group}
\[\mathbf{k}_{\SL}(M) = \mathbf{k}_{SU(2)}(M).\]

Next, we find a universal bundle for $SU(2)$. This is furnished by
the Hopf bundle, $\pfbundle{SU(2) \cong S^3}{S^7}{\quad}{S^4}$.
See \cite[III. \S 5]{hus:ht} or \cite[\S 20]{ste:tfb} for a
detailed description. Since $\pi_i(S^7) = 0$ for $i \leq 6$, this
bundle is $6$-universal, and so $\mathbf{k}_{SU(2)}(M) = [M, S^4]
= H^4(M,\Integer)$. The last equality here is given by the Hopf
theorem \cite[II. \S 8]{hus:ht}.

Finally, we claim that $H^4(M,\Integer) = 0$ for any noncompact
$4$-manifold. This is not a trivial claim. All previous analyses
of this problem, for example \cite{ger:sss-tgr1} and
\cite{ish:sffds-t}, gloss over this point, stating that it is
obvious. While it is obvious that $H_4(M,\Integer) = 0$ follows
from the noncompactness, because every $4$-chain is finite and so
must have a boundary in a noncompact manifold, this does not
immediately imply that $H^4(M,\Integer) = 0$. To obtain this
result, we need a version of Poincar\'e duality suited to
orientable noncompact manifolds. This is given by
\[H^p(M,\Integer) \cong H_{4-p}^{\mathfrak{lf}}(M,\Integer).\]
Here $H_j^{\mathfrak{lf}}$ are the locally finite singular
homology groups. See \cite{mas:sht,mas:bcat}.
%-% are these references the right ones? (at this point I'm bluffing!)
It is easy to see that $H_0^{\smash{\mathfrak{lf}}}(M,\Integer) =
0$, and so $H^4(M,\Integer)=0$ as required.\footnote{A locally
finite $0$-cycle is a discrete set of points in $M$, counted with
multiplicities. For each such $s$ we may choose a ray from $s$ to
$\infty$, so that these rays are all disjoint. Thus a locally
finite $0$-cycle is the boundary of a locally finite $1$-chain.
Thanks to Dr. J. Hillman for this argument, and the suggestion to
use this species of Poincar\'{e} duality.}

We now find that all $SU(2)$ bundles over $M$ are equivalent, and
so all $\SL$ bundles are equivalent. In particular, the trivial
bundle $M \times \SL$ always exists, and so all $\SL$ bundles must
be trivial bundles. This result has also been proved in
\cite{ger:sss-tgr1}. The proof given here simply fills in some of
the gaps of the discussions in \cite{ger:sss-tgr1} and
\cite{ish:sffds-t}.

This has an important corollary, due to Geroch
\cite{ger:sss-tgr1}.
\begin{cor}
If a $\LL$ principal fibre bundle $P$ over a noncompact
$4$-manifold has a spinor structure, then the spinor structure is
trivial, as a bundle, and moreover the $\LL$ bundle itself is
trivial, \[P \cong M \times \LL.\] Thus an orthonormal $\LL$
structure has a spinor structure if and only if the orthonormal
bundle is parallelisable, that is, there is a global orthonormal
frame field.
\end{cor}
\begin{proof}
Say $Q$ is any spinor structure for $P$. Then $Q$, as a bundle, is
trivial, so $Q = M \times \SL$. The spinor map $u:Q \To P$ is a
principal fibre bundle morphism, and so gives a trivialisation of
the bundle $P$. Since $P$ is trivial, it has a global cross
section, which is exactly the global orthonormal frame field.
Conversely, if $P$ is a trivial bundle, then the condition of the
Existence Theorem is automatically satisfied, and so $P$ has a
spinor structure.
\end{proof}

Thus in the case that $G = \LL$, the `weak triviality' condition
of the Existence Theorem, roughly that $\pi_1(P) \cong \pi_1(M)
\times \pi_1(G)$, is equivalent to the triviality of $P$, that is,
$P = M \times G$.

Given that all the spinor structures for such an orthonormal
structure are the same, namely trivial, as bundles, how is it that
they differ as spinor structures? The spinor maps differ, and we
can construct each of them directly from the corresponding
homomorphism.

\begin{defn*}
A homomorphism $\varphi: \pi_1(M) \To \pi_1(G)$ is
\defnemph{realisable} if there is a smooth map $\zeta: M \To G$ so
$\zeta_* = \varphi$. That is, the map between the fundamental
groups induced by $\zeta$ is exactly $\varphi$.
\end{defn*}

A possibility at this point is that \emph{all} such homomorphisms
are realisable. A counterexample is provided by $M = \RP^4$, $G =
SO(3) \cong \RP^3$. Here $\pi_1(M) \cong \pi_1(G) \cong
\Integer_2$, and the isomorphism is not realisable.\footnote{To
see this, we need some algebraic topology. Suppose $\zeta:\RP^4
\To \RP^3$ induces an isomorphism of the fundamental groups. Then,
via the natural Hurewicz isomorphism \cite[II \S6]{hus:ht} (see
also the proof of Lemma \ref{lem:homomorphisms-cohomology-group}),
the homomorphism between the first homology groups induced by $f$
is an isomorphism. Next the evaluation homomorphism
$H^n(X,\Integer)$ to $\Hom(H_n(X),\Integer)$ is also natural, and
for $n=1$ it is an isomorphism, by the Universal Coefficient
Theorem \cite[Ch. 5, \S 5]{spa:at}, and so $\zeta$ induces an
isomorphism of the first cohomology groups. Since $\zeta$ is
continuous, it actually induces a ring homomorphism of the
cohomology rings equipped with the cup product \cite[Ch. 5, \S
6]{spa:at}, $f^* : H^n(\RP^3,\Integer) \To H^n(\RP^4,\Integer)$.
Finally, if $\alpha$ is the generator of $H^1(\RP^3,\Integer)$,
then $\alpha \smallsmile \alpha \smallsmile \alpha \smallsmile
\alpha \in H^4(\RP^3,\Integer) = 0$, but $f^*(\alpha)$ is the
generator of $H^1(\RP^4,\Integer)$, and $f^*(\alpha \smallsmile
\alpha \smallsmile \alpha \smallsmile \alpha) = f^*(\alpha)
\smallsmile f^*(\alpha) \smallsmile f^*(\alpha) \smallsmile
f^*(\alpha)$ is nontrivial in $H^4(\RP^4,\Integer)$ \cite[Ch. 5,
\S 8]{spa:at}. This is a contradiction, and so the isomorphism
between $\pi_1(\RP^4)$ and $\pi_1(\RP^3)$ is not realisable. I
would like to thank Dr. J. Hillman for suggesting this argument.}

\begin{prop}\label{prop:trivial-bundles-realisable-maps}
Suppose $P = M \times G$. There is always a trivial spinor
structure, $Q = M \times \tilde{G}$, with $u:Q \To P$ defined by
$u(m,\tilde{g}) = (m,\rho(\tilde{g}))$. Suppose $\varphi:\pi_1(M)
\To \pi_1(G)$ is a homomorphism, and $u':Q' \To P$ is the
corresponding spinor structure relative to the trivial spinor
structure, according to the Classification Theorem.

Then
\begin{enumerate}
\item the homomorphism $\varphi$ is realisable if and only if $Q'$
is trivial, and
\item in this case, if $\zeta:M \To G$ induces the
homomorphism $\varphi$, then there is a bundle equivalence $a: Q
\To Q'$ so that $u' \compose a = u_\zeta$ defined by
\[u_\zeta(m,\tilde{g}) = (m,\zeta(m) \rho(\tilde{g})).\]
\end{enumerate}
\end{prop}
\begin{proof}
To begin, we see from Lemma \ref{lem:homotopy-centre} that
\begin{align}
u'_*(\pi_1(Q))
    & = \setc{([\alpha],[e])\cat([e],\varphi([\alpha]))}{[\alpha] \in \pi_1(M)} \nonumber \\
    & = \setc{([\alpha],\varphi([\alpha]))}{[\alpha] \in \pi_1(M)}. \label{eq:u-dash-image}
\end{align}

Firstly suppose $Q'$ is trivial as a bundle. Thus there is a
bundle equivalence $a: Q \To Q'$. Define $u_\zeta = u' \compose a
: Q \To P$. Since $(\pi_P \compose u_\zeta)(m,\tilde{g}) = m$,
there is some function $\kappa: M \times \tilde{G}$ so
\[u_\zeta(m,\tilde{g}) = (m,\kappa(m,\tilde{g})).\] Now
\begin{align*}
u_\zeta(m,\tilde{g}) & = u_\zeta(m,\tilde{e} \tilde{g}) \\
    & = u_\zeta(m,\tilde{e}) \rho(\tilde{g}) \\
    & = (m,\kappa(m,\tilde{e})) \rho(\tilde{g})
\end{align*}
and thus $\kappa(m,\tilde{g}) =
\kappa(m,\tilde{e})\rho(\tilde{g}).$ Define $\zeta: M \To G$ by
$\zeta(m) = \kappa(m,\tilde{e})$. Therefore $u_\zeta(m,\tilde{g})
= (m,\zeta(m) \rho(\tilde{g}))$.

Now, as $a$ is a diffeomorphism it induces an isomorphism of
fundamental groups, and so the image of $u_{\zeta *}$ is the same
as the image of $u'_*$. Thus as a spinor structure $u_\zeta : Q
\To P$ is equivalent to $u': Q' \To P$.

We now need to prove that $\zeta_* = \varphi$. To see this, note
that the general element of $\pi_1(Q)$ is $([\alpha],[\tilde{e}])$
where $[\alpha] \in \pi_1(M)$ and $[\tilde{e}]$ is the constant
path at $\tilde{e} \in \tilde{G}$. The map $u_{\zeta*}$ acts on
this as $u_{\zeta*}([\alpha],[\tilde{e}]) =
([\alpha],[\zeta\compose\alpha])=([\alpha],\zeta_* [\alpha])$, and
so
\[u_{\zeta*}(\pi_1(Q)) = \setc{([\alpha],\zeta_*[\alpha])}{[\alpha] \in
\pi_1(M)}.\] Comparing this with Equation \eqref{eq:u-dash-image}
demonstrates the $\varphi = \zeta_*$, completing this half of the
proof.

Conversely, suppose $\varphi:\pi_1(M) \To \pi_1(G)$ is realisable
as the map $\zeta:M \To G$. Define $u_\zeta$ as above, and,
following the same argument, we still have
\begin{align*}
 u_{\zeta*}(\pi_1(Q)) & = \setc{([\alpha],\zeta_*[\alpha])}{[\alpha] \in \pi_1(M)} \\
    & = \setc{([\alpha],\varphi([\alpha]))}{[\alpha] \in \pi_1(M)} \\
    & = u'_*(\pi_1(Q)).
\end{align*}
Thus $u_\zeta$ is equivalent to $u'$ as a spinor structure, and
thus the bundles $Q$ and $Q'$ are equivalent as bundles, and so
$Q'$ is trivial.

Finally, we briefly describe the freedom available in choosing
$\zeta$. Suppose $\zeta_1$ and $\zeta_2$ both induce the
homomorphism $\varphi$. It is easy to prove, using Lemma
\ref{lem:pi1G-commutative}, that $\mu = \zeta_1^{-1} \zeta_2$,
defined via the group product, induces the trivial homomorphism
$\pi_1(M) \To \pi_1(G)$. Such maps are exactly the smooth maps of
the form $\rho \compose \nu$, where $\nu:M \To \tilde{G}$
\cite[\S2.4]{spa:at}. That is, $\zeta_1$ and $\zeta_2$ differ by a
map which lifts to a map into $\tilde{G}$. It is easily seen that,
in the above argument, if $\zeta_1$ and $\zeta_2$ are derived from
two bundle equivalence $a_1, a_2: Q \To Q'$, then these bundle
equivalences differ by a map $\nu : M \To \tilde{G}$, and
$\zeta_1^{-1} \zeta_2 = \rho \compose \nu$.
\end{proof}

This result is restricted to the special situation in which the
spinor structures are trivial as bundles. A generalisation would
be a desirable, and one is suggested by this last result. We have
seen that all the spinor structures being trivial is equivalent to
all the homomorphisms $\pi_1(M) \To \pi_1(G)$ being realisable.
The basis of the last proof was that a bundle equivalence showed
that the spinor maps `differed' by exactly the realisation of the
homomorphism, and, conversely, that a realisation of a
homomorphism enabled us to define a spinor structure on the first
bundle which was equivalent to the second spinor structure.
Perhaps the bundle type of a spinor structure is determined by
whether the corresponding homomorphism is realisable? This idea is
formalised as the following.

\begin{conj}
Suppose $u: Q \To P$ is a spinor structure, $\varphi : \pi_1(M)
\To \pi_1(G)$ is a homomorphism, and $u' : Q' \To P$ is the
corresponding spinor structure. Then $Q$ and $Q'$ are equivalent
as bundles if and only if $\varphi$ is realisable.
\end{conj}

I suspect this conjecture is in fact correct, but the tools
developed here appear to be insufficient.\footnote{See also Remark
1.14 in \S II of \cite{law:sg}. An argument closely related to
this conjecture is
given there. % and if you're really interested in proving this conjecture, look there!
If $G=SO(n)$, and $\dim M < n$, then it is easy to prove that
every homomorphism is realisable. It is proved in \cite{law:sg}
that in this case this implies that all the spinor structures are
equivalent as bundles.} If $\varphi$ is realisable as a map into
the centre of $G$, $M \To Z(G)$, then it is relatively easy to
show that $Q$ and $Q'$ are equivalent.

It turns out that the homomorphisms $\pi_1(M) \To \pi_1(G)$ form a
commutative group $\mathcal{H}$, and the realisable homomorphisms
form a subgroup $\mathcal{R}$. If this conjecture is true, the
various bundles appearing as spinor structures would be in one to
one correspondence with the factor group
$\mathcal{H}/\mathcal{R}$. It would be interesting in this case to
find a way of constructing the bundles directly from this factor
group.\footnote{We could of course construct them indirectly, by
actually constructing the spinor structure, according to
\S\ref{sssec:sufficiency}, associated with an element of a coset
in $\mathcal{H}/\mathcal{R}$, and then forgetting about the spinor
map and looking only at the underlying bundle. (This prompts a
joke, paraphrasing S. Eilenberg. Q: `How does a mathematician eat
Chinese with 3 chopsticks?' A: `They put one down and eat Chinese
with 2 chopsticks.')} We will leave these questions open, however.

Finally, why is it interesting in the first place to be able to
classify the various spinor structures according to the type of
bundle? Fundamentally, it is because bundle equivalences allow us
to compare inequivalent spinor structures. This idea will be used
in Proposition \ref{prop:inequivalent-connections} to compare the
connections associated with different spinor structures. We will
see later in \S\ref{sec:spinor-algebra} that bundle equivalences
are to `spinor fields' what diffeomorphisms are to vector fields.

\subsection{Inequivalent spinor
connections}\label{sec:inequivalent-connections} At this stage we
have a complete classification of the spinor structures, and a
rule for generating a connection associated with each spinor
structure. A natural question to ask is whether we can compare the
resulting connections, and, in that case, whether they are
genuinely different. It turns out that for two spinor structures
with the same underlying principal fibre bundle the \emph{bundle
equivalences} are maps which might potentially identify two spinor
connections as being the same. When two spinor structures are
trivial as bundles, we can carrying out this comparison, using the
explicit relation between the spinor structures given by the
realisation of the classifying homomorphism.

Suppose $u: Q \To P$ and $u' : Q' \To P$ are two spinor
structures. Suppose $\omega$ is a connection on $P$, and
$\hat{\omega}$ is the connection on $Q$ described by the above
construction, and $\hat{\omega}'$ is the connection on $Q'$. Now,
if $Q$ and $Q'$ are different as bundles, then there is no obvious
sense in which we can compare the spinor connections
$\hat{\omega}$ and $\hat{\omega}'$. On the other hand, suppose
there is a bundle equivalence (but not a spinor structure
equivalence) $a: Q \To Q'$, so $\pi_Q' \compose a = \pi_Q$, as in
Figure \ref{fig:the-not-quite-commuting-diagram}. This map gives
us a way of comparing the two spinor connections, because we can
pull-back the connection form $\hat{\omega}'$ on $Q'$ via $a$ to a
form on $Q$. It is trivial to check that $a^* \hat{\omega}'$ is in
fact a connection on $Q$, because $a$ is a principal bundle
morphism. This suggests the possibility that for a cleverly chosen
bundle equivalence $a$, we might have $a^*\hat{\omega}' =
\hat{\omega}$, in which case we could say that the spinor
connections, although defined via different spinor structures, are
`the same'.

\begin{figure}[!htp]
\begin{equation*}
\xymatrix@R+6pt{
 Q \ar@{-->}[rr]^a \ar[dr]^u \ar@(d,ul)[ddr]_{\pi_Q} &                  & Q' \ar[dl]_{u'} \ar@(d,ur)[ddl]^{\pi_{Q'}}\\
                                                     & P \ar[d]^{\pi_P} &    \\
                                                     &  M                &
}
\end{equation*}
\caption{\label{fig:the-not-quite-commuting-diagram}Two
inequivalent spinor structures and a bundle equivalence $a:Q \To
Q'$. (This is not a commuting diagram, since $a$ is not a spinor
structure equivalence here.)}
\end{figure}

We now restrict ourselves to the circumstance of two spinor
structures which are trivial as bundles. Proposition
\ref{prop:trivial-bundles-realisable-maps} shows that the most
general situation consists of a trivial spinor structure, $u: M
\times \tilde{G} \To M \times G$, where $u(m,\tilde{g}) =
(m,\rho(\tilde{g}))$, and a related spinor structure $u_\zeta : M
\times \tilde{G} \To M \times G$ given by $u_\zeta(m,\tilde{g}) =
(m,\zeta(m) \rho(\tilde{g}))$. If the two spinor structures are
inequivalent, then $\zeta$ induces the corresponding homomorphism
classifying $u_\zeta$ relative to $u$, and so $\zeta_*$ is a
nontrivial homomorphism.

Suppose $\omega$ is a connection on $P = M \times G$. By the
elevator property, $\omega$ is determined by its values at the
points $(m,e)$. If $v \in T_m M$, and $X \in T_e G$, we can write
\begin{align*}
\omega_{(m,e)}(v,X) & = \omega_{(m,e)}(v,0) + \omega_{(m,e)}(0,X) && \text{(by linearity)}\\
    & = \omega_{(m,e)}(v,0) + \psi_{(m,e)*} X && \text{(since $X \in V_{(m,e)}$)}\\
    & = \omega_{(m,e)}(v,0) + X.
\end{align*}
The last line follows because $\psi_{(m,e)}(m,g) = g$, so
$\psi_{(m,e)*}$ is the identity on the Lie algebra.

Next, we consider the two connection forms on $Q$ obtained as
$\hat{\omega} = \rho_{* e}^{-1} u^* \omega$ and $\hat{\omega}' =
\rho_{* e}^{-1} u_\zeta^* \omega$. The first of these has a very
simple form in the trivialisation, namely
\begin{align*}
 (u^* \omega)_{(m,\tilde{e})}(v,X) & = \omega_{(m,e)}(u_*(v,X)) \\
    & = \omega_{(m,e)}(v,\rho_{* e} X) \\
    & = \omega_{(m,e)}(v,0) + \rho_{* e} X,
\end{align*}
and so $\hat{\omega}_{(m,\tilde{e})}(v,X) = \rho_{* e}^{-1}
\omega_{(m,e)}(v,0) + X$. Finding a corresponding expression for
$\hat{\omega}'$ is slightly more work.

\begin{prop}\label{prop:inequivalent-connections}
The connection form on $Q$ obtained via $u_\zeta$ is defined by
\[\hat{\omega}'_{(m,\tilde{e})}(v,X) = \rho_{* e}^{-1} \Ad(\zeta(m)^{-1})
\omega_{(m,e)}(v,0) + \rho_{* e}^{-1} \zeta(m)^{-1} \zeta_* v +
X.\]
\end{prop}
\begin{rem}
Note that in this equation $\zeta_*$ denotes the \emph{derivative}
of $\zeta$, not the induced homomorphism between fundamental
groups!
\end{rem}
\begin{proof}
Suppose $n:\J \To M$ is a path, with $n(0) = m$, and $v =
\dot{n}(0)$ and $\tilde{g}:\J \To G$ is a path, with $\tilde{g}(0)
= \tilde{e}$, and $X = \dot{\tilde{g}}(0)$. Then
\begin{align*}
 u_{\zeta *}(v, X) & = \dat{t}{0} u_\zeta\left(n(t),\tilde{g}(t) \right) \\
    & = \dat{t}{0} \left( n(t), \zeta(n(t)) \rho(\tilde{g}(t)) \right ) \\
    & = \left( v, \left(\dat{t}{0} \zeta(n(t))\right) \rho(\tilde{g}(0)) +
            \zeta(n(0)) \left(\dat{t}{0} \rho(\tilde{g}(t)) \right)\right) \\
    & = (v, \zeta_* v + \zeta(m) \rho_{* e} X)
\end{align*}
In this last line here by $\zeta(m) \rho_{* e} X$ we really mean
`the derivative of left multiplication by $\zeta(m)$ acting on
$\rho_{* e} X$'. Thus while $\rho_{* e} X \in \LA{G} = T_e G$,
$\zeta(m) \rho_{* e} \in T_{\zeta(m)} G$. Next, we calculate
\begin{align*}
(u_\zeta^* \omega)_{(m,\tilde{e})}(v,X)
    & = \omega_{(m,\zeta(m))}(u_{\zeta*}(v,X)) \\
    & = \omega_{(m,\zeta(m))}(v, \zeta_* v + \zeta(m) \rho_{* e} X) \\
    & = \omega_{(m,\zeta(m))}(v,0) + \omega_{(m,\zeta(m))}(0, \zeta_*
        v + \zeta(m) \rho_{* e} X) \\
    & =  \Ad(\zeta(m)^{-1}) \omega_{(m,e)}(v,0) + \psi_{(m,\zeta(m))*}(\zeta_* v + \zeta(m) \rho_{* e} X) \\
    & = \Ad(\zeta(m)^{-1}) \omega_{(m,e)}(v,0) + \zeta(m)^{-1} \zeta_* v
        + \rho_{* e} X.
\end{align*}
While $\zeta_* v \in T_{\zeta(m)} G$, we have $\zeta(m)^{-1}
\zeta_* v \in T_e G = \LA{G}$, as is appropriate.
\end{proof}

It is impossible to claim, on the basis of these calculations,
that for \emph{every} allowed choice of $\zeta$ we have $u^*
\omega \neq u_\zeta^* \omega$. However, it seems likely that this
is the case, and certainly in the case that $\omega$ is `flat'
with respect to the trivialisation this is easy to prove. With
such an $\omega$, $(u_\zeta^* - u^*)\omega$ reduces to
$\zeta(m)^{-1} \zeta_* v$, and this cannot be zero everywhere if
$\zeta$ is to induce a nontrivial homomorphism. We will continue
this example, giving a calculation in a concrete case in the
discussion of spinor classification and the Dirac equation, in \S
\ref{ssec:implications}.

Recall that the freedom in choosing $\zeta$ is exactly a function
$\nu:M \To \tilde{G}$, as described at the end of the proof of
Proposition \ref{prop:trivial-bundles-realisable-maps}. Thus if
there is no $\zeta$ so $u^* \omega = u_\zeta^* \omega$ then we can
say that the connections obtained from the two spinor structures
are always \emph{inequivalent}, where the natural notion of
equivalence is that two connections on a $\tilde{G}$ bundle are
equivalent if one is the pull-back of the other by a function
$\nu: M \To \tilde{G}$.

\newpage
\part{Implications for the Dirac Equation and Physics}
\label{part:implications-for-physics}

We now restrict ourselves entirely to the situation of an $\LL$
reduction of the frame bundle of a noncompact $4$-dimensional
manifold, that is, to the situation of general relativity.

To begin, therefore, we give an explicit description of the
covering group of $\LL$, that, is $\SL$, and of the covering map.
We discuss some of the finite dimensional representation theory of
$\SL$, and show that the $\LL$ tensors can be embedded in an
appropriate way into the $\SL$ tensors. This embedding extends to
an embedding of the world tensors of a $\LL$ bundle into the `spin
tensors' of an $\SL$ spinor structure. Further, if we have a
connection on the $\LL$ bundle, we obtain a spinor connection on
the $\SL$ bundle, and so covariant derivatives for both types of
tensors. As we would hope, the two covariant derivatives agree on
the embedded world tensors.

Thus, in this particular case, we arrive at a powerful `spinor
algebra', which includes as a subset the normal world tensor
algebra. These ideas have had applications in mathematical
physics, particularly in \cite{pen:ss-t1} and Witten's proof of
the Positive Energy Theorem in general relativity (see
\cite{par:wppet,wit:nppet}). We use the spinor algebra solely to
demonstrate a simple formulation of the Dirac equation, usually
presented somewhat mysteriously. This formulation, based as it is
on spinor structures, immediately carries across to curved
space-times.

However, as we have seen, on nontrivial manifolds there is a
choice of spinor structures, and in order to talk about the Dirac
equation we must make such a choice. Further, we have seen that
for $\LL$ bundles, the spinor structures are particularly simple,
and classified easily, precisely because all spinor structures are
trivial as bundles. We give an example of how the Dirac equation
can depend on the choice of spinor structure.

\section{The covering homomorphism $\SL \To \LL$}
\label{ssec:covering-map}

In this section we give the formulas for the double covering map
$$\rho:\SL \To \LL.$$ The following description is well known and can be found
in many places, and we follow the conventions of \cite{hug:itt}
and \cite{pen:ss-t}. However the details are presented here
because in \S\ref{ssec:embedding-world-tensors} they give the
relationship between the tensor and spinor algebras for $\LL$.
Further, this covering map allows us to calculate the fundamental
groups of $SO(3)$ and $\LL$ in
\S\ref{ssec:fundamental-group-SO(n)} and
\S\ref{ssec:fundamental-group-SO_0(1,3)} respectively. These
results then guarantee that $\SL$ is the simply connected covering
group of $\LL$.

\begin{prop}
\label{prop:coveringmap} There is a $2$ to $1$ covering
homomorphism $\rho$ from $\SL$ to $\LL$.
\end{prop}
\begin{proof}
Let $V$ be the vector space of self-adjoint $2$ by $2$ complex
matrices. We let the group $\SL$ act on $V$ by
\begin{equation}
\label{eq:V-representation} A: M \mapsto A M A^*,
\end{equation}
for $M \in V$ and $A \in \SL$. This is a well defined map of $V$
to itself, since \[(A M A^*)^* = A M^* A^* = A M A^*.\] That is,
the action of $\SL$ preserves the self-adjointness of $M$. This
group action is a representation of $\SL$ on $V$.

We next make an identification of $\Real^4$ with $V$, via the map
 \begin{eqnarray*}
 \imath : \Real^4 & \To & V \\
 \imath  (t,x,y,z) & = & \frac{1}{\sqrt{2}}
 \begin{pmatrix}
 t+ z & x + i y \\
 x- i y  & t-z \\
 \end{pmatrix}.
 \end{eqnarray*}
This map is clearly a linear isomorphism. Further, the usual
Lorentzian metric on $\Real^{1+3}$ can be expressed as
\[t^2 - x^2 - y^2 - z^2 = 2 \det \frac{1}{\sqrt{2}}
 \begin{pmatrix}
 t+z & x + i y \\
 x- i y  & t-z \\
 \end{pmatrix}.\]
That is, $||v||_{1,3} = 2 \det \imath v$ for all $v \in \Real^4$.
Additionally, the $t$ component is recovered easily, as
\begin{equation}
\label{eq:tcomponent} t = \frac{1}{\sqrt{2}} \trace
\frac{1}{\sqrt{2}}
 \begin{pmatrix}
 t+z & x + i y \\
 x- i y  & t-z \\
 \end{pmatrix}
  = \frac{1}{\sqrt{2}} \trace \imath (t,x,y,z).
\end{equation}

Using this identification, we define a map $\rho$ from $\SL$ to
$\textrm{End}(\Real^4)$ by
\begin{equation}\label{eq:defnrho}
\rho(A) v = \imath^{-1} (A \imath (v) A^*).
\end{equation}
Now, $|| \rho(A) v ||_{1,3} = 2 \det(A \imath (v) A^*)=2 \det
\imath (v)=||v||_{1,3}$. Thus the image of $\rho$ is contained in
$O(1,3)$, the group of isometries of $\Real^{1+3}$. Finally, since
$\SL$ is connected, and $\rho$ is clearly continuous, the image of
$\rho$ must be connected, and so lies within $\LL$. That is,
$\rho:\SL \To \LL$.

It is straightforward to see that $\rho$ is a group homomorphism,
since
 \begin{eqnarray*}
 \rho(A B) v & = & \imath^{-1} (A B \imath (v) B^* A^*) \\
             & = & \imath^{-1} (A \imath (\imath^{-1} (B \imath (v) B^*)) A^*) \\
             & = & \rho(A) \imath^{-1} (B  \imath (v) B^*) \\
             & =& \rho(A) \rho(B) v.
 \end{eqnarray*}
Using this, we can think of $\rho$ as defining a representation of
$\SL$ on $\Real^{1+3}$.

The map $\imath$ is not just a linear isomorphism. It intertwines
the representation of $\SL$ on $V$ and the representation via
$\rho$ on $\Real^{1+3}$. This follows trivially from the
definition of $\rho$ in Equation \eqref{eq:defnrho}, but it is
nevertheless important.

Next we prove that $\rho$ is surjective, by exhibiting elements of
$\SL$ which are mapped to arbitrary rotations about the three
coordinate axes, and elements of $\SL$ which are mapped to
arbitrary boosts in the $z$ direction. Firstly, the rotations are
given by
\begin{subequations}
\begin{eqnarray}
 \rho
 \begin{pmatrix}
 e^{i \theta /2} & 0 \\
 0  & e^{-i \theta /2} \\
 \end{pmatrix}
  & = &
 \psmallmatrix{
 1&0&0&0\\
 0&\cos \theta & - \sin \theta & 0\\
 0&\sin \theta & \cos \theta & 0 \\
 0&0 & 0 & 1 \\
 } \label{eq:z-rotation}\\
  \rho
 \begin{pmatrix}
 \cos \theta /2 & - \sin \theta/2 \\
 \sin \theta /2 & \cos \theta /2 \\
 \end{pmatrix}
  & = &
 \psmallmatrix{
 1&0&0&0\\
 0&\cos \theta & 0&\sin \theta\\
 0& 0 & 1 & 0 \\
 0&-\sin \theta & 0&\cos \theta \\
 } \\ \rho
 \begin{pmatrix}
 \cos \theta /2 & i \sin \theta /2 \\
 i \sin \theta/2 & \cos \theta /2\\
 \end{pmatrix}
  & = &
 \psmallmatrix{
 1&0&0&0\\
 0& 1 & 0 & 0 \\
 0&0&\cos \theta & - \sin \theta\\
 0&0&\sin \theta & \cos \theta\\
 }.
\end{eqnarray}
\end{subequations}
Any rotation can be expressed as a product of rotations of these
forms. Notice that each of these matrices actually lies within
$SU(2)$.
%\footnote{
% kill this footnote?
%Interestingly, the elements of $SU(2)$ which give the general
%rotation through Euler angles \cite[p. 143]{gol:cm} $\theta$,
%$\phi$, $\psi$ are
%\[\pm
% \begin{pmatrix}
% \cos \frac{\theta}{2} e^{i(\phi+\psi)/2}  &-\sin \frac{\theta}{2} e^{i(\phi-\psi)/2} \\
% \sin \frac{\theta}{2} e^{-i(\phi-\psi)/2} & \cos \frac{\theta}{2} e^{-i(\phi+\psi)/2} \\
% \end{pmatrix}.\]
%(See \cite[\S1.2]{pen:ss-t1}.) The entries here are the
%Cayley--Klein parameters \cite[p. 155]{gol:cm}. The form of this
%matrix is considerably more concise than the form of the matrix in
%$SO(3)$ producing the same rotation \cite[p. 147]{gol:cm}.}
The boosts in the $z$ direction are given by
\[
\rho
 \begin{pmatrix}
 e^{ k /2} & 0 \\
 0  & e^{- k /2} \\
 \end{pmatrix}
   =
\psmallmatrix{
 \cosh k & 0 & 0 & \sinh k \\
 0 & 1 & 0&0\\
 0 & 0 & 1&0\\
 \sinh k & 0 & 0 & \cosh k \\
}
\]
Further, a boost along any axis can be written as the product of a
rotation taking that axis to the $z$ axis, a boost along the $z$
axis, and the inverse rotation. Since any element of $\LL$ can be
written as a product of rotations and boosts
\cite[\S1.2]{pen:ss-t1}, the map $\rho$ is surjective.

Finally, to establish that $\rho$ is $2$ to $1$, we find the
kernel. Assume $A \in \ker(\rho)$. In particular $A$ must preserve
the $t$ component of any vector in $\Real^4$, and so using
Equation \eqref{eq:tcomponent} we obtain
\begin{equation}
\label{eq:tracepreserving}\trace(M) = \trace(A M A^*) = \trace(A
A^* M)
\end{equation}
for any $M \in V$. Write
\[A A^* =
 \begin{pmatrix}
 \alpha & \beta \\
 \gamma & \delta \\
 \end{pmatrix}
.\] Substituting the following matrices $M_1 =
\psmallmatrix{1&0\\0&0}$, $M_2 = \psmallmatrix{0&0\\0&1}$, $M_3 =
\psmallmatrix{0&1\\1&0}$, $M_4 = \psmallmatrix{0&i\\-i&0}$ in $V$
for $M$ in Equation \eqref{eq:tracepreserving} we obtain the
following conditions on $A$:
\[\alpha = 1,\;\beta+\gamma=0,\; \beta-\gamma=0,\; \delta=1.\]
This simply states that $A A^* =I$, and so $A$ is a unitary
matrix, $A \in SU(2)$. Thus there are $\alpha, \beta \in \Complex$
such that $|\alpha|^2 + |\beta|^2 = 1$, so that \[A =
 \begin{pmatrix}
 \alpha & \beta \\
 - \cc{\beta} & \cc{\alpha} \\
 \end{pmatrix}
  \in SU(2).\]

We next calculate $\rho(A)(0,0,0,\sqrt{2})$ as
\[A
 \begin{pmatrix}
 1 & 0 \\
 0 & -1 \\
 \end{pmatrix}
  A^* =
 \begin{pmatrix}
 |\alpha|^2 - |\beta|^2 & - 2 \alpha \beta \\
 - 2 \cc{\alpha \beta} & - |\alpha|^2 + |\beta|^2\\
 \end{pmatrix}.\]
If $A \in \ker(\rho)$, then $\alpha \beta = 0$, and so either
$\alpha$ or $\beta$ is zero. If $\alpha = 0$, then
\[\rho(A)(0,0,0,\sqrt{2}) =
 \begin{pmatrix}
  - |\beta|^2 & 0 \\
 0 &  + |\beta|^2\\
 \end{pmatrix},\]
and so $-|\beta|^2 = 1$, which is impossible. Thus $\beta = 0$,
and $|\alpha| = 1$. Next, we calculate \[\rho(A)(0,\sqrt{2},0,0) =
 \begin{pmatrix}
 0 & 1 \\
 1 & 0 \\
 \end{pmatrix}
  A^* =
 \begin{pmatrix}
 0 & \alpha^2 \\
 \cc{\alpha}^2 & 0 \\
 \end{pmatrix},\]
and the condition $\alpha^2 =1$ implies that $A = \pm I$. Both of
these possibilities are clearly in the kernel of $\rho$, since
$(\pm I) A (\pm I^*) = A$, and so the kernel is exactly $\set{\pm
I}$.

Any such surjective homomorphism with a discrete kernel is always
a covering map. In fact, strictly speaking, since both $\SL$ and
$\LL$ are $6$ dimensional, and connected, it is not necessary to
explicitly demonstrate that $\rho$ is surjective, given the other
results. For the purpose of understanding the geometry, however,
it is useful to have exhibited the matrices for the rotations and
boost above.
\end{proof}

\section{Spinor algebra}\label{sec:spinor-algebra}
For the purpose of this section, and the next, we consider a
certain fixed $\LL$ reduction, $\Lambda M$, of the frame bundle of
a manifold $M$ such that a spinor structure exists. If there is
not a unique spinor structure, we choose one in particular, so
$\Sigma M$ is an $\SL$ principal fibre bundle, and $u: \Sigma M
\To \Lambda M$ is the spinor map.

We define the spinor algebra in terms of vector bundles associated
with a spinor structure, and the matrix representation of $\SL$.
This is analogous to the construction of the world tensor algebra
described in \S\ref{ssec:world-tensors} using the associated
vector bundles of the frame bundle. It may be useful while reading
this section to refer occasionally to \S\ref{sec:tensor-algebra}.
Now the construction of the global tensor algebra via the
principal fibre bundle proves its worth. In the earlier discussion
of the world tensor algebra, we could have made a more direct
route by avoiding discussion of the frame bundle, and building up
the world tensor algebra from the tangent bundle, which has an
intrinsic geometric meaning. We have preferred to emphasise the
less direct route, making the frame bundle central, and the
tangent bundle secondary, as discussed in
\S\ref{sec:tensor-algebra}. This is because there is no analogous
direct route now. That is, the vector bundles associated to the
spinor structure must be generated by the associated bundle
construction. It is in part for this reason that the theory of
spinor structures as principal fibres bundles is worth
developing---because it provides an effective approach to the
global $\SL$ tensor algebras.

Further, recall from
\S\ref{ssec:tensor-algebra-orthonormal-bundle} that we could
construct the world tensor algebra as vector bundles associated to
\emph{either} the frame bundle or a reduction to an orthonormal
bundle. This is not the case here, precisely because the matrix
representation of $\SL = \tilde{SO}_0(1,3)$ does not extend to a
representation of $\GLncover{4}$. This is because the matrix
representation of $\SL$ does not descend to a representation of
$\LL$, being faithful, whereas every finite dimensional
representation of $\GLncover{4}$ descends to a representation of
$GL^+(4,\Real)$, as we saw in \S\ref{sec:metric-independence}.
That is, in order to discuss the spinor algebra, we must
\emph{first} choose a particular reduction of the frame bundle.
There is no such thing as a spinor algebra for a $\GLncover{4}$
spinor structure of the $GL^+(4,\Real)$ oriented frame bundle.

The $\SL$ spinor algebra is well known.\footnote{Related spinor
algebras for other groups, in particular the double coverings
$SU(2) \To SO(3)$, $SU(2) \times SU(2) \To SO(4)$ and
$\SL(2,\Quaternion) \To SO_0(1,5)$ (see Theorem 8.4 of \cite[\S
I.8]{law:sg}, which also mistakenly claims that
$\tilde{SL}(4,\Real)$ is a double cover of $SO_0(3,3)$) can be
described using the same approach as we take here. Clifford
algebras \cite{ati:cm,law:sg,tay:nha} are closely related to these
of constructions. The theory of Clifford algebras provides a
representation of a double cover of each of the orthogonal groups.
It does not, however, treat the four fold simply connected covers
of $SO_0(p,3)$ for $p,q \geq 3$. Finally, we point out that there
are `pinors' associated with the disconnected covers of $O(p,q)$.
This joke, such as it is, can be blamed on J.--P. Serre
\cite{ati:cm}. }
%-% on the other hand, I know how to prove that all
%-% the finite dimensional representations of the four fold cover descend to the two fold cover.
In particular an accessible summary of the material is
\cite{hug:itt}, and it is discussed extremely thoroughly in
\cite[\S 2.5]{pen:ss-t1}. We follow the conventions of these two
books. The novelty in this section is using the associated vector
bundle construction to pass from the local to the global tensor
algebra. This enables us to discuss most of the algebra in the
simple, local, setting, guaranteeing that this work then carries
across to the global setting.

We begin, as before, by describing a local tensor algebra. This
time there is an additional complication---as well as the dual
representation, we need the complex conjugate representation. The
group $\SL$ acts on $\Complex^2$ by matrix multiplication. If we
wish to refer to components in $\Complex^2$ we will use numerical
indices $0$ and $1$. We will write $S$ for the vector space
$\Complex^2$ carrying the matrix representation $\lambda$ of
$\SL$. As before, $S^*$ denotes $\Complex^2$ carrying the dual
representation $\lambda^*$. We introduce $\cc{S}$ carrying the
complex conjugate representation $\cc{\lambda}$, given by the
complex conjugate of a matrix in $\SL$ acting on $\Complex^2$, and
$\cc{S^*}$ carrying the dual complex conjugation representation,
$\cc{\lambda^*}$. The dual complex conjugate representation is
exactly the same as the complex conjugate dual representation, and
so there are no further basic representations.

Developing the geometric tensor algebra based upon these
representations is straightforward, and only a slight
generalisation of previous work. Specifically, the general tensor
representation $\spinors{k}{k'}{l}{l'}$ has valence
$\valences{k}{k'}{l}{l'}$, and the elements are multilinear maps
\begin{equation*}
\underbrace{\left(S \times \dots \times S\right)}_{l\textrm{
times}} \times \underbrace{\left(S^* \times \cdots \times
S^*\right)}_{k\textrm{ times}} \times \underbrace{\left(\cc{S}
\times \dots \times \cc{S}\right)}_{l'\textrm{ times}} \times
\underbrace{\left(\cc{S^*} \times \dots \times
\cc{S^*}\right)}_{k'\textrm{ times}} \To \Complex.
\end{equation*}
These objects are called \emph{spinors}. The action of $\SL$ on
spinors is defined as in the general setting in
\S\ref{sec:local-tensor-algebra}. It has been proved
\cite{nai:lrlg} that every irreducible representation appears as a
subrepresentation of these, in particular as a space of completely
symmetric spinors.

We next introduce the local abstract index tensor algebra, by
specifying the labelling sets. We use uppercase \rroman letters
for spinor arguments requiring an element of $S$ or $S^*$, and
uppercase primed \rroman letters for those requiring an element of
$\cc{S}$ or $\cc{S^*}$. Again, we write, for example,
$\mathcal{S}^{\ci{A}}{}_{\ci{B}}{}^{\ci{C'}}{}_{\ci{D'}}$ for
those abstract index spinors with labels $\ci{A}, \ci{B}, \ci{C'},
\ci{D'}$ and whose underlying geometric spinor is in
$\spinors{1}{1}{1}{1}$.

The four representations can be written out in abstract index
notation, using the idea that for $s \in \SL$, $\lambda(s):S \To
S$, and so $\lambda(s)$ can be considered as a map $S \times S^*
\To \Complex$, and so an element of $\spinors{1}{0}{1}{0}$. We
write it as $s^{\ci{A}}{}_{\ci{B}}$, and thus $\lambda(s)u$
appears as $s^{\ci{A}}{}_{\ci{B}} u^{\ci{B}}$ in abstract index
notation. Similarly $\lambda^*(s)w$ appears as
$(s^{-1})^{\ci{A}}{}_{\ci{B}} u_{\ci{A}}$, $\cc{\lambda}(s)v$
appears as $\cc{s}^{\ci{A'}}{}_{\ci{B'}} v^{\ci{B'}}$ and
$\cc{\lambda^*}(s)z$ as $(\cc{s^{-1}})^{\ci{A'}}{}_{\ci{B'}}
z_{\ci{A'}}$. The higher valence spinor representations in
abstract index notation then have the obvious forms suggested by
these.

The usual tensor operations carry across, \emph{mutatis mutandis}.
We can take tensor products of spinors, permute indices, but only
within each of the four types, and perform contractions.
Contractions must be between a superscript and subscript pair of
unprimed indices, or such a pair of primed indices. Attempting to
apply these operations to invalid pairs of indices has no
geometrical meaning in the underlying tensor algebra.

Because none of these operations on spinors interchange unprimed
and unprimed indices, we can freely rearrange them relative to
each other, as long as the ordering of unprimed indices and the
ordering of primed indices is preserved. Thus the spinor
$A^{\ci{A}}{}_{\ci{B}}{}^{\ci{C'}}{}_{\ci{D'}}$ denotes exactly
the same object as $A^{\ci{A C' }}{}_{\ci{B D' }}$.

We can also take complex conjugates of spinors. Complex
conjugation interchanges $S$ and $\cc{S}$, and also $S^*$ and
$\cc{S^*}$. To take the complex conjugate of a spinor, we take the
complex conjugate of its arguments, and of the resulting complex
number. For example, if $T \in \spinors{2}{0}{0}{0}$, and $w, z
\in \cc{S^*}$, then $\cc{T} \in \spinors{0}{2}{0}{0}$ and
$\cc{T}(w,z) = \cc{T(\cc{w},\cc{z})}$. Thus the operation of
complex conjugation maps spinors in $\spinors{k}{k'}{l}{l'}$ to
spinors in $\spinors{k'}{k}{l'}{l}$. For example,
\[\cc{{T^{\ci{A B}}}_{\ci{C D'}}} = {\cc{T}^{\ci{A' B'}}}_{\ci{C' D}}.\]
Complex conjugation intertwines the relevant representations, as,
for example, $\cc{s^{\ci{A}}{}_{\ci{B}} u^{\ci{B}}} =
\cc{s}^{\ci{A'}}{}_{\ci{B'}} \cc{u}^{\ci{B'}}$. If $k = k'$ and
$l=l'$, complex conjugation becomes an involution of
$\spinors{k}{k}{l}{l}$, and so we can pick out the \emph{real}
spinors,\footnote{As Penrose points out \cite[\S3.1]{pen:ss-t1},
we can call these spinors real rather than self-adjoint, because
the abstract index tensor product is commutative.} which are
invariant under complex conjugation.

With the local spinor algebra and its operations thus set out, we
invoke the associated vector bundle construction to provide the
global abstract index spinor algebra. Spinor operations have a
convenient notation, but the appearance of indices does not imply
use of specific local components.
%-% perhaps more discussion should go here?
Thus a global spin vector, for example, is an object of the form
$v^{\ai{A}} = [\bdelt{p},v^{\ci{A}}]$ where $\bdelt{p} \in \Sigma
M$, and $v \in S = \spinors{1}{0}{0}{0}$.

It is worth pointing out here an important respect in which
spinors differ from world tensors. Given a smooth map $f$ between
manifolds $M$ and $N$, we can push forward a tangent vector on $M$
to a tangent vector on $N$. From our viewpoint, this is because
such a smooth map, by its derivative, induces a principal bundle
morphism between the frame bundles $F M$ and $F N$. This is not
the case for orthonormal bundles and spinor bundles, unless $f$ is
an isometry.\footnote{Compare \cite[\S 13.1]{wal:gr} and \cite[\S
5]{hug:itt}.} Thus we cannot push forward or pull back a spinor by
a diffeomorphism. A bundle equivalence between spinor structures,
as in \S\ref{sec:classifying-as-bundles}, however, can effect this
operation.

Just as in the analysis of the group $O(p,q)$ in
\S\ref{sec:SO-groups} there was an important \emph{invariant
tensor} $\eta_{\ci{a b}}$, there is a similar tensor for $\SL$.
This is the `volume form' $\eps_{\ci{A B}}$, defined by
\begin{equation}\label{eq:varepsilon-definition}
\eps_{\ci{A B}} w^{\ci{A}} z^{\ci{B}} = w^0 z^1 - w^1 z^0
\end{equation}
for all $w,z \in S$. The right hand side refers to the components
of $w$ and $z$ in $\Complex^2$. In fact, up to a complex scalar
multiple, there is only one such antisymmetric valence
$\valences{0}{0}{2}{0}$ tensor. It is easy to see $\eps_{\ci{A
B}}$ is an invariant tensor for $\SL$, since it transforms as
\begin{equation}\label{eq:epsilon-transforms}
s^{\ci{C}}{}_{\ci{A}} s^{\ci{D}}{}_{\ci{B}} \eps_{\ci{C D}} =
(\det s) \eps_{\ci{A B}} = \eps_{\ci{A B}}.
\end{equation}
There is also a valence $\valences{0}{0}{0}{2}$ spinor
$\ceps_{\ci{A' B'}}$ obtained by complex conjugation.

Using the volume form, we define index raising and lowering
conventions for the spinor algebra, as in
\S\ref{ssec:index-manipulation}. We make the identifications
\[z^{\ci{A}} \mapsto z_{\ci{A}} = z^{\ci{B}} \eps_{\ci{B
A}},\] and see from the definition of $\eps_{\ci{A B}}$ in
Equation \eqref{eq:varepsilon-definition} that this is invertible,
so there is a $\eps^{\ci{A B}}$ so that
\[z_{\ci{A}} \mapsto z^{\ci{A}} = \eps^{\ci{A B}} z_{\ci{B}}.\]
Later we will need the component form of these relations, for
elements of $\Complex^2$. These are
\begin{equation}\label{eq:manipulate-components}
z_0 = - z^1, \quad z_1 = z^0.
\end{equation}

That $\eps^{\ci{A B}}$ is the inverse can be written alternatively
as $\eps_{\ci{A B}} \eps^{\ci{C B}} = \delta^{\ci{C}}_{\ci{A}}$ or
$\eps^{\ci{A B}} \eps_{\ci{A C}} = \delta^{\ci{B}}_{\ci{C}}$. This
notation for the inverse is itself compatible with the index
raising and lowering convention. We must be careful in applying
these operations, because $\eps_{\ci{A B}}$ is antisymmetric, that
is, $\eps_{\ci{A B}} = - \eps_{\ci{B A}}$ and $\eps^{\ci{A B}} = -
\eps^{\ci{B A}}$. This has the effect that, for example,
$z_{\ci{A}} = z^{\ci{B}} \eps_{\ci{B A}} = -z^{\ci{B}} \eps_{\ci{A
B}}$. The mnemonic for correct index manipulation is `adjacent
indices - descending to the right' \cite[p. 14]{hug:itt}. As with
raising and lowering the indices of world tensors, we now have to
keep track of the relative ordering of superscript and subscript
indices, so that they may be unambiguously raised and lowered. The
complex conjugate $\ceps_{\ci{A' B'}}$ allows analogous index
raising and lowering conventions for primed indices.

Now, since $\eps_{\ci{A B}}$ is $\SL$ invariant, it defines a
natural valence $\valences{0}{0}{2}{0}$ global spinor, by
\[\eps_{\ai{A B}} = [\bdelt{p},\eps_{\ci{A B}}]\] for any
$\bdelt{p} \in \Sigma M$. Thus the raising and lowering
conventions carrying immediately across to the global spinor
algebra, just as for the world tensors.

\subsection{Embedding of the world tensors in the spin tensors}
\label{ssec:embedding-world-tensors} The world tensors can be
embedded as the \emph{real spin tensors} in the spinor
algebra.\footnote{This is discussed in \cite[\S 3]{hug:itt},
\cite[\S3.1]{pen:ss-t1} and \cite[\S13.1]{wal:gr} with varying
levels of detail, from a strictly algebraic viewpoint.} To see
this, we first look at the representation of $\SL$ on tensors in
$\spinors{1}{1}{0}{0} = S \otimes \cc{S}$. An element $s \in \SL$
transforms $T^{\ci{A}\ci{A'}}$ to
\begin{equation}
\label{eq:SccS-representation} {s^{\ci{A}}}_{\ci{B}}
{\cc{s}^{\ci{A'}}}_{\ci{B'}} T^{\ci{B}\ci{B'}}.
\end{equation}
If $T^{\ci{A}\ci{A'}}$ is real, so that $\cc{T^{\ci{A}\ci{A'}}} =
\cc{T}^{\ci{A'}\ci{A}} = \cc{T}^{\ci{A}\ci{A'}}$, then $s$ acting
on $T^{\ci{A}\ci{A'}}$ is also real, since
\begin{eqnarray*}
\cc{{s^{\ci{A}}}_{\ci{B}} {\cc{s}^{\ci{A'}}}_{\ci{B'}}
T^{\ci{B}\ci{B'}}} & = & {\cc{s}^{\ci{A'}}}_{\ci{B'}}
{\cc{\cc{s}}^{\ci{A}}}_{\ci{B}} \cc{T}^{\ci{B'}\ci{B}} \\
 & = & {s^{\ci{A}}}_{\ci{B}} {\cc{s}^{\ci{A'}}}_{\ci{B'}}
T^{\ci{B}\ci{B'}}.
\end{eqnarray*}
Thus this representation is reducible, and in particular has a
subrepresentation defined on the real subspace of $S \otimes
\cc{S}$.

We first make an identification of $V$, as defined in
\S\ref{ssec:covering-map} as the vector space of self-adjoint $2$
by $2$ matrices, with $\Re(S \otimes \cc{S})$. This identification
is
\begin{equation}
\label{eq:jmath-definition} \jmath \left(M=\begin{pmatrix}
T^{0 0'} & T^{0 1'} \\
T^{1 0'} & T^{1 1'}
\end{pmatrix}\right) = T^{\ci{A}\ci{A'}}.
\end{equation}
It is clear that the reality condition on $T^{\ci{A}\ci{A'}}$ is
the same as the conditions on the components $T^{0 0'}, T^{1 1'}
\in \Real$ and $T^{0 1'} = \cc{T^{1 0'}}$. The map $\jmath$ simply
relates the two presentations of the vector space, as matrices in
$V$, and as tensors in $\Re(S \otimes \cc{S})$. Next we check that
this identification intertwines the representations of $\SL$ on
$V$ and on $\Re(S \otimes \cc{S})$. This is seen easily,
translating the matrix form of Equation
\eqref{eq:V-representation} to the component notation of Equation
\eqref{eq:SccS-representation}. Thus for $s \in \SL$, and $M \in
V$,
\begin{eqnarray*}
s (\jmath M) & = & {s^{\ci{A}}}_{\ci{B}}
{\cc{s}^{\ci{A'}}}_{\ci{B'}} T^{\ci{B}\ci{B'}} \\
            & = & {s^{\ci{A}}}_{\ci{B}}
T^{\ci{B}\ci{B'}}
{\cc{s}^{\ci{A'}}}_{\ci{B'}} \\
            & = & \jmath (A M A^*) \\
            & = & \jmath (s M),
\end{eqnarray*}
where $A$ is the matrix associated with $s$, with components
$$A=\begin{pmatrix}
{s^0}_0 & {s^0}_1 \\ {s^1}_0 & {s^1}_1
\end{pmatrix},$$ and $M$
is the matrix with components as in Equation
\eqref{eq:jmath-definition}.

Combining this with our earlier result about $\imath$ intertwining
the representations $\rho$ on $\Real^{1+3}$ and the representation
of Equation \eqref{eq:V-representation} on $V$, we obtain the
result

\begin{prop}
\label{prop:intertwine} The composition $(\jmath \compose \imath)$
is a linear isomorphism between $\Real^{1+3}$ and $\Re(S \otimes
\cc{S})$, intertwining the representations, so that for $s$ in
$\SL$,
\[ (\jmath \compose \imath) \rho(s) = s (\jmath \compose \imath).\]
\end{prop}

For notational convenience, we will often omit this map, writing
instead \[(\jmath \compose \imath) T^{\ci{a}} = T^{\ci{A A'}}.\]
Although we have introduced this map abstractly, in terms of
components it is quite simple. If the components of $T^{\ci{a}}$
in $\Real^4$ are $T^1, T^2, T^3, T^4$, and the components of
$T^{\ci{A A'}}$ are $T^{00'}, T^{01'}, T^{10'}, T^{11'}$, then the
components are related according to
%\begin{align}
%T^1 & = \frac{T^{00'} + T^{11'}}{\sqrt{2}}  &&& T^{00'} & = \frac{T^1 + T^4}{\sqrt{2}} \nonumber \\
%T^2 & = \frac{T^{01'} + T^{10'}}{\sqrt{2}}  &&& T^{01'} & = \frac{T^2 + i T^3}{\sqrt{2}} \nonumber \\
%T^3 & = \frac{T^{01'} - T^{10'}}{i \sqrt{2}} & \text{and} && T^{10'} & = \frac{T^2 - i T^3}{\sqrt{2}} \label{eq:components-related}\\
%T^4 & = \frac{T^{00'} - T^{11'}}{\sqrt{2}} &&& T^{11'} & =
%\frac{T^1 - T^4}{\sqrt{2}} \nonumber
%\end{align}
%
\begin{align*}
 T^1 & = \frac{T^{00'} + T^{11'}}{\sqrt{2}}, &&&
 T^2 & = \frac{T^{01'} + T^{10'}}{\sqrt{2}}, &&&
 T^3 & = \frac{T^{01'} - T^{10'}}{i \sqrt{2}}, &&&
 T^4 & = \frac{T^{00'} - T^{11'}}{\sqrt{2}}
\end{align*}
and
\begin{align}
 T^{00'} & = \frac{T^1 + T^4}{\sqrt{2}}, &&&
 T^{01'} & = \frac{T^2 + i T^3}{\sqrt{2}}, &&&
 T^{10'} & = \frac{T^2 - i T^3}{\sqrt{2}}, &&&
 T^{11'} & = \frac{T^1 - T^4}{\sqrt{2}} \label{eq:components-related}
\end{align}

This linear isomorphism induces a linear isomorphism of all the
tensor products, which again intertwines the representations.
Since $\Re(A) \otimes \Re(B) = \Re(A \otimes B)$, this identifies
all of the local tensors over $\Real^{1+3}$ with all of the real
spinors. Again, we will often not explicitly write this map, and
understand that if $T^{\ci{a}_1 \dots \ci{a}_k}{}_{\ci{b}_1 \dots
\ci{b}_l}$ is a tensor, then $T^{\ci{A}_1 \ci{A'}_1 \dots \ci{A}_k
\ci{A'}_k}{}_{\ci{B}_1 \ci{B'}_1 \dots \ci{B}_l \ci{B'}_l}$ is the
corresponding real spinor. Again, the components are related
according to the obvious extension of Equation
\eqref{eq:components-related}.

From this, we easily obtain the desired embedding on the tangent
bundle into the spinors.

\begin{prop}\label{prop:embedding-world-tensors}
Let $\bdelt{b} \in \Lambda M$ be an element of the $\LL$
orthonormal frame bundle, $T^{\ci{a}}$ be an element of
$\Real^{1+3}$, and $u$ be the spinor map. Then the map
\[ T^{\ai{a}} = \left[ \bdelt{b}, T^{\ci{a}} \right] \mapsto \left[
u^{-1}(\bdelt{b}), (\jmath \compose \imath) T^{\ci{a}} = T^{\ci{A
A'}} \right] = T^{\ai{A A'}}\] is an associated bundle isomorphism
%-%do we need to explain what associated bundle isomorphism means?
$\mathcal{T}^{\ai{a}} \To \Re(\mathcal{S}^{\ai{A A'}})$ from $TM$,
the tangent bundle, to the bundle of real valence
$\valences{1}{1}{0}{0}$ spin tensors.
\end{prop}
\begin{proof}
We need to check that the map is well defined. This requires two
steps. Firstly, an element of the frame bundle $\bdelt{b}$ will
have two inverse images under $u$. However these will differ by
the generator of the kernel of the covering map $\rho$, so the two
inverse images are of the form $\bdelt{c}$ and $\bdelt{c}(-I)$.
Now,
\[ \left[ \bdelt{c}(-I) , T^{\ci{A}\ci{A'}} \right] = \left[
\bdelt{c} , (-I) T^{\ci{A}\ci{A'}} (-I) \right] = \left[ \bdelt{c}
, T^{\ci{A}\ci{A'}} \right],\] and so this ambiguity is removed.

Secondly, if $g \in \LL$, then $\left[ \bdelt{b}g, T^{\ci{a}}
\right] = \left[ \bdelt{b},  g T^{\ci{a}} \right]$. We need to
check that $\left[ u^{-1}(\bdelt{b} g), (\jmath \compose \imath)
T^{\ci{a}} \right] = \left[ u^{-1}(\bdelt{b}), (\jmath \compose
\imath) (g T^{\ci{a}})  \right]$. Choose $s \in \SL$ so that
$\rho(s) = g$, and apply Proposition \ref{prop:intertwine}. Then
\begin{eqnarray*}
\left[ u^{-1}(\bdelt{b} g), (\jmath \compose \imath) T^{\ci{a}}
\right] & = & \left[ u^{-1}(\bdelt{b}) s, (\jmath \compose \imath)
T^{\ci{a}} \right]\\ &= &\left[ u^{-1}(\bdelt{b}) , s (\jmath
\compose \imath) T^{\ci{a}} \right] \\&=& \left[ u^{-1}(\bdelt{b})
, (\jmath \compose \imath) (\rho(s) T^{\ci{a}}) \right] \\&=&
\left[ u^{-1}(\bdelt{b}) , (\jmath \compose \imath) (g T^{\ci{a}})
\right],
\end{eqnarray*} as required.

That the map is a linear isomorphism between the bundles follows
immediately from the fact that $(\jmath \compose \imath)$ is a
linear isomorphism between the underlying vector spaces.
\end{proof}

Rather than giving this map an explicit name, we identify the
objects $T^{\ai{a}}$ and $T^{\ai{A A'}}$, keeping the same kernel
letter and substituting the pair of spinor indices $\ai{A}$,
$\ai{A'}$ for the world vector index $\ai{a}$. When we write
equations with mixed indices, that is, both lowercase and
uppercase indices, it is best to consider this as \emph{notation}
for an equation with solely uppercase indices, that is, an
equation solely in terms of spinors associated with the $\SL$
bundle.\footnote{In this case it is not appropriate to use the
product bundle defined in \S\ref{ssec:product-bundles}.}

Again, this map extends in an obvious way to identify tensor
products, embedding the world tensors into global spinor algebra.
Because the underlying linear isomorphism intertwines the
representations, all the tensor operations are compatible with
these identifications. Thus for example we can write
\[g_{\ai{a b}} = g_{\ai{A A' B B'}}, \quad t^{\ai{a}} p_{\ai{a b}} = t^{\ai{A A'}} p_{\ai{A A' B
B'}},\quad s^{\ai{a B}} = s^{\ai{A A' B}},\] and so forth.

\subsection{Relationship between $\eps_{\ai{A B}}$ and
$g_{\ai{a b}}$}
%spinor structure determines conformal class of metric
Notice that at this stage there are two independent conventions
for raising and lowering indices. We can manipulate tensor indices
using $\eta_{\ai{a b}}$, and spinor indices using $\eps_{\ai{A
B}}$ or $\ceps_{\ai{A' B'}}$. Since we have now proposed an
embedding of the world tensors into the spinors, we must check
that these conventions are equivalent---that is, that raising a
lowercase index using $\eta_{\ci{a b}}$ is the same as raising
separately the two corresponding uppercase indices using
$\eps_{\ci{A B}}$ and $\ceps_{\ci{A' B'}}$. This is confirmed in
the following.

\begin{prop}
The volume form and inner product are related as
\[\eta_{\ci{a b}} = \eta_{\ci{A A' B B'}} = \eps_{\ci{A B}}\ceps_{\ci{A'
B'}}.\]
\end{prop}
\begin{proof}
We simply calculate in components, using Equation
\eqref{eq:components-related}.
\begin{align*}
\eta_{\ci{a b}} x^{\ci{a}} y^{\ci{b}} & = x^1 y^1 - x^2 y^2 - x^3 y^3 - x^4 y^4 \\
    & =
    \frac{(x^{00'} + x^{11'})(y^{00'}+y^{11'})}{\sqrt{2}} -
        \frac{(x^{01'} + x^{10'})(y^{01'}+y^{10'})}{\sqrt{2}} + \\
        & \qquad \frac{(x^{01'} - x^{10'})(y^{01'}-y^{10'})}{\sqrt{2}} -
        \frac{(x^{00'} - x^{11'})(y^{00'}-y^{11'})}{\sqrt{2}}
     \\
    & = x^{00'}y^{11'} + x^{11'}y^{00'} - x^{01'}y^{10'} -
    x^{10'}y^{01'} \\
    & = \ceps_{\ci{A' B'}} x^{0\ci{A'} }y^{1 \ci{B'}} - x^{1
    \ci{A'}} y^{0\ci{B'}} \\
    & = \eps_{\ci{A B}} \ceps_{\ci{A' B'}} x^{\ci{A
    A'}} y^{\ci{B B'}}.
\end{align*}
Thus $\eta_{\ci{a b}} = \eps_{\ci{A B}}\ceps_{\ci{A' B'}}$.
\end{proof}

It follows straight from this that the index manipulation
conventions agree on the embedded tensors over $\Real^{1+3}$, and
also that the corresponding result holds for the global tensors,
\[ g_{\ai{a b}} = \eps_{\ai{A B}} \ceps_{\ai{A' B'}}.\]

\section{The $\SL$ spinor connection}
An $\LL$ connection on the orthonormal frame bundle $\Lambda M$
lifts as in \S\ref{sec:lifting-connection} to an $\SL$ connection
on the spinor bundle $\Sigma M$. We will show that the connection
obtained in this way is compatible with the embedding of the world
tensors into the spin tensors described in
\S\ref{sec:spinor-algebra}. In particular, we have the following.

\begin{prop}
Suppose $t^{\ai{b}}$ is a tangent vector field on $M$, and
$t^{\ai{B B'}}$ is the corresponding real spinor, according to
Proposition \ref{prop:embedding-world-tensors}. Let $\omega$
denote a connection on the orthonormal bundle $\Lambda M$, and
$\hat{\omega}$ be the connection on the spinor bundle $\Sigma M$
described in Proposition \ref{prop:lifting-connection}. Further,
let $\nabla_{\ai{a}}$ and $\hat{\nabla}_{\ai{a}}$ be the
corresponding covariant derivatives. Then
\[ \nabla_{\ai{a}} y^{\ai{b}} = \hat{\nabla}_{\ai{a}} y^{\ai{B B'}}. \]
\end{prop}
\begin{proof}
We choose adapted local trivialisations of $\Sigma M$ and $\Lambda
M$. Let $\hat{\psi}:U \times \SL \To \Sigma M$ be a local
trivialisation, and let $\psi:U \times \LL$ be defined by $\psi=u
\compose \hat{\psi}$. Fix $m_0 \in U$, and say
$\bdelt{q_0}=\hat{\psi}(m_0,e)$, and $\bdelt{p_0}=\psi(m_0,e)$.

Let $x^{\ai{a}}$ be a vector field defined on $U$, and let $m:\I
\To U$ be the integral curve of $x^{\ai{a}}$ starting at $m_0$. We
can form two parallel transports of the path $m$, via $\omega$ and
$\hat{\omega}$, to obtain $\tilde{m}_\bdelt{p_0}$ and
$\tilde{m}_\bdelt{q_0}$. In the local trivialisation these
parallel transports are $\bdelt{p}(t) = (m(t),g(t)) =
\psi^{-1}(\tilde{m}_\bdelt{p_0}(t))$ and
$\bdelt{q}(t)=(m(t),\tilde{g}(t)=
\hat{\psi}^{-1}(\tilde{m}_\bdelt{q_0}(t))$. Here $g:\I \To \LL$
and $\tilde{g}:\I \To \SL$. Now, in accordance with Proposition
\ref{prop:spinor-connection-parallel-transport}, $u \compose
\tilde{m}_\bdelt{q_0} = \tilde{m}_\bdelt{p_0}$, and so
\begin{align*}
\bdelt{p}(t) & = \psi^{-1}(\tilde{m}_\bdelt{p_0}(t)) \\
             & = \psi^{-1}(u \compose \tilde{m}_\bdelt{q_0}(t)) \\
             & = (\psi^{-1}\compose u \compose \hat{\psi})(\bdelt{q}(t)) \\
             & = (m(t),\rho(\tilde{g}(t))).
\end{align*}
Thus $g(t)=\rho(\tilde{g}(t))$.

Finally now we calculate the covariant derivative, using Equation
\eqref{eq:covariant-derivative-with-g}.
\begin{align*}
x^{\ai{a}} \hat{\nabla}_{\ai{a}} y^{\ai{B B'}}(m_0)
  & = \left[(m_0,e), x^{\ai{a}} (\dd t^{\ci{B} \ci{B'}})_{\ai{a}}(m_0) -
                \frac{d}{dt}(\tilde{g}(t)(y^{\ci{B} \ci{B'}}))
                \right] \\
  & = \biggl[(m_0,\rho(e)),  (\jmath \compose \imath)^{-1} \left(x^{\ai{a}} (\dd t^{\ci{B} \ci{B'}})_{\ai{a}}(m_0) -
                \frac{d}{dt}(\tilde{g}(t)(y^{\ci{B} \ci{B'}}))
                \right)
                \biggr] \\
  & = \left[(m_0,e), x^{\ai{a}} (\dd t^{\ci{b}})_{\ai{a}}(m_0) -
                \frac{d}{dt}((\jmath \compose \imath)^{-1} \tilde{g}(t)(y^{\ci{B} \ci{B'}}))
                \right] \\
  & = \left[(m_0,e), x^{\ai{a}} (\dd t^{\ci{b}})_{\ai{a}}(m_0) -
                \frac{d}{dt}(\rho(\tilde{g}(t))(y^{\ci{b}}))
                \right] \\
  & = \left[(m_0,e), x^{\ai{a}} (\dd t^{\ci{b}})_{\ai{a}}(m_0) -
                \frac{d}{dt}(g(t)(y^{\ci{b}}))
                \right] \\
  & = x^{\ai{a}} \nabla_{\ai{a}} y^{\ai{b}}(m_0).\qedhere
\end{align*}
\end{proof}

It is clear that this argument extends to show that the two
covariant derivatives agree on any of the embedded world tensors.
Following this result, we use the same notation $\nabla_{\ai{a}}$
to denote both covariant derivatives, because they agree on the
embedded world tensors. Further, we can easily apply early results
to obtain the following important proposition.
\begin{prop}\label{prop:varepsilon-covariantly-parallel}
Let $\nabla_{\ai{a}}$ be a covariant derivative associated to a
connection form $\hat{\omega}$ on the spinor bundle. Then
\[\nabla_{\ai{a}} \eps_{\ai{B C}} = 0.\]
\end{prop}
\begin{proof}
The tensor $\eps_{\ci{B} \ci{C}}$ is an invariant tensor for
$\SL$, according to Equation \eqref{eq:epsilon-transforms}, and so
its associated tensor field is covariantly parallel, by
Proposition \ref{prop:invariant-vector-fields-parallel}. Thus
\[\nabla_{\ai{a}} \eps_{\ai{B C}} = 0.\qedhere\]
\end{proof}

Using this we can unambiguously raise and lower the spinor indices
of the covariant derivative operator. Thus $\nabla_{\ai{A A'}} =
\nabla_{\ai{a}}$, and $\eps^{\ai{A B}} \nabla_{\ai{B A'}} =
\nabla^{\ai{A}}{}_{\ai{A'}} = \nabla_{\ai{B A'}} \eps^{\ai{A B}}$.
Further, the covariant derivative is consistent with our raising
and lowering conventions, as we proved in
\S\ref{ssec:metric-connections} for the world tensors, in the
sense that if $y_{\ai{B}}$ is a valence $\valences{0}{0}{1}{0}$
spinor, then
\[ \nabla_{\ai{a}} y^{\ai{B}} = \nabla_{\ai{a}} \eps^{\ai{B
C}} y_{\ai{C}} =  \eps^{\ai{B C}} \nabla_{\ai{a}} y_{\ai{C}}.\]
Here we have used the Leibniz rule and the above Proposition. This
will be important in our next and final topic.

\section{The Dirac Equation}\label{sec:dirac-equation}

\newcommand{\gammazero}{\frac{1}{\sqrt{2}} %
%\left(\begin{array}{rrrr}
\psmallmatrix{
0 & 0 & 0 & -1 \\
0 & 0 & 1 & 0  \\
0 & -1& 0 & 0  \\
1 & 0 & 0 & 0 }}
%\end{array}\right)
\newcommand{\gammaone}{\frac{1}{\sqrt{2}}\psmallmatrix{
0 & 0 & 1 & 0  \\
0 & 0 & 0 & -1 \\
1 & 0 & 0 & 0  \\
0 & -1& 0 & 0 }}
\newcommand{\gammatwo}{\frac{1}{\sqrt{2}}\psmallmatrix{
0 & 0 & -i& 0  \\
0 & 0 & 0 & -i \\
i & 0 & 0 & 0  \\
0 & i & 0 & 0 }}
\newcommand{\gammathree}{\frac{1}{\sqrt{2}}\psmallmatrix{
0 & 0 & 0 & -1 \\
0 & 0 & -1& 0  \\
0 & -1& 0 & 0  \\
-1& 0 & 0 & 0 }}

\newcommand{\psivector}{\begin{pmatrix}
\phi_0 \\ \phi_1 \\ \chi_{0'} \\ \chi_{1'}\end{pmatrix}}
\newcommand{\smallpsivector}{\psmallmatrix{\phi_0 \\ \phi_1 \\ \chi_{0'} \\
\chi_{1'}}}

%strictly speaking, Dirac is eponymous, not his equations.
%perhaps the equations are eponymic?
Introducing his eponymous equations in 1928 Dirac \cite{dir:qte1}
made a significant step forward in physics. The Dirac theory of
electrons described the quantum mechanical behaviour of massive
spin $\frac{1}{2}$ particles, in a relativistic setting. In fact,
the Dirac equation constituted the very first physical theory
incorporating both special relativity and quantum mechanics. Dirac
introduced his equation in a series of two papers, based on
physical reasoning, yet with a strong appreciation of the
mathematical form. In fact, Dirac once said `physical laws should
have mathematical beauty' \cite{dal:bmfrs}. The natural setting of
the Dirac equation is in special relativity, on Minkowskian
space-time.

The Dirac equation as it usually appears in the physics literature
\cite{dir:qte1,sch:irqft} is a partial differential equation
\begin{equation}\label{eq:dirac-gamma}
\sum_{\mu=1}^4 \partial_{\mu} \gamma^{\mu} \psi = \psi
\end{equation}
where $\psi$ is a $4$ component complex vector, and each of the
$\gamma^{\mu}$ is a $4$ by $4$ matrix. Dirac also specified rules
for the transformation of $\psi$ under $\SL$. Under such a
transformation of $\psi$, while at the same time transforming
$\partial_{\mu}$ according to the corresponding element of $\LL$,
it is possible to show that the Dirac equation is invariant. Using
this presentation of the Dirac equation this is a very cumbersome
process. Soon, this invariance will be transparent. The gamma
matrices are chosen to satisfy the Clifford--Dirac equations,
\begin{equation}\label{eq:clifford-dirac}
\frac{1}{2}(\gamma^{\mu} \gamma^{\nu} + \gamma^{\nu} \gamma^{\mu})
=
\begin{cases} \phantom{-}0 & \text{if $\mu \neq
\nu$} \\
-1 & \text{if $\mu=\nu=1$} \\
\phantom{-}1 & \text{if $\mu=\nu = 2,3,$ or $4$}
\end{cases}
\end{equation}

There are no `standard' gamma matrices---depending on the context
and application some set of four matrices satisfying the
Clifford--Dirac equations are used. For the purpose of this work,
we will consider the Dirac equation written using the following
gamma matrices.
\begin{align*}
\gamma^1 = \gammazero, &&& \gamma^2 = \gammaone, \\
\gamma^3 = \gammatwo, &&& \gamma^4 = \gammathree.
\end{align*}

If we write the components of $\psi$ as \[\psi = \psivector\] then
the Dirac equation written out in full reads
\begin{multline}
\partial_1 \gammazero \smallpsivector + \partial_2 \gammaone \smallpsivector
+\\+\partial_3 \gammatwo \smallpsivector + \partial_4 \gammathree
\smallpsivector = \smallpsivector.
\end{multline}

The Dirac equation can be written in the language of the $\SL$
spinor algebra we have developed \cite{ber:t-csgr,pen:ss-t1}. Its
appearance becomes very simple, and the gamma matrices and the
Clifford--Dirac identities disappear entirely.

\begin{prop}
In flat Minkowski space, $M = \Real \times \Real^3$, the gamma
matrix Dirac equation is equivalent to the following pair of
spinor equations,
\begin{subequations}
\label{eq:dirac-eq}
\begin{eqnarray}
 \nabla^{\ai{A}}{}_{\ai{A'}} \phi_{\ai{A}} & = \chi_{\ai{A'}} \label{eq:dirac-eq1}\\
 \nabla_{\ai{A}}{}^{\ai{A'}} \chi_{\ai{A'}} & = \phi_{\ai{A}}. \label{eq:dirac-eq2}
\end{eqnarray}
\end{subequations}
\end{prop}
\begin{proof}
Since we work in Minkowskian coordinates, the covariant derivative
is just a partial derivative. Equations \eqref{eq:dirac-eq1} and
\eqref{eq:dirac-eq2} are equivalent to
\begin{align*}
\partial^{A A'} \phi_A & = \chi^{A'}\\
\partial^{A A'} \chi_{A'} & = \phi^A,
\end{align*}
and so
\begin{align*}
\partial^{A 0'} \phi_A & = \chi^{0'} = \chi_{1'}\\
\partial^{A 1'} \phi_A & = \chi^{1'} = - \chi_{0'}\\
\partial^{0 A'} \chi_{A'} & = \phi^0 = \phi_1\\
\partial^{1 A'} \chi_{A'} & = \phi^1 = - \phi_0,
\end{align*}
where we have used Equation \eqref{eq:manipulate-components}.
Next, we fulfill the summation of $A$ or $A'$, and write these
equations in matrix form.
\begin{align*}
\begin{pmatrix}
\partial^{00'} & \partial^{10'} & 0 & 0 \\
\partial^{01'} & \partial^{11'} & 0 & 0 \\
0 & 0 & \partial^{00'} & \partial^{01'} \\
0 & 0 & \partial^{10'} & \partial^{11'}
\end{pmatrix}
\psivector & = \begin{pmatrix} \chi_{1'} \\ - \chi_{0'} \\ \phi_1 \\
- \phi_0 \end{pmatrix}, \\
 \intertext{or, equivalently}
\begin{pmatrix}
0 & 0 & -\partial^{10'} & -\partial^{11'} \\
0 & 0 & \partial^{00'} & \partial^{01'} \\
-\partial^{01'} & -\partial^{11'} & 0 & 0 \\
\partial^{00'} & \partial^{10'} & 0 & 0
\end{pmatrix}
\psivector & = \psivector.
\end{align*}
Using Equation \eqref{eq:components-related} to rewrite the
partial derivative operators with tensor indices, we obtain
\begin{small}
\begin{equation}
\frac{1}{\sqrt{2}}
\begin{pmatrix}
0 & 0 & -\partial^2 + i \partial^3 & -\partial^1 + \partial^4 \\
0 & 0 & \partial^1 + \partial^4 & \partial^2 + i \partial^3 \\
-\partial^2 - i \partial^3 & -\partial^1 + \partial^4 & 0 & 0 \\
\partial^1 + \partial^4 & \partial^2 - i \partial^3 & 0 & 0
\end{pmatrix}
\psivector = \psivector.
\end{equation}
\end{small}
Finally, using $\partial^1 = \partial_1$ and $\partial^i = -
\partial_i$ for $i = 2,3,4$, we see that this agrees with the
explicitly written Dirac equation above.
\end{proof}

This result indicates that the spinor equations
\eqref{eq:dirac-eq} are an appropriate generalisation of the Dirac
equation. Using the framework of spinor structures for
pseudo-Riemannian manifolds, these spinor differential equations
describe the behaviour of Dirac particles on any $(1+3)$
dimensional Lorentzian manifold. That is, assuming a spinor
structure exists and a particular spinor structure has been
chosen, we have a natural extension of the Dirac equation to the
setting of general relativity.

With the formalism of the $\SL$ spinor algebra available, the
somewhat arbitrary gamma matrices are replaced by a very simple
set of differential equations. Similarly, the awkward
transformation laws of the original Dirac equation are avoided
entirely---the expressions in Equation \eqref{eq:dirac-eq} consist
solely of intrinsic \emph{geometric} objects.

\subsection{Implications of the choice of spinor
structure}\label{ssec:implications} The results of
\S\ref{sec:classifying-as-bundles} are all available in the
current context, and so any spinor structure is defined on the
trivial bundle $M \times \SL$. We have seen previously that the
choice of spinor structure is reflected in the spinor connection,
and this section discusses the `physical implications' of the
choice of spinor structure. Physicists have previously
investigated this idea in various ways
\cite{avi:lgivf,for:vpncs,ish:sffds-t}, with various degrees of
rigour!

For simplicity, we will consider a particularly straightforward
example. The example will demonstrate many of the theoretical
ideas discussed throughout the length of this thesis. Let $M =
\Real^3 \times S^1$, and give this the obvious metric tensor such
that the `$S^1$ direction' is spacelike. The orthonormal structure
is $P = M \times \LL$, and a simple connection form is defined by
$\omega_{(m,e)}(v,X) = X$, for $v \in T_m M$, and $X \in
\LAM{SO}{1,3}$.

We can easily calculate the fundamental group of $P$, as $\pi_1(P)
= \pi_1(\Real^3 \times S^1) \times \pi_1(\LL) = \Integer \times
\Integer_2$, and so spinor structures exist. More precisely, there
are two, corresponding to the two homomorphisms $\pi_1(M) =
\Integer \To \Integer_2 = \pi_1(\LL)$, $n \mapsto 0$ and $n
\mapsto n \pmod 2$. These spinor structures can both be
constructed on the trivial bundle $Q = M \times \SL$, as in
Proposition \ref{prop:trivial-bundles-realisable-maps}. Define
$u:Q \To P$ and $u_\zeta : Q \To P$ by
\[u(m,\tilde{g}) = (m,\rho(\tilde{g})) \quad \text{and} \quad
u_{\zeta}(m,\tilde{g}) = (m,\zeta(m) \rho(\tilde{g}))\] where
$\zeta$ is any smooth function $M \To \LL$ which induces the
nontrivial homomorphism $n \mapsto n \pmod 2$, for example
\[\zeta(t,x,y,\theta) =   \psmallmatrix{
 1&0&0&0\\
 0&\cos \theta & - \sin \theta & 0\\
 0&\sin \theta & \cos \theta & 0 \\
 0&0 & 0 & 1 \\
}.\] Next, we consider the resulting connections on $Q$. Firstly,
according to Proposition \ref{prop:inequivalent-connections}, the
connection obtained via $u$ is simply
$\hat{\omega}_{(m,\tilde{e})}(v,X) = X$, for $v \in T_m M$ and $X
\in \LAM{SL}{2,\Complex}$. The connection obtained via $u_\zeta$
is \[\hat{\omega}'_{(m,\tilde{e})}(v,X) = X + \rho_{e*}^{-1}
\zeta(m)^{-1} \zeta_* v.\] It is immediately clear that if $\zeta$
induces a nontrivial homomorphism, then $\zeta_* \neq 0$, and so
there is no possible choice of $\zeta$ so that these connections
are the same. Consider in particular the $\zeta$ defined above. If
$m=(t,x,y,\theta)$, $v=(\tau, w,z,\psi)\in T_m M$, then
\begin{align*}
\zeta_{m *} v & = \dat{t}{0}
 \psmallmatrix{
  1&0&0&0\\
  0&\cos (\theta+t\psi) & - \sin (\theta+t\psi) & 0\\
  0&\sin (\theta+t\psi) & \cos (\theta+t\psi) & 0 \\
  0&0 & 0 & 1 \\
 } \\
\intertext{and}
 \zeta(m)^{-1} \zeta_{m *} v & = \dat{t}{0}
  \psmallmatrix{
   1&0&0&0\\
   0&\cos \theta & - \sin \theta & 0\\
   0&\sin \theta & \cos \theta & 0 \\
   0&0 & 0 & 1 \\
  }^{-1}
  \psmallmatrix{
   1&0&0&0\\
   0&\cos (\theta+t\psi) & - \sin (\theta+t\psi) & 0\\
   0&\sin (\theta+t\psi) & \cos (\theta+t\psi) & 0 \\
   0&0 & 0 & 1 \\
  } \\
  & = \dat{t}{0} \psmallmatrix{
 1&0&0&0\\
 0&\cos t\psi & - \sin t\psi & 0\\
 0&\sin t\psi & \cos t\psi & 0 \\
 0&0 & 0 & 1 \\
 } = \psi
\psmallmatrix{
 1&0&0&0\\
 0&0 & - 1 & 0\\
 0&1 & 0 & 0 \\
 0&0 & 0 & 1 \\
 }.
\end{align*}
Further, using the derivative of Equation \eqref{eq:z-rotation} to
calculate $\rho_{* e}^{-1}$, we see
\[\rho_{* e}^{-1} \zeta(m)^{-1} \zeta_{m *} v = \frac{i \psi}{2}
\begin{pmatrix}
1 & 0 \\
0 & -1
\end{pmatrix} \in \LAM{sl}{2,\Complex}.\]

Next, we want to compare the Dirac equations corresponding to
these two connections. To do this, chose the obvious cross section
of $Q$, $\sigma(m) = (m,\tilde{e})$. Then
$\sigma^*\hat{\omega}(v)=\hat{\omega}(v,0) = 0$, and $\sigma^*
\hat{\omega}'(v) = \hat{\omega}'(v,0) = \frac{i \psi}{2}
\psmallmatrix{1 & 0 \\ 0 & -1} \in \LAM{sl}{2,\Complex}$.
According to \S\ref{sssec:local-representatives}, these local
representatives have the forms $\sigma^*\hat{\omega}
\leftrightarrow K_{\ai{a}}{}^{\ci{B}}{}_{\ci{C}}$ and
$\sigma^*\hat{\omega}' \leftrightarrow
L_{\ai{a}}{}^{\ci{B}}{}_{\ci{C}}$ in index notation. However, from
the above calculations we see that
$K_{\ai{a}}{}^{\ci{B}}{}_{\ci{C}} = 0$, and the only nonzero
components of $L_{\ai{a}}{}^{\ci{B}}{}_{\ci{C}}$ are
\begin{equation*}
L_4{}^0{}_0 = \frac{i}{2} \quad \text{ and } \quad L_4{}^1{}_1 =
-\frac{i}{2}.
\end{equation*}

The difference between the two connections
$\hat{\omega}'-\hat{\omega}$ then defines a tensor, according to
the prescription of \S\ref{sssec:difference-between-connections},
\[L_{\ai{a}}{}^{\ai{B}}{}_{\ai{C}}(m) = \left[\sigma(m),
L_{\ai{a}}{}^{\ci{B}}{}_{\ci{C}}(m)\right].\] Further, if
$\nabla_{\ai{a}}$ is the covariant derivative associated with
$\hat{\omega}$, and $\nabla'{}_{\ai{a}}$ is the covariant
derivative associated with $\hat{\omega'}$, then according to the
expression for the covariant derivative in Equation
\eqref{eq:covariant-derivative2} the difference between these
covariant derivatives acting on, say, $\phi_{\ai{C}}$ is given by
exactly
\[(\nabla'{}_{\ai{a}} - \nabla_{\ai{a}}) \phi_{\ai{C}} =
- L_{\ai{a}}{}^{\ai{B}}{}_{\ai{C}} \phi_{\ai{B}}.\]
%-%explain this properly?
Next, if we write the Dirac equation associated with the
connection $\hat{\omega}'$,
\begin{eqnarray*}
 \nabla'{}^{\ai{A}}{}_{\ai{A'}} \phi_{\ai{A}} & = \chi_{\ai{A'}}\\
 \nabla'{}_{\ai{A}}{}^{\ai{A'}} \chi_{\ai{A'}} & = \phi_{\ai{A}},
\end{eqnarray*}
we can re-express this as
\begin{eqnarray*}
 \nabla^{\ai{A}}{}_{\ai{A'}} \phi_{\ai{A}} - L^{\ai{A}}{}_{\ai{A'}}{}^{\ai{B}}{}_{\ai{A}} \phi_{\ai{B}} & = \chi_{\ai{A'}}\\
 \nabla_{\ai{A}}{}^{\ai{A'}} \chi_{\ai{A'}} - \cc{L}_{\ai{A}}{}^{\ai{A' B'}}{}_{\ai{A'}} \chi_{\ai{B'}}& =
 \phi_{\ai{A}}.
\end{eqnarray*}

This calculation shows that, in general, choosing a different
spinor structure modifies the Dirac equation by the addition of a
tensor term. In the particular example we are calculating with, we
can simplify this tensor. Using Equation
\eqref{eq:components-related}, the only nonzero components of
$L_{\ci{A A'}}{}^{\ci{B}}{}_{\ci{C}}$ are
\begin{align*}
 L_{0 0'}{}^{0}{}_{0} & = -\frac{i}{2\sqrt{2}} &&& L_{11'}{}^{0}{}_{0} & = \frac{i}{2\sqrt{2}}\\
 L_{0 0'}{}^{1}{}_{1} & = \frac{i}{2\sqrt{2}}  &&& L_{11'}{}^{1}{}_{1} & = -\frac{i}{2\sqrt{2}},
\end{align*}
and so applying Equation \eqref{eq:manipulate-components} and
contracting, $L^{\ci{A}}{}_{\ci{A'}}{}^{\ci{B}}{}_{\ci{A}} =
D_{\ci{A'}}{}^{\ci{B}}$, where
\begin{align*}
D_{0'}{}^{0} = D_{1'}{}^{1} = 0 \quad \text{ and } \quad
D_{1'}{}^{0} = - D_{0'}{}^{1} = \frac{i}{2\sqrt{2}}.
\end{align*}
With this tensor, the Dirac equation for the connection
$\hat{\omega}'$ reads
\begin{eqnarray*}
 \nabla^{\ai{A}}{}_{\ai{A'}} \phi_{\ai{A}} - D_{\ai{A'}}{}^{\ai{A}} \phi_{\ai{A}} & = \chi_{\ai{A'}}\\
 \nabla_{\ai{A}}{}^{\ai{A'}} \chi_{\ai{A'}} - \cc{D}_{\ai{A}}{}^{\ai{A'}} \chi_{\ai{A'}}& =
 \phi_{\ai{A}}.
\end{eqnarray*}
This represents only the very start of an analysis of the Dirac
equation for different spinor structures. One could for example
write down the `plane wave solutions' on the manifold $\Real^3
\times S^1$ for the two different spinor connections. At the very
least, we have shown that one can not unambiguously ignore the
choice of spinor structures available when setting up the
mathematical framework for the Dirac equation on topologically
nontrivial manifolds.

\newpage
\part*{Conclusion}

We have discussed how Riemannian geometry, including the theory of
covariant derivatives and tensor calculus, fits into the general
setting of principal fibre bundles. As it turns out, the theory of
spinor structures for Riemannian geometry extends naturally to the
general setting, and a large part of the work here has been in
establishing the appropriate classifications for abstract spinor
structures. With a constructive classification in hand, we have
investigated several questions about spinor structures:
\begin{itemize}
\item What happens if we reduce or enlarge the structure group?
\item Are the underlying principal fibre bundles all the same?
\item How many different spinor connections are there?
\end{itemize}
We have also given an explicit description of an important
physical application of spinor structures---describing the
behaviour of relativistic particles in quantum mechanics using the
Dirac equation. The questions above, and their answers, shed light
on the interaction of topology and the physics of the Dirac
equation.

On several topics in this thesis we have certainly not said the
last word. One avenue for further work would be to prove or refute
the conjecture in \S \ref{sec:classifying-as-bundles}, classifying
the principal fibre bundles underlying the various spinor
structures. If it were true, then it would be interesting to find
a direct construction of the class of possible bundles. The other
obvious direction is in continuing the analysis of the Dirac
equation for different spinor structures. In particular, it may be
possible to prove quite generally that the spinor connections are
always inequivalent. Building on the mathematical foundation
provided here, a detailed physical picture describing the
differences between the solutions of the Dirac equation for each
of the inequivalent spinor connection needs to be developed. The
results here suggest that in giving a mathematical description of
the physical universe, to begin we must describe the topology and
metric structure, and \emph{also make a choice between the
available spinor structures}, because this global topological
choice has physical implications.

\vfill \begin{center} \Huge $\tau \grave{o} \,\,\,\tau
\acute{\epsilon} \lambda o \varsigma$
\end{center}

\newpage
%\part*{Appendices}
\stepcounter{part}

\addtocontents{toc}{\vspace{0.5cm}}

\appendix

\section{The fundamental group of $SO_0(p,q)$}
\label{sec:fundamental-groups-SO}

Firstly, if $G$ is any Lie group, then $\pi_1(G)$ is commutative.

\begin{lem}\label{lem:pi1G-commutative}
Suppose $a:\I \To G$ and $b:\I \To G$ are loops in a Lie group
$G$. Then $[{a} \cat {b}] = [a b] = [{b} \cat {a}]$, where $r s$
is the loop $t \mapsto a(t) b(t)$.
\end{lem}
\begin{proof}
Consider the homotopy $H:\I \times \I \To G$
\begin{equation}
    H(s,t)=\left\{
        \begin{array}{ll}
             b(\frac{2t}{1+s}) & \textrm{if $ 0 \leq t \leq \frac{1-s}{2}$,} \\
             a(2 \frac{t-1}{s+1} + 1) b(\frac{2t}{1+s}) & \textrm{if $\frac{1-s}{2} < t \leq \frac{1+s}{2}$,} \\
             a(2 \frac{t-1}{s+1} + 1) & \textrm{if $\frac{1+s}{2} < t \leq 1$} \\
        \end{array}
        \right.
\end{equation}
This proves $[{a} \cat {b}] = [a b]$. A similar homotopy
establishes the other half.
\end{proof}

Note the resemblance of this result to Lemma
\ref{lem:homotopy-centre}, which is essentially a generalisation.

We now give a list of the fundamental groups for all the special
orthogonal groups in each dimension. The argument uses the
explicit description of the covering map $\rho : \SL \To \LL$ from
\S\ref{ssec:covering-map}. The discussion will rely on knowledge
of covering space theory and the long exact sequence of homotopy
groups for fibrations.

\subsection{The fundamental group of $SO(n)$}
\label{ssec:fundamental-group-SO(n)} We begin with the trivial
cases. When $n=1$, the special orthogonal group is trivial. When
$n=2$, it is just the circle group, so $\pi_1(SO(2)) = \Integer$.

Next, we deal with $n=3$ using two corollaries of Proposition
\ref{prop:coveringmap}.

\begin{cor}
The restriction of the covering map $\rho$ in Proposition
\ref{prop:coveringmap} to $SU(2)$ is a $2$ to $1$ covering
homomorphism from $SU(2)$ to $SO(3)$.
\end{cor}
\begin{proof}
This follows immediately from the argument given in the proof of
Proposition \ref{prop:coveringmap}. As seen there, elements of
$SU(2)$ are `trace preserving', and so fix the $t$ component. Thus
$\rho$ maps $SU(2)$ into $SO(3)$, and this restriction is clearly
onto, because the pre-images of the rotations, as exhibited, all
lie in $SU(2)$. Finally, the kernel of $\rho$ lies in $SU(2)$, and
so the restricted map is also $2$ to $1$.
\end{proof}

\begin{cor}
The fundamental group of $SO(3)$ is $\Integer_2$.
\end{cor}
\begin{proof}
Since $SU(2)$ is topologically $S^3$ it is simply connected. Thus
it is the universal covering space for $SO(3)$. Covering space
theory for Lie groups \cite[\S16.30.2]{die:ta3} (and see
\S\ref{ssec:covering-lie-groups}) states that the fundamental
group of the base space is the kernel of the universal covering
map. Thus $\pi_1(SO(3)) \cong \ker(\rho) \cong \Integer_2$.
\end{proof}
The explicit formulas given in \S\ref{ssec:covering-map} show that
the homotopy class of any $2\pi$ rotation is the generator of the
fundamental group of $SO(3)$.

We will now offer an inductive argument that $\pi_1(SO(n)) =
\Integer_2$ for any $n \geq 3$, and that the inclusion of $SO(3)$
into $SO(n)$, acting on the first $3$ coordinates, induces an
isomorphism $\pi_1(SO(3)) \cong \pi_1(SO(n))$. Thus the generator
is a $2 \pi$ rotation.

The group $SO(n+1)$ acts transitively on the sphere $S^n \subset
\Real^{n+1}$. The stabiliser of the point $z=(0,\ldots,0,1) \in
S^n$ is $SO(n)$, acting on the first $n$ coordinates of
$\Real^{n+1}$. We write $i:SO(n) \To SO(n+1)$ for this inclusion.
The group $SO(n)$ is a closed subgroup of the Lie group $SO(n+1)$,
and so a Lie subgroup. We can thus apply the result of \cite[\S
7.5]{ste:tfb} to see that
\[\pfbundle{SO(n)}{SO(n+1)}{p}{S^n}\]
is a principal fibre bundle. Here $p$ can be thought of as either
the action of $SO(n+1)$ on the point $z$, or the quotient map of
$SO(n)$ acting on $SO(n+1)$.

Next, we write down the long exact sequence of homotopy groups for
a principal fibre bundle \cite[II \S17]{ste:tfb}, which in part
reads
\[\dotsb \To \pi_{m+1}(S^n) \To \pi_m(SO(n)) \xrightarrow{i_*}
\pi_m(SO(n+1)) \xrightarrow{p_*} \pi_m(S^n) \To \dotsb\] Now, if
$0 < m < n-1$, $\pi_{m+1}(S^n) = \pi_m(S^n) = 0$, and so the
section of the exact sequence above becomes
\[0 \To \pi_m(SO(n)) \xrightarrow{i_*}
\pi_m(SO(n+1)) \To 0.\] Thus $i_* : \pi_m(SO(n)) \To
\pi_m(SO(n+1))$ is an isomorphism, and in particular for $n>2$
\[\pi_1(SO(n)) \cong \pi_1(SO(n+1)).\] Finally, by induction,
$\pi_1(SO(n)) \cong \Integer_2$ for all $n \geq 3$.

\subsection{The fundamental group of $SO_0(p,q)$}\label{ssec:fundamental-group-SO(p,q)}
We refer to \cite[Proposition 1.122]{kna:lgbi}, which proves that
there is a homeomorphism $S(O(p) \times O(q)) \times \LA{p} \To
SO(p,q)$, where $S(O(p) \times O(q))$ denotes the subgroup of
$O(p) \times O(q)$ of matrices with unit determinant, and $\LA{p}$
is the linear space of Hermitian matrices in $\LAM{so}{p,q}$.
Restricting this map to the connected components of the
identities, we obtain a homeomorphism $SO(p) \times SO(q) \times
\LA{p} \To SO_0(p,q)$. Since $\LA{p}$ is necessarily homotopically
trivial, we obtain a homotopy equivalence between $SO(p) \times
SO(q)$ and $SO_0(p,q)$. In turn this gives an isomorphism of the
fundamental groups, and so using the results of
\S\ref{ssec:fundamental-group-SO(n)} we find
\begin{equation*}
\pi_1(SO_0(p,q)) \cong \left\{
\begin{tabular}{c|ccc}
     &   $p=1$  & $p=2$ & $p \geq 3$ \\
 \hline
 $q=1$ &   $<e>$  & $\Integer_{\phantom{2}}$ & $\Integer_2$ \\
 $q=2$ & $\Integer_{\phantom{2}}$ & $\Integer \times \Integer_{\phantom{2}}$ & $\Integer_2 \times
 \Integer_{\phantom{2}}$ \\
 $q \geq 3$ & $\Integer_2$ & $\Integer \times \Integer_2$ & $\Integer_2
 \times \Integer_2$
\end{tabular}
  \right.
\end{equation*}

\subsection{The universal cover of $\LL$ is
$\SL$} \label{ssec:fundamental-group-SO_0(1,3)}We have established
in the previous section that $\pi_1(\LL) = \Integer_2$, and so to
describe the universal covering group we need only find some two
fold covering group. This is of course given by the covering map
$\rho: \SL \To \LL$ of \S\ref{ssec:covering-map}.

\section{Maximal compact subgroups}
\label{sec:maximal-compact-subgroups}

\begin{prop}\label{prop:maximal-compact-subgroups}
Let $\LA{p}$ be the real vector space of Hermitian matrices in
$\LAM{sl}{n,\Complex}$ and $\LA{r}$ be the vector space of
symmetric matrices in $\LAM{sl}{n,\Real}$. There is
\begin{enumerate}
\item a homeomorphism $SU(n) \times \LA{p} \To
SL(n,\Complex)$ and
\item a homeomorphism $SO(n) \times \LA{r} \To
SL(n,\Real)$.
\end{enumerate}
\end{prop}
\begin{proof}
See Proposition 1.122 in \cite{kna:lgbi}. Related results can be
achieved directly by methods of linear algebra, as in
\cite[\S8.4]{hof:la}, or quite generally by means of the global
Iwasawa decomposition \cite[VI \S3]{hel:dglgss}.
\end{proof}

Thus $SL(n,\Complex)$ is homotopy equivalent to $SU(n)$, and
$SL(n,\Real)$ is homotopy equivalent to $SO(n)$. We say that
$SU(n)$ and $SO(n)$ are the respective \emph{maximal compact
subgroups}. Further, $GL^+(n,\Real) \cong SL(n,\Real) \times
\Real^+$, and so also $GL^+(n,\Real)$ is homotopy equivalent to
$SO(n)$. This equivalence is given by the inclusion $\iota: SO(n)
\To GL^+(n,\Real)$. In particular, this inclusion induces an
isomorphism of the fundamental groups, $\iota_* :\pi_1(SO(n)) \To
\pi_1(GL^+(n,\Real))$.

Using the result of \S\ref{ssec:fundamental-group-SO(n)}, we have
now proved that $\pi_1(GL^+(n,\Real)) = \Integer_2$, for all $n
\geq 3$.

%\section{Clifford algebras and the
%representation theory of $\tilde{SO}_0(p,q)$}
%
%something about $\text{Spin}_0(p,q)$, and how
%representations push down?

\section{Technical results}

\subsection{Proof of Proposition
\ref{prop:connections-exist}}\label{ssec:proof-of-proposition-connections-exist}
We now give the proof that every principal fibre bundle allows a
connection.

It is for the purposes of this construction that we require the
base manifold to be paracompact. This is not too burdensome, and
is nearly always included in the definition of a smooth manifold.
\begin{defn*}
A manifold is said to be \defnemph{paracompact} if every open
covering of the manifold has a locally finite refinement \cite[p.
16]{cho:amp}.
\end{defn*}
Thus if $\left( U_\alpha \right)_{\alpha \in \mathcal{A}}$ is an
open covering of $M$, there is a covering $\left( V_\alpha
\right)_{\alpha \in \mathcal{A}}$ so $V_\alpha \subset U_\alpha$
for each $\alpha \in \mathcal{A}$, and each point on $M$ is
contained in only finitely many $V_\alpha$.

If the manifold is connected this is equivalent to there being a
countable basis for the topology \cite[Appendix 2]{kob:fdg1}. On
such manifolds we can construct partitions of unity.
\begin{defn*}
Given an open covering $\left( U_\alpha \right)_{\alpha \in
\mathcal{A}}$ of $M$, a \defnemph{partition of unity} subordinate
to this covering is a collection of smooth functions $\left(
f_\alpha \right)_{\alpha \in \mathcal{A}}$ on $M$ so
\begin{enumerate}
\item $0 \leq f_\alpha \leq 1$ for each $\alpha \in \mathcal{A}$,
\item the support of $f_\alpha$, that is, the closure of $\setc{ m
\in M}{f_\alpha(m) \neq 0}$, is contained in $U_\alpha$ for each
$\alpha \in \mathcal{A}$, and
\item $\sum_{\alpha \in \mathcal{A}} f_\alpha = 1$.
\end{enumerate}
\end{defn*}

\begin{lem}
Let $\left( U_\alpha \right)_{\alpha \in \mathcal{A}}$ be a
locally finite open covering of $M$ so that each $U_\alpha$ is
relatively compact. Then there exists a partition of unity $\left(
f_\alpha \right)_{\alpha \in \mathcal{A}}$ subordinate to this
covering.
\end{lem}
\begin{proof}
See \cite[Appendix 3]{kob:fdg1}.
\end{proof}

\begin{lem}
\label{lem:fancy-open-covering} Given a $G$ bundle $\xi =
\pfbundle{G}{P}{\pi}{M}$ on a paracompact manifold $M$, there
exists a locally finite open covering $\left( U_\alpha
\right)_{\alpha \in \mathcal{A}}$ of $M$ by local trivialisations
$(U_\alpha, \varphi_\alpha)$ so that each $U_\alpha$ is relatively
compact. (Thus for each $\alpha \in \mathcal{A}$,
$\varphi_\alpha:\pi^{-1}(U_\alpha) \To U_\alpha \times G$ is a
diffeomorphism, and $\closure{U_\alpha}$ is compact.)
\end{lem}
\begin{proof}
Firstly, associate with each point $m \in M$ an open set $m \in
V_m \subset M$ such that $G$ is trivial over $V_m$. Next, choose
for each point $m$ an coordinate chart $(W'_m, \psi'_m)$, with $m
\in W'_m$. Since $\psi'_m(W'_m)$ is an open set in $\Real^n$,
there is an $\eps_m$ so $B_{\eps_m}(\psi'_m(m))$, the open ball of
radius $\eps_m$ about $\psi'_m(m)$, is contained in
$\psi'_m(W'_m)$. Next, let $W_m =
{\psi'_m}^{-1}(B_{\frac{\eps_m}{2}}(\psi'_m(m)))$, and $\psi_m =
\restrict{\psi'_m}{W_m}$. Since $W_m$ is homeomorphic to
$B_{\frac{\eps_m}{2}}(\psi'_m(m)) \subset B_{\eps_m}(\psi'_m(m))$,
it is relatively compact. Thus the collection $(W_m,\psi_m)_{m\in
M}$ is a covering of $M$ by relatively compact coordinate charts.

Let $U_m = V_m \cap W_m$. The bundle is locally trivial over these
sets, which are also relatively compact and coordinate charts. Any
open subset of such a set also satisfies these properties. These
sets form a covering of $M$, and so applying our assumption of
paracompactness, we obtain a locally finite open covering $\left(
U_\alpha \right)_{\alpha \in \mathcal{A}}$ which is a refinement
of the covering $\left( U_m \right)_{m \in M}$, and so consists of
relatively compact local trivialisations.
\end{proof}

\begin{proof}[Proof of the Proposition]
The proof here follows that in \cite{cho:amp}. A similar proof
appears in \cite{die:ta4}. Let $\left( U_\alpha \right)_{\alpha
\in \mathcal{A}}$ be an open covering of $M$ as described in Lemma
\ref{lem:fancy-open-covering}, and let $\left( f_\alpha
\right)_{\alpha \in \mathcal{A}}$ be a partition of unity
subordinate to this open covering. We will define a connection on
$P$ using the Lie algebra valued form description, defining a
connection on $\pi^{-1}(U_\alpha)$ for each $\alpha \in
\mathcal{A}$ and patching these together using the partition of
unity.

We now define a connection form on the each of the sub-bundles
$\pi^{-1}(U_\alpha)$. Put simply, we choose the obvious
\emph{flat} connection relative to the local trivialisation
$\varphi_\alpha$. Given $\bdelt{p} \in \pi^{-1}(U_\alpha)$,
$\varphi_\alpha(\bdelt{p})=(m,g)$, say. If $u \in T_\bdelt{p} P$,
then $\varphi_{\alpha *} u \in T_{(m,g)}U_\alpha \times G$. This
tangent space splits, since $T_{(m,g)}U_\alpha \times G = T_m
U_\alpha \times T_g G$. Thus we can always write $u$ as $u = v +
w$, where $\varphi_{\alpha *} v \in T_m U_\alpha$, and
$\varphi_{\alpha *} w \in T_g G$. Moreover, given this
decomposition, $g_* u = g_* v + g_* w$, and this represents a
similar decomposition, since the action of $g$ commutes with the
trivialisation. We then define $\omega^\alpha_\bdelt{p}(u) =
\psi_{\bdelt{p} *}w$. The map $\psi_{\bdelt{p}}$ is defined as
before in \S\ref{ssec:connection-forms}, by
$\psi_\bdelt{p}(\bdelt{p'})=\tau(\bdelt{p}, \bdelt{p'})$. The
first property we require of a connection form, that it maps
vertical vectors into the Lie algebra according to
$\psi_{\bdelt{p} *}$, is satisfied since the vertical vectors $u$
are those such that $v = 0$, and so this definition gives
$\omega^\alpha_\bdelt{p}(u) = \psi_{\bdelt{p} *}u$. Next, we
calculate
\begin{eqnarray*}
(\psi_{\bdelt{p} g} \compose g)(\bdelt{p'}) & = & \psi_{\bdelt{p} g}(\bdelt{p'} g) \\
                                         & = & \tau(\bdelt{p} g, \bdelt{p'} g) \\
                                         & = & g^{-1} \tau(\bdelt{p}, \bdelt{p'}) g \\
                                         & = & g^{-1} \psi_\bdelt{p}(\bdelt{p'}) g,
\end{eqnarray*}
and so $(\psi_{\bdelt{p} g} \compose g)_* = \Ad(g^{-1})
\psi_{\bdelt{p} *}$. Then
\begin{eqnarray*}
\omega^\alpha_{\bdelt{p} g}(g_* u) & = & \psi_{\bdelt{p} g *}g_* w \\
                           & = & \Ad(g^{-1}) \psi_{\bdelt{p} *} w \\
                           & = & \Ad(g^{-1}) \omega^\alpha_\bdelt{p}(u),
\end{eqnarray*}
and so $\omega^\alpha$ is in fact a connection form on
$\pi^{-1}(U_\alpha)$.

Finally, we obtain a connection form on the entire bundle simply
by writing $\omega = \sum_{\alpha \in \mathcal{A}} f_\alpha
\omega^\alpha$.
\end{proof}

\subsection{Extending a connection on a reduced
bundle}\label{ssec:extending-connections} In this section we show
that if $\xi = \pfbundle{H}{P}{\pi_P}{M}$ is a reduction of $\eta
= \pfbundle{G}{Q}{\pi_{Q}}{M}$, with reduction map $r:P \To Q$,
and $\omega$ is a connection form on $\xi$, there is a
straightforward prescription for extending $\omega$ to a
connection form $\tilde{\omega}$ on $\eta$. The proof is very
straightforward---after giving a prescription for the extension,
we check that it is well defined, and gives a form satisfying the
connection axioms of \S \ref{ssec:connection-forms}. First, we
need a preliminary result.

\begin{lem}\label{lem:horizontal-lifting-elevator-property}
The horizontal lifting map has a related `elevator property',
\begin{equation}
\sigma_{\bdelt{p} g} = g_* \sigma_{\bdelt{p}}.
\end{equation}
\end{lem}
\begin{proof}
Firstly, $\omega_{\bdelt{p}g}(u) = \omega_{\bdelt{p}g}(g_*
{g^{-1}}_* u) = \Ad(g^{-1}) \omega_{\bdelt{p}}({g^{-1}}_* u)$, by
the elevator property for $\omega$, and so $\ker \omega_{\bdelt{p}
g} = g_* \ker \omega_{\bdelt{p}}$. Next, since $\pi(\bdelt{p} g) =
\pi(\bdelt{p})$, we have $\pi_* g_* = \pi_*$. If
$\sigma_{\bdelt{p}}(u) = y \in \ker \omega_{\bdelt{p}}$, then
$\pi_* y = u$ and if $\sigma_{\bdelt{pg }}(u) = v \in g_* \ker
\omega_{\bdelt{p}}$ then $\pi_*v = u$. Certainly $g_* y \in g_*
\ker \omega_{\bdelt{p}}$, and $\pi_* g_* y = \pi_* y = u$, so
$\sigma_{\bdelt{p}g}(u) = g_* y = g_* \sigma_{\bdelt{p}}(u)$, as
required.
\end{proof}

We now define $\tilde{\omega}$ on $r(P) \subset Q$. This
definition relies on the horizontal lifting map for $\omega$,
defined in \S\ref{ssec:horizontal-lifting-map}. Let
\begin{equation}\label{eq:omega-restricted}
\tilde{\omega}_{r(\bdelt{p})}(u) = \tilde{\psi}_{r(\bdelt{p}) *}(u
- r_* \sigma_{\bdelt{p}} \pi_{Q *} u).
\end{equation}
The motivation for this definition comes from Lemma
\ref{lem:connection-form-from-horizontal-lifting-map}.

To extend $\tilde{\omega}$ to all of $Q$, we note that any
$\bdelt{q} \in Q$ can be written in the form $\bdelt{q} =
r(\bdelt{p}) g$ for some $\bdelt{p} \in P$ and $g \in G$. We then
define
\begin{equation}\label{eq:omega-full}
\tilde{\omega}_{\bdelt{q}}(u) = \Ad(g^{-1})
\tilde{\omega}_{r(\bdelt{p})}({g^{-1}}_* u).
\end{equation}

\begin{prop}
This prescription for $\tilde{\omega}$ is well defined, and gives
a connection for on $Q$.
\end{prop}
\begin{proof}
Suppose $\bdelt{q}$ is written in two ways, as $\bdelt{q} =
r(\bdelt{p}) g$ and $\bdelt{q} = r(\bdelt{p'}) g'$, so $\bdelt{p'}
= \bdelt{p} h$ and $g' = h^{-1} g$ for some $h \in H$. Then
\begin{align}
 \tilde{\omega}_{\bdelt{q}}(u) & = \tilde{\omega}_{r(\bdelt{p'}) g'}(u) && \nonumber\\
    & = \Ad({g'}^{-1}) \tilde{\omega}_{r(\bdelt{p'})}({{g'}^{-1}}_* u) && \text{by \eqref{eq:omega-full}}\nonumber\\
    & = \Ad(g^{-1} h) \tilde{\omega}_{r(\bdelt{p}) h}(h_* {g^{-1}}_* u) && \nonumber\\
    & = \Ad(g^{-1} h) \tilde{\psi}_{r(\bdelt{p})h*}(h_* {g^{-1}}_*u - r_* \sigma_{\bdelt{p} h} \pi_{Q *} h_* {g^{-1}}_* u) && \text{by \eqref{eq:omega-restricted}}\nonumber\\
    & = \Ad(g^{-1} h) \tilde{\psi}_{r(\bdelt{p})h*}(h_* {g^{-1}}_*u - r_* \sigma_{\bdelt{p} h} \pi_{Q *} {g^{-1}}_* u) && \nonumber\\
    & = \Ad(g^{-1} h) \tilde{\psi}_{r(\bdelt{p})h*}(h_* {g^{-1}}_*u - h_* r_* \sigma_{\bdelt{p}} \pi_{Q *} {g^{-1}}_* u) && \text{by Lemma \ref{lem:horizontal-lifting-elevator-property}}\nonumber\\
    & = \Ad(g^{-1} h) \tilde{\psi}_{r(\bdelt{p})h*}h_*({g^{-1}}_*u - r_* \sigma_{\bdelt{p}} \pi_{Q *} {g^{-1}}_* u).
    && \label{eq:omega-consistency}
\end{align}
Now $\tilde{\psi}_\bdelt{q}(\bdelt{q'}) =
\tau(\bdelt{q},\bdelt{q'})$, and so $(\tilde{\psi}_{\bdelt{q}
h}\compose h)(\bdelt{q'}) = \tau(\bdelt{q}h,\bdelt{q'}h) =
h^{-1}\tau(\bdelt{q},\bdelt{q'})h$. Thus
\begin{equation}\label{eq:psi-adjoint-relation}
\tilde{\psi}_{r(\bdelt{p})h*}h_* = \Ad(h^{-1})
\tilde{\psi}_{r(\bdelt{p}) *}.
\end{equation}
This holds also for any $g \in G$. Applying this to Equation
\eqref{eq:omega-consistency}, we obtain that
\begin{eqnarray*}
 \tilde{\omega}_{r(\bdelt{p'}) g'}(u)
  & = & \Ad(g^{-1}) \tilde{\psi}_{r(\bdelt{p}) *} ({g^{-1}}_*u - r_* \sigma_{\bdelt{p}} \pi_{Q *} {g^{-1}}_* u)\\
  & = & \Ad(g^{-1}) \tilde{\omega}_{r(\bdelt{p})}({g^{-1}}_* u).
\end{eqnarray*}
Thus the value of $\tilde{\omega}_{\bdelt{q}}(u)$ is independent
of the particular presentation $\bdelt{q}=r(\bdelt{p}) g$ chosen.
The definition of $\tilde{\omega}$ in Equation
\eqref{eq:omega-full} guarantees that the elevator property is
satisfied.

Finally, we to check that vertical vectors are mapped into the Lie
algebra according to Definition \ref{defn:connection-form}. If $u$
is vertical, so $\pi_{Q *} u = 0$, then $\pi_{Q *} {g^{-1}}_* u =
0$ for every $g \in G$. Thus
\begin{align*}
\tilde{\omega}_{\bdelt{q}}(u) & = \Ad(g^{-1})
    \tilde{\omega}_{r(\bdelt{p})}({g^{-1}}_* u) \\
    & = \Ad(g^{-1}) \tilde{\psi}_{r(\bdelt{p}) *}({g^{-1}}_* u - r_* \sigma_{\bdelt{p}} \pi_{Q *} {g^{-1}}_* u) && \text{by \eqref{eq:omega-restricted}}\\
    & = \Ad(g^{-1}) \tilde{\psi}_{r(\bdelt{p}) *} {g^{-1}}_* u \\
    & = \Ad(g^{-1}) \tilde{\psi}_{r(\bdelt{p})g g^{-1} *} {g^{-1}}_* u \\
    & = \Ad(g^{-1}) \Ad(g) \tilde{\psi}_{r(\bdelt{p})g *} u && \text{applying \eqref{eq:psi-adjoint-relation}}\\
    & = \tilde{\psi}_{r(\bdelt{p})g *} u,
\end{align*}
as required.
\end{proof}

Using Lemma \ref{lem:connection-form-from-horizontal-lifting-map}
and the definition of $\tilde{\omega}$ at points $r(\bdelt{p})$ we
see that $\sigma_{r(\bdelt{p})} = r_* \sigma_{\bdelt{p}}$. This
means that the parallel transports for the two connections agree,
in the sense that the parallel transport of the point
$r(\bdelt{p}) \in Q$ along $\alpha : \I \To M$ is exactly $r$
composed with the parallel transport of $\bdelt{p} \in P$ along
$\alpha$.
%-%prove this more carefully? as in Proposition \ref{prop:spinor-connection-parallel-transport}.

For example, we can use this fact, that connections on reduced
bundles can be extended, to extend a connection on an orthonormal
bundle to a connection on the full frame bundle.

\subsection{Proof of Lemma \ref{lem:homotopy-centre}}\label{ssec:proof-lem:homotopy-centre}
\begin{figure}[!hbp]
\begin{center}
\includegraphics[width=0.9\textwidth]{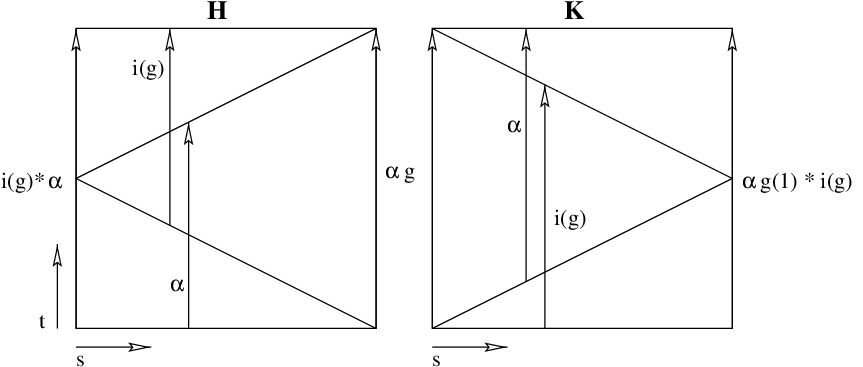}
\end{center}
\caption{Schematic indication of the construction of the
homotopies $H$ and $K$.\label{fig:homotopy-construction}}
\end{figure}

\begin{proof}
We define two homotopies, $H,K:\I \times \I \To P$, as indicated
in Figure \ref{fig:homotopy-construction}. Let
\[
    H(s,t)=\left\{
        \begin{array}{ll}
             \alpha(\frac{2t}{1+s}) & \textrm{if $ 0 \leq t \leq \frac{1-s}{2}$,} \\
             \alpha(\frac{2t}{1+s}) g(2 \frac{t-1}{s+1} + 1) & \textrm{if $\frac{1-s}{2} < t \leq \frac{1+s}{2}$,} \\
             \bdelt{p_0} g(2 \frac{t-1}{s+1} + 1) & \textrm{if $\frac{1+s}{2} < t \leq 1$} \\
        \end{array}
        \right.
        ,
\]
and
\[
    K(s,t)=\left\{
        \begin{array}{ll}
             \bdelt{p_0} g(\frac{2t}{2-s}) & \textrm{if $ 0 \leq t \leq \frac{s}{2}$,} \\
             \alpha(2 \frac{t-1}{2-s} + 1) g(\frac{2t}{2-s}) & \textrm{if $\frac{s}{2} < t \leq 1-\frac{s}{2}$,} \\
             \alpha(2 \frac{t-1}{2-s} + 1) g(1)& \textrm{if $1-\frac{s}{2} < t \leq 1$} \\
        \end{array}
        \right.
        .
\]
It is easy to see that these piecewise definitions give continuous
maps. Further,
\begin{align*}
    H(0,t) & = \left\{
        \begin{array}{ll}
             \alpha(2t) & \textrm{if $0 \leq t \leq \frac{1}{2}$} \\
             \bdelt{p_0} g(2t-1) & \textrm{if $\frac{1}{2} < t \leq 1$} \\
        \end{array}
        \right. \\
            & = (i(g) \cat \alpha)(t),
\end{align*}
$H(1,t) = \alpha(t) g(t) = K(0,t)$, and
\begin{align*}
    K(1,t) & = \left\{
        \begin{array}{ll}
             \bdelt{p_0} g(2t)& \textrm{if $0 \leq t \leq \frac{1}{2}$} \\
             \alpha(2t-1) g(1)& \textrm{if $\frac{1}{2} < t \leq 1$} \\
        \end{array}
        \right. \\
            & = ({\alpha g(1)} \cat {i(g)})(t).
\end{align*}
Thus $K$ establishes the first homotopy, and in the case $g \in
\Omega P$, $\alpha \in \Omega P$, $g(1)=e$ so $H$ and $K$ give the
required homotopies for the second part.
\end{proof}

\addtocontents{toc}{\vspace{0.5cm}}
\

% ----------------------------------------------------------------
\newpage
\bibliographystyle{amsplain}
\bibliography{bibliography}
\end{document}